\documentclass[11pt]{article}

\usepackage{booktabs}
\usepackage[english]{babel}
\usepackage{amsmath,amssymb,amsbsy,amstext, amsthm, simplewick}
\usepackage{hyperref}
\hypersetup{
  colorlinks   = true, 
  urlcolor     = blue, 
  linkcolor    = black, 
  citecolor   = black 
}
\usepackage{graphicx}
\usepackage{amsfonts}
\usepackage{amssymb}
\usepackage[small]{caption}
\usepackage{upgreek}
  \usepackage{framed}
\usepackage{enumitem}
\usepackage{dsfont}
\usepackage{bm}

\usepackage[numbers,sort&compress]{natbib}
\let\OLDthebibliography\thebibliography
\renewcommand\thebibliography[1]{
  \OLDthebibliography{#1}
  \setlength{\parskip}{4pt}
  \setlength{\itemsep}{0pt plus 0.3ex}
}

\usepackage{wrapfig}
\usepackage{subfig}
\usepackage[makeroom]{cancel}
\usepackage{soul}
\usepackage{bbold}
\usepackage{lipsum}
\usepackage{mdframed}
\usepackage{mathtools}
\usepackage[dvipsnames]{xcolor}
\usepackage[export]{adjustbox}

\usepackage{caption}
\usepackage{floatrow}
\allowdisplaybreaks

\makeatletter
\g@addto@macro\bfseries{\boldmath}
\makeatother

\newcommand\refeqq[1]{Eq.~(\ref{#1})}
\newcommand\refeqs[1]{Eqs.~(\ref{#1})}
\newcommand\refta[1]{Tab.~\ref{#1}}
\newcommand\refse[1]{Sec.~\ref{#1}}
\newcommand\refses[1]{Secs.~\ref{#1}}
\newcommand\citere[1]{Ref.~\cite{#1}}
\newcommand\citeres[1]{Refs.~\cite{#1}}
\newcommand\refap[1]{App.~\ref{#1}}
\def\reffi#1{\mbox{Fig.~\ref{#1}}}

\newcommand{\gev}{\ \mathrm{GeV}}
\newcommand{\tev}{\ \mathrm{TeV}}

\definecolor{lightgreen}{cmyk}{0.2, 0, 0.2, 0.2}
\definecolor{lightgray}{cmyk}{0.1,0.2,0,0.1}
\definecolor{lightgray2}{cmyk}{0.1,0.1,0,0.1}

\makeatletter
\newlength{\apb@width}
\newcommand{\autoparbox}[2][c]{\settowidth{\apb@width}{#2}\parbox[#1]{\apb@width}{#2}}

\newcommand{\Cen}[2]{%
  \ifmeasuring@
    #2%
  \else
    \makebox[\ifcase\expandafter #1\maxcolumn@widths\fi]{$\displaystyle#2$}%
  \fi
}
\makeatother

\newcommand{\beq}{\begin{equation}\begin{aligned}}
\newcommand{\eeq}{\end{aligned}\end{equation}}

\numberwithin{equation}{section}

\def\beq{\begin{equation}}
\def\eeq{\end{equation}}

\def\Beq{\begin{equation}\begin{aligned}}
\def\Eeq{\end{aligned}\end{equation}}

\def\bea{\begin{eqnarray}}
\def\eea{\end{eqnarray}}

\def\beq{\begin{equation}}
\def\eeq{\end{equation}}
\def\bea{\begin{eqnarray}}
\def\eea{\end{eqnarray}}

\def\brinv{{\mathrm{BR}_{\rm inv}}}
\def\cuni{{c_{\rm uni}}}

\DeclareRobustCommand{\SkipTocEntry}[4]{}
\DeclareSymbolFont{extraup}{U}{zavm}{m}{n}
\DeclareMathSymbol{\varheart}{\mathalpha}{extraup}{86}
\DeclareMathSymbol{\vardiamond}{\mathalpha}{extraup}{87}

\begin{document}

\hypersetup{pageanchor=false}

\begin{titlepage}

\setcounter{page}{1} \baselineskip=15.5pt \thispagestyle{empty}

\begin{flushright}
\mbox{}
DESY-22-128
\end{flushright}

\bigskip\

\vspace{0.3cm}
\begin{center}

{\fontsize{20.74}{24}
\bfseries  Higgs-boson visible and invisible constraints \\ \vspace{0.2cm} on hidden sectors}

\end{center}

\vspace{0.2cm}

\begin{center}
{\fontsize{12}{30}\selectfont Thomas Biek\"otter\footnote{thomas.biekoetter@desy.de} \& Mathias Pierre\footnote{mathias.pierre@desy.de}}
\end{center}
\begin{center}

\vskip 7pt

\textsl{Deutsches Elektronen-Synchrotron DESY, Notkestr. 85, 22607 Hamburg, Germany}\\
\vskip 7pt

\end{center}

\vspace{0.3cm}
\centerline{\bf ABSTRACT}
\vspace{0.3cm}
We investigate the impact of  
interactions between
hidden sectors and the
discovered Higgs boson $h_{125}$,
allowing for additional invisible decay channels of $h_{125}$.
We perform $\chi^2$-fits to the measurements of
the Higgs-boson cross sections 
as a function of the invisible branching ratio
and different combinations of coupling modifiers, where the latter quantify 
modifications of the couplings of
$h_{125}$ compared to the predictions of the Standard Model.
We present generic results in terms of exclusion limits
on the coupling modifiers and the
invisible branching ratio of $h_{125}$. 
Additionally, we apply our results to
a variety of concrete 
model realizations containing 
a hidden sector:
dark matter within Higgs- and singlet-portal
scenarios, models featuring (pseudo) Nambu-Goldstone bosons
and two Higgs doublet extensions.
One of the main conclusions of our work is that in
a wide class of models
the \textit{indirect} constraints resulting from
the measurements of the cross sections of $h_{125}$
provide substantially stronger constraints on the
invisible Higgs-boson branching ratio 
compared to the \textit{direct} limits obtained from 
searches for the invisible decay of $h_{125}$.
However, we demonstrate that the presence of
an invisible decay mode of $h_{125}$ can also
open up parameter space regions which otherwise
would be excluded as a result of the
indirect constraints. As a byproduct of our analysis,
we show that in light of the new results
from the LZ collaboration a fermionic DM
candidate within the simplest Higgs-portal
scenario is ruled out under standard assumptions.

\vspace{0.2in}

 \end{titlepage}

\hypersetup{pageanchor=true}


\tableofcontents


\section{Introduction}

In 2012 a particle was
discovered at the Large Hadron
Collider (LHC) by both the ATLAS~\cite{Aad:2012tfa}
and CMS
collaborations~\cite{Chatrchyan:2012xdj}
with a mass of about $125\gev$ that
within the current experimental uncertainties
behaves in agreement with the predictions
for a Higgs boson of
the Standard Model (SM) of particle
physics~\cite{CMS:2022dwd,ATLAS:2022vkf}.
As a consequence, theories
beyond the SM (BSM) in which the shortcomings
of the SM can be addressed have to contain a particle
that plays the role of the discovered Higgs
boson~$h_{125}$.
This is especially relevant for BSM theories
in which $h_{125}$ is coupled to so far
unknown \textit{hidden sectors}.
If the hidden sector
contains particles with masses below
$125\gev$, modifications of the
properties of
$h_{125}$ can be present as a result
of exotic decays into
lighter BSM states. 
Here it is important to note that,
in addition to the fact that the
exotic decays can be searched for
at colliders,
additional decay modes also suppress the
ordinary decay modes of $h_{125}$
into SM particles.
Therefore, the discovered Higgs boson
acts as a probe for new physics
beyond the SM (BSM), and the LHC plays
a vital role (and will do so for many more years to come) in order to shed light on
the existence of light hidden sectors
that might have escaped discovery until today.

A remarkable amount of experimental
knowledge about the recently
discovered Higgs boson has been
gathered at the LHC within the last
ten years, where nature was kind to us
by choosing a Higgs-boson mass
for which several production
modes and decay channels
of $h_{125}$ as predicted by the SM
can be observed at the
LHC~\cite{LHCHiggsCrossSectionWorkingGroup:2013rie}.
The Higgs boson has been discovered
in the gluon-fusion production mode (ggH) with subsequent
decay into di-photon pairs and
via its
decay into pairs of off-shell vector bosons
giving rise to four-lepton final
states~\cite{Aad:2012tfa,
Chatrchyan:2012xdj}.\footnote{It is
remarkabe that the main discovery channel
$gg \to h_{125} \to \gamma\gamma$ involves two
loop-induced processes according to the
fact that the Higgs-boson does not carry
color- or electric charge. As a consequence, the
Higgs boson is also a valuable probe for
indirect effects of BSM particles
that could leave their footprint
exclusively via
radiative corrections to such loop-induced
couplings of $h_{125}$
(see e.g.~\citere{Lineros:2020eit}).
This possibility
is, however,
not further considered in this paper.}
The fine mass resolution of the discovery
channels permit a measurement of the
mass of $h_{125}$ at the sub-percent
level~\cite{CMS:2020xrn,ATLAS:2022net}.
As of today, in addition to the ggH production mode, the production via
the (tree-level) couplings to vector bosons in the
vector-boson fusion (VBF)
mode and via production
in association with a $W$- or a $Z$-boson (VH)
have been observed with a statistical
significance of $5\sigma$ or more~\cite{CMS:2022dwd,
ATLAS:2022vkf}.
Finally, although indirect experimental evidence for
the coupling of $h_{125}$ to top-quarks was
already present as a consequence of the
measurement of ggH production,
the presence of a large
Yukawa coupling of $h_{125}$ to top quarks
has also been directly confirmed by means of
the observation of Higgs-boson production in association with
a top-quark pair
(ttH)~\cite{CMS:2022dwd,ATLAS:2022vkf}.
With regards to the decays of
$h_{125}$, in addition to the decay modes
into pairs of photons and off-shell vector bosons
as stated above,
more recently also the decays
of $h_{125}$ into pairs of
bottom quarks and
into pairs of tau leptons
have been observed
with a statistical significance at the
level of $5\sigma$~\cite{CMS:2022dwd,ATLAS:2022vkf},
such that the presence of couplings
to the third-generation
quarks and charged leptons
has been established.
Finally, there is first experimental evidence,
although not yet statistically significant enough
for a discovery, for the presence of rare decays
of $h_{125}$ into pairs of muons~\cite{ATLAS:2021wwb} and
of the decays into
a photon and a $Z$-boson~\cite{ATLAS:2020qcv,CMS:2022ahq}.
There is so far no statistically significant
indication for exotic decay modes into
hidden sectors, but
the current experimental uncertainties
leave room for such decays
at the level of a few percent, as will
be the main topic of this paper.

\par \medskip

One of the most puzzling open problems of  modern physics concerns the presence of a dark matter (DM) component in our universe, whose relative contribution to the total energy budget has been measured to a great accuracy by the Planck collaboration to be $\Omega_\text{DM} h^2 = 0.11933 \pm  0.00091$~\cite{Aghanim:2018eyx}. According to the standard Weakly Interacting Massive Particles (WIMP) paradigm, such a dark component could have been produced by the well-known \textit{freeze-out} mechanism, by decoupling from the SM thermal bath while becoming non-relativistic. This scenario has been extensively 
studied in the literature and strong constraints from dark matter direct detection experiment, such as LUX~\cite{LUX:2016ggv}, Xenon1T~\cite{Aprile:2018dbl}, PandaX~\cite{PandaX-4T:2021bab} and
LUX-ZEPLIN (LZ)~\cite{LZnew},
have pushed the most classic models towards corners of allowed parameter space~\cite{Arcadi:2017kky,Mambrini,Escudero:2016gzx,Casas:2017jjg,Ellis:2017ndg,Arcadi:2019lka,Arcadi:2021mag}.
The Higgs boson plays a central role in many theories attempting to solve the dark matter problem in this context as any generic theory
containing a fundamental scalar could allow for this state to couple to the Higgs doublet via a quartic coupling. As the possibilities for a WIMP to carry a charge with respect to the SM gauge group are very constrained and limited~\cite{Cirelli:2005uq,Bottaro:2021snn,Bottaro:2022one}, this coupling is one of the only options for a renormalizable coupling of SM singlet states to one of the SM fields, while respecting the SM gauge
invariance~\cite{Patt:2006fw}.
This coupling plays a major role in the context of \textit{Higgs portal} scenarios where a dark matter candidate interacts with the SM only via coupling to the Higgs sector~\cite{Djouadi:2012zc,Arcadi:2017kky,Casas:2017jjg,Arcadi:2019lka,Arcadi:2021mag}.

\begin{table}
\centering
\begin{tabular}{l|c|c|c|c||c}
Collaboration & $\sqrt{s}$ [TeV] & Data $[\mathrm{fb}^{-1}]$ &
  $h_{125}$ production &
  Exp.~[\%] & \textbf{Obs.~[\%]} \\
\hline
\hline
CMS~\cite{CMS:2022qva} &
$8 + 13$ & $19.7 + 140$ &
  VBF & 10 & $\mathbf{18^*}$ \\
CMS~\cite{CMS:2018yfx} & $7+8+13$ & $4.9 + 19.7 + 38.2$ &
  VBF + VH + ggHj & 15 & \textbf{19} \\
CMS~\cite{CMS:2017zts} & 13 & 35.9 & VH + ggHj &
  40 & \textbf{53} \\
CMS~\cite{CMS:2017nxf} & 13 & 35.9 & ZH & 44 & \textbf{45} \\
\hline
ATLAS~\cite{ATLAS-CONF-2020-052} & $7+8+13$ & $4.7+20.3+139$ &
  VBF + ttH &
  11 & $\mathbf{11^*}$ \\
ATLAS~\cite{ATLAS:2019cid} & $7+8+13$ & $4.7+20.3+36.1$ &
  VBF + VH & 17 & \textbf{26} \\
ATLAS~\cite{ATLAS:2022yvh} &
  $13$ & $139$ & VBF & 10.3 & \textbf{14.5} \\
ATLAS~\cite{ATLAS:2017nyv} & 13 & 36.1 & ZH & 39 & \textbf{67} \\
ATLAS~\cite{ATLAS:2018nda} & 13 & 36.1 & VH & 58 & \textbf{83}
\end{tabular}
\caption{Expected (exp) and observed (obs)
experimental upper limits at
the 95\% confidence level on the
invisible branching ratio
of $h_{125}$ as reported by ATLAS and CMS, obtained
from direct searches for
the invisible decay of $h_{125}$ at the LHC
at $\sqrt{s} = 13\tev$. The currently strongest
upper limit reported by each collaboration
are marked with a star.
The production modes of $h_{125}$ targeted in
the different
searches are vector-boson fusion (VBF), production in
association with a $Z$-boson (ZH) or $Z$- and $W$-bosons
(VH), and gluon-fusion production in association with
a jet (ggHj).}
\label{tab:directlims}
\end{table}

Assuming, as discussed above,
the presence of a hidden sector
which exclusively couples to the visible
sector via the Higgs portal,
Higgs-boson decays into particles of
the hidden sector can be kinematically allowed.
Here it should be noted that one motivation
to investigate hidden sectors with
BSM states in this mass range is that the
strongest direct detection constraints can be evaded
for dark-matter masses at the
level of a few~GeV or below.
If the BSM particles are stable or
sufficiently long-lived in order to
escape the detector,
so-called invisible (inv) decays of
$h_{125}$ are present which can be
searched for at the LHC in final
states with large missing transverse energy.
Direct searches for
the invisible decay of $h_{125}$ have been
performed by both the ATLAS and the CMS
collaboration utilizing various different production
modes. We summarize in
\refta{tab:directlims} the current
LHC searches for the invisible decay
of $h_{125}$ that include part or all
of the Run~2 dataset at $13\tev$.
No significant excesses over the SM
background have been observed, such that
upper limits on the branching ratio
for the invisible decay of $h_{125}$,
denoted $\brinv$ from hereon, have been
determined (see last
column of \refta{tab:directlims}). Assuming that $h_{125}$ couples
to the SM fermions and gauge bosons
according to the predictions of
the SM, the currently strongest limit on $\brinv$
was reported by ATLAS~\cite{ATLAS-CONF-2020-052},
\begin{equation}
\brinv < 11\% \ \textrm{ at 95\% confidence level (CL)},
\label{eq:dirlim}
\end{equation}
which was obtained by combining the
datasets collected at 7, 8 and $13\tev$, and
by utilizing the VBF and ttH production modes.
In the following, we will refer to the constraint
shown in \refeqq{eq:dirlim}
as the \textit{direct limit}
on $\brinv$, according to the fact that it
is extracted from directly searching for
the invisible decay of~$h_{125}$.

In many phenomenological
analyses of BSM theories in which
a Higgs-boson invisible decay plays a role,
only the above mentioned direct limit on
$\brinv$ is applied in order to exclude
parameter space regions of the theory that
are in disagreement with the
Higgs-boson measurements at the LHC
(see e.g.~\citeres{Escudero:2016gzx,Arcadi:2017kky,
Ellis:2017ndg,Yaguna:2021rds,Okada:2020zxo,
Hara:2021lrj}).
However,
additional, although more model-dependent,
constraints on $\brinv$
arise from the cross-section measurements of
$h_{125}$, considering that additional
BSM decay modes give rise to a suppression
of the ordinary decay modes of $h_{125}$.
By performing global scans to the LHC
Higgs-boson measurements in terms of
$\brinv$ and so-called coupling modifiers
that quantify deviations of the couplings
of~$h_{125}$
with respect to the SM predictions, one
can set limits on $\brinv$ as a function
of the coupling
modifiers (see,
e.g.~\citeres{Espinosa:2012vu,Espinosa:2012im,
Belanger:2013kya,
Bechtle:2014ewa,Kraml:2019sis,
Bechtle:2020uwn} for
earlier analyses of this kind).
Since in this approach the measurements of the ordinary
decay modes of $h_{125}$ are used
to constrain $\brinv$, in contrast to
directly searching for the invisible
decay of~$h_{125}$, we will refer to
the limits on $\brinv$ resulting from the
global analysis of the
cross-section measurements as the
\textit{indirect constraints} in the following.

In this paper our goal is to exploit the
complementarity of the direct and the indirect
constraints on $\brinv$.
To this end, the first step of our analysis
is to determine the indirect constraints
on $\brinv$ via global fits to the cross-section
measurements of $h_{125}$, where we
will make use of the public code
\texttt{HiggsSignals~v.3}~\cite{hsnew}.
We perform such global scans assuming that
the couplings of $h_{125}$ remain
unchanged compared to the couplings of
a SM Higgs boson, but also assuming more complicated
Higgs-portal models in which, in addition to the
presence of an invisible decay mode, there
are modifications of the couplings of
$h_{125}$ to the fermions and gauge bosons.\footnote{Similar
analyses have been performed, for instance,
in \citeres{Espinosa:2012im,
Belanger:2013xza,Bernon:2014vta,
Bechtle:2020uwn},
and more recently both
CMS and ATLAS presented constraints on
$\brinv$ obtained in the so-called
$\kappa$-framework~\cite{CMS:2022dwd,ATLAS:2022vkf}.}
If coupling modifications are considered,
the upper limit on $\brinv$
is a function of 
the coupling modifiers.
Having determined the indirect constrains
on $\brinv$, we then compare them
to the direct limit on $\brinv$ in order to
investigate which of the two kind of constraints
results in a stronger exclusion 
in a variety
of different BSM scenarios.
To give a brief outlook on the key results
that we have found, we emphasize already here that
in many of the Higgs-portal scenarios considered
in the literature the indirect constraints
can provide substantially stronger exclusions
and should therefore 
not be overlooked.
However, in models in which the properties
of $h_{125}$ are not as predicted by the
SM, we also found parameter space regions
that are in conflict with the
LHC Higgs-boson measurements if there is no invisible
decay mode of $h_{125}$, whereas the same
parameter regions are well in agreement
with the LHC measurements
if a value of $\brinv$ at the level of
a few percent is present.

\par \medskip
The outline of the paper is as follow.
In Sec.~\ref{sec:limitsfromhs} we describe
the setup and approach used to derive
the indirect constraints on $\brinv$ as a function
of the couplings modifiers, thus
accounting for deviations with respect to the
SM expectation of the couplings of
$h_{125}$ to the SM fields.
We present generic constraints in terms of
the coupling modifiers and the invisible
branching fraction for a different set of
assumption regarding the structure of
the coupling modifiers.
Going beyond this generic framework,
we analyse in \refse{sec:application}
the constraints in a variety of
concrete BSM scenarios: Higgs- and
singlet-mediated dark matter models,
constructions featuring (pseudo)
Nambu-Goldstone bosons, and extended
Higgs sectors featuring a second Higgs
doublet.
In \refse{sec:conclu}, we summarize our
main results and conclude.

\section{Indirect limits on the invisible decay
of the Higgs boson \texorpdfstring{\boldmath{$h_{125}$}}{h125}}
\label{sec:limitsfromhs}
As mentioned above, in the first step of
our analysis we will remain agnostic about
the precise nature of the hidden sector.
Instead of specifying
a concrete model, we perform $\chi^2$-fits
to the measurements of the Higgs boson $h_{125}$
as a function of the branching ratio for
invisible decay modes of $h_{125}$ and,
in addition, as a function of coupling
modifiers that quantify modifications of the
couplings of $h_{125}$ compared to the
SM predictions.
These $\chi^2$-analyses, for which we make use of the
public code
\texttt{HiggsSignals v.3}~\cite{Bechtle:2013xfa,
Bechtle:2020uwn,hsnew}
(see the discussion
below for details), will be discussed in this
section. The $\chi^2$-analyses provide us
with $68\%$ and $95\%$ confidence-level
(CL) upper limits on
$\mathrm{BR}_{\rm inv}$ as a function of the
coupling modifiers. These upper limits
can be compared
to the upper limits on the
invisible branching ratio
resulting from direct-searches for the decays
of $h_{125}$ into invisible final state.
At a later stage of our analysis, discussed
in \refse{sec:application},
we will apply the constraints
on the invisible braching ratio $\mathrm{BR}_{\rm inv}$ that we obtained from
the global $\chi^2$-fit to a range of commonly
studied hidden sectors possessing
dark matter candidates, extended scalar content (2HDM) and pseudo
Nambu-Goldstone bosons,
in which the Higgs portal plays
a major role.

Before starting the discussion of the
results of the $\chi^2$-analyses, we briefly
discuss the strategy that is implemented in
the public code \texttt{HiggsSignals}, and which
is used here in order to compare the predicted
cross sections and signal rates of $h_{125}$
to the experimental measurements.
For a more detailed description of the code,
we refer the reader to \citere{hsnew} (see also
\citere{Bechtle:2020uwn}
for a more detailed discussion
on the statistical interpretation of the
$\chi^2$-values provided by
\texttt{HiggsSignals}).
The code contains a large set of experimental
data from LHC
measurements at $8\tev$ and (mostly) $13\tev$
center-of-mass energy.\footnote{The data repository
of \texttt{HiggsSignals} can be found
at \url{https://gitlab.com/higgsbounds/hsdataset}.}
The measurements are
implemented not only in the form of inclusive
cross sections (or signal
rates), but also in the form of the
simplified template cross
sections~\cite{Berger:2019wnu} which
were designed to allow
for a combination of measurements of different
decay channels of $h_{125}$.
In total, the \texttt{HiggsSignals} dataset
currently comprises 24 independent measurements
from the CMS and the ATLAS collaborations.
For the purpose of our paper,
as will be discussed in detail below, the most
important measurements are the ones
utilizing the
di-photon final state due to the sensitivity
of the di-photon branching ratio on
the presence of BSM decay modes
of $h_{125}$.

On the theory side, in order to compare the
model predictions to the experimental data,
\texttt{HiggsSignals} requires as input
either directly the signal rates
(or alternatively the cross
sections and branching
ratios) of $h_{125}$ in the various different
production or decay modes, or the user has the
option to provide effective coupling modifiers
from which the cross sections and branching ratios
are derived internally by a re-scaling of the
SM predictions.
The coupling modifiers are defined as the couplings
of the particle state at $125\gev$ in the given
BSM theory normalized to the couplings of a
SM Higgs boson at the same mass
(see below for details). Thus, in order
to check the SM against the experimental data,
one would choose all modifiers to be equal
to one, whereas in models that
feature modifications
of the properties of $h_{125}$ the couplings
modifiers deviate from one.
In our analysis, we will combine both
input formats. This is possible since the
most recent update of \texttt{HiggsSignals},
which is now incorporated into the public
code \texttt{HiggsTools}~\cite{hsnew}.
We make use of the input in terms of the
coupling modifiers in order to obtain
the LHC cross sections of $h_{125}$
and in order to calculate the partial
decay widths of all conventional decay
modes of $h_{125}$ into SM particles.
Subsequently, in order to set the desired
value of the branching ratio for the
additional decay mode into invisible
final states, we give the respective
partial decay width $\Gamma_{\rm inv}$
as input.
The partial width
$\Gamma_{\rm inv}$ that is required
to set a desired value of $\mathrm{BR}_{\rm inv}$
can be calculated as
\begin{equation}
\Gamma_{\rm inv} = \frac
{\mathrm{BR}_{\rm inv}}
{1 - \mathrm{BR}_{\rm inv}} \sum_{i}
    \Gamma_{i}^{\rm SM}\,,
\label{eq:gammainv}
\end{equation}
where $\Gamma^{\rm SM}_i$ stands for the
individual partial widths for decays of
$h_{125}$ into SM final states, which were
calculated previously as a function
of the coupling modifiers. Accordingly,
the sum runs over all relevant
SM decay modes of $h_{125}$, i.e.~$i =
\{b\bar b, gg, WW^*,
ZZ^*, c\bar c, s \bar s,
\tau^+ \tau^-, 
\gamma\gamma, \mu^+ \mu^-, 
Z\gamma\}$.
The total width of $h_{125}$ is then given by
\begin{equation}
\Gamma_{\rm tot} =
\Gamma_{\rm inv} + \sum_{i}
    \Gamma_{i}^{\rm SM} \ .
\end{equation}
As already discussed in the introduction,
the fact that the total width of $h_{125}$
is modified via the presence of the
additional contribution $\Gamma_{\rm inv}$
gives rise to the fact that $\mathrm{BR}_{\rm inv}$
can be constrained via the measured cross sections
and signal rates of $h_{125}$.

In order to capture the coupling
modifications that arise in as many
UV-complete models as possible while
maintaining a manageable number of free
parameters, we define four independent
coupling modifiers~$c_i$,
with $i={V,u,d,\ell}$.
The coefficients $c_i$
modify the couplings of
the discovered Higgs boson
$h_{125}$ to the SM mass
eigenstates according to
\begin{equation}
    \mathcal{L}\,=\,\sum_{f=u,d,\ell} c_f  \left( \dfrac{m_f}{v} \right)h_{125} \,\bar f f +   c_V \left(\dfrac{2 m_W^2}{v} \right) h_{125} \,W^{+ \mu} W^{-}_{\mu}+   c_V \left(\dfrac{m_Z^2}{v} \right) h_{125} \,Z^{\mu} Z_{\mu}+ \mathcal{L}_\text{inv}\,,
    \label{eq:defcouplingmodifiers}
\end{equation} 
where $c_V$,  $c_u$, $c_d$ and
$c_\ell$ are, respectively,
the coupling modifier of the massive
gauge bosons, to up-type quarks,
to down-type quarks and charged 
leptons.\footnote{Consequently,
we do not take into account 
flavour-dependent modifications to
the Higgs-boson couplings.
We also do not consider possible sources
of CP violation, such that $h_{125}$ is
assumed to be a purely CP-even state.}  $ \mathcal{L}_\text{inv}$ is the
contribution that contains the
portal couplings to the hidden sector
responsible for the presence of the invisible
branching ratio $\mathrm{BR}_{\rm inv}$, which
we will further specify when we consider
concrete BSM theories in \refse{sec:application}.
For the loop-induced
couplings of $h_{125}$ to photons and
gluons, we assume that there are no sizable
BSM contributions in additions to
the loop diagrams
with SM particles in the loops, reflecting the
fact that light hidden-sector particles can only
couple very weakly via
the SM gauge interactions in order to be
physically viable.
Hence, the respective coupling coefficients
$c_{\gamma\gamma}$ and $c_{gg}$ can be
calculated as a function of $c_V$, $c_u$ and $c_d$.\footnote{The contributions from the
charged leptons can safely be neglected due
to the suppression from the smaller Yukawa
couplings $Y_\ell \ll Y_t, Y_b$.}
The structure of the coupling modifiers in
the various BSM theories considered in this work
are summarized in Tab.~\ref{tab:my_table}.

\begin{table}
    \centering
    \begin{tabular}{l|c|c|c|c||c}
     \text{Scenario}   & $c_u$   & $c_d$ & $c_\ell$ & $c_V$ & Discussion \\
        \hline
        \hline
        SM-like couplings & 1 & 1 & 1 & 1 &
        \refse{sec:smlikecpls} \\
        \text{Universal couplings}  & $c_\text{uni}$  & $c_\text{uni}$ & $c_\text{uni}$ & $c_\text{uni}$ & Sec.~\ref{sec:uni}  \\
           \text{Non-universal couplings}       & $c_f$ & $c_f$ & $c_f$  & $c_V $ &  Sec.~\ref{sec:nonunicoupling}  \\
           \text{Non-universal couplings}       & $c_u$ & $c_d$ & $c_\ell = c_d$  & $1$ &  Sec.~\ref{sec:nonunicoupling}  \\
        \hline
        \text{Higgs portal DM}  & 1 & 1 & 1 & 1 &  Sec.~\ref{sec:Higgsportal} \\
        \text{Singlet portal DM}  &  $c_\text{uni} \leq 1$ &  $c_\text{uni} \leq 1$ &  $c_\text{uni} \leq 1$ &  $c_\text{uni} \leq 1$ &  Sec.~\ref{sec:singletportal} \\
        \text{(P)NGB}  &  $c_\text{uni} \leq 1$ &  $c_\text{uni} \leq 1$ &  $c_\text{uni} \leq 1$ &  $c_\text{uni} \leq 1$ &  Sec.~\ref{sec:PNGB}\\
        \text{2HDM Type I}  & $c_f$ & $c_f$ & $c_f$ & $c_V \leq 1$ &  Sec.~\ref{sec:2HDM}  \\
        \text{2HDM Type II} & $c_u$& $c_d$& $c_\ell = c_d$ &$c_V \leq 1$ & Sec.~\ref{sec:2HDM}  \\
        \text{2HDM Type III} & $c_u$& $c_d = c_u$& $c_\ell = c_u$&$c_V \leq 1$ & Sec.~\ref{sec:2HDM} \\
        \text{2HDM Type IV}  & $c_u$ & $c_d$&$c_\ell = c_u$ &$c_V \leq 1$ & Sec.~\ref{sec:2HDM} \\
    \end{tabular}
    \caption{\small Summary of the
    coupling
    modifcications in terms of the
    modifiers defined in
    \refeqq{eq:defcouplingmodifiers}
    for the various models considered in
    this work and references to the
    (sub)sections of the manuscript
    where each specific case is discussed.}
    \label{tab:my_table}
\end{table}

As already mentioned above,
in order to derive the indirect limits on
$\mathrm{BR}_{\rm inv}$ from the measurements
related to $h_{125}$ we perform a $\chi^2$-analysis
utilizing the public code \texttt{HiggsSignals}. We determine in each scan
\begin{equation}
\Delta \chi^2
\left(c_i, \mathrm{BR}_{\rm inv}\right) =
\chi^2\left(c_i, \mathrm{BR}_{\rm inv}\right)
- \chi^2_{\rm SM} \ ,
\label{eq:chisqdef}
\end{equation}
where $\chi^2(c_i, \mathrm{BR}_{\rm inv})$ is
the fit result for the respective BSM scenario
given as a function of the coupling modifiers
$c_i$ and $\mathrm{BR}_{\rm inv}$, and $\chi^2_{\rm SM}$
is the fit result assuming properties
of $h_{125}$
as predicted by the SM, i.e.~$\mathrm{BR}_{\rm inv} = 0$ and $c_i = 1$.\footnote{Strictly
speaking, also in the SM one
finds $\mathrm{BR}_{\rm inv} > 0$
due to the decay mode $h_{125} \to
Z Z^* \to \nu \nu \bar \nu \bar \nu$.
However, the resulting branching
ratio is of the order
of $0.1\%$~\cite{LHCHiggsCrossSectionWorkingGroup:2011wcg}.
Considering the current experimental
precisions of the signal-rate measurements of
$h_{125}$,
this value
can be approximated by zero for all practical
purposes.}
In order to set limits on the coupling modifiers
or model parameters, we demand that the BSM scenario
is not disfavoured compared to the SM fit result at
the 95\%~CL or more.
In a one-dimensional
parameter estimation,
this translates into the
condition~\cite{ParticleDataGroup:2020ssz}
\begin{equation}
\Delta \chi^2
\leq 3.84 \ .
\label{eq:1dimcond}
\end{equation}
In a joint estimation of two free parameters,
the respective condition is
\begin{equation}
\Delta \chi^2 \leq
5.99 \ .
\end{equation}
In addition to the allowed regions
of the coupling modifiers obtained
using the 95\%~CL conditions, we will
show in our plots also allowed regions using
the 68\%~($1 \sigma$)~CL for
illustrative reasons, which corresponds to the
conditions $\Delta \chi^2 \leq 1$ and
$\Delta \chi^2 \leq 2.30$ in one-dimensional
and two-dimensional fits,
respectively~\cite{ParticleDataGroup:2020ssz}.

By defining the allowed/excluded regions for the coupling
modifiers based on the $\chi^2$-definition shown
in \refeqq{eq:chisqdef}, where we utilize the SM
as a reference model, the obtained limits only
correspond to the 95\%~CL limits if
the SM result $\chi^2_{\rm SM}$ is a good approximation
of the best-fit result of the BSM scenario
under consideration, i.e.~$\chi^2_{\rm min} =
\min(\chi^2) \simeq
\chi^2_{\rm SM}$. 
As we will discuss
below, 
we encountered situations in which
$\chi^2_{\rm min}$ was considerably smaller than
$\chi^2_{\rm SM}$ for certain parameter
configurations.\footnote{We emphasize that
$\chi^2_{\rm min} < \chi^2_{\rm SM}$ does not
necessarily signify a global statistical
preference of the respective BSM scenario
compared to the SM, as such a conclusion can
only be drawn by
also taking into account
the different numbers of degrees of freedom
of both models. For the context of our
analysis, in which we are interested only in the
the constraints on the coupling modifiers
and $\mathrm{BR}_{\rm inv}$
that can be derived within a
certain BSM scenario,
the global preference of different BSM scenarios
against the SM (and also against each other)
is not relevant.}
In such a situation, the more common
approach (in a frequentist analysis)
would be to construct the confidence intervals
of the free parameters by comparing to the
parameter point that features the best-fit value
$\chi^2_{\rm min}$ instead of comparing to
$\chi^2_{\rm SM} > \chi^2_{\rm min}$,
i.e.~constructing the limits based on
$\Delta \chi^2 = \chi^2 - \chi^2_{\rm min}$.
This approach would yield stronger constraints
on the parameters and therefore smaller allowed
regions in the $c_i$ parameter space
as compared to our approach, such
that the latter
should be
regarded as more conservative.
We found that
the experimental measurements responsible
for values of
$\Delta \chi^2 < 0$
are mainly the ones
related to the ttH production
of $h_{125}$, where the signal extraction
is affected by systematic
uncertainties regarding the theoretical
predictions for the background
estimation~\cite{ATLAS-CONF-2019-045,ATLAS:2021qou}.
Taking this into account,
it is reasonable to be conservative
in the determination of the exclusion limits,
thus sticking
to the definition of $\Delta \chi^2$
as shown in \refeqq{eq:chisqdef}, even though
$\chi^2_{\rm min}$ is smaller than
$\chi^2_{\rm SM}$ in some scenarios.
We also note that in our plots we indicate the
$\Delta \chi^2$-distribution together with
the value of $\chi^2_{\rm min}$
for all parameter
points in addition to the 68\%~CL and 95\%~CL
exclusions, such that the
reader can also apply a different criterion
in order to define the allowed regions.
Finally, it should also be noted that the opposite case
with $\chi^2_{\rm min} > \chi^2_{\rm SM}$ is not
possible for the different BSM scenarios considered
in the following, because
in the fits of the BSM theories in terms
of the coupling modifiers we always include
the parameter space point $c_i = 1$
and $\mathrm{BR}_{\rm inv} = 0$ in which
the particle state $h_{125}$ resembles exactly
the properties of a SM Higgs boson.

As was already mentioned above, we start
our analysis be performing the $\chi^2$-fits
in different
BSM scenarios in a generic fashion in terms
of the coupling modifiers $c_i$. The different
scenarios are discussed in the following
subsections.
We summarize in \refta{tab:my_table}
the different constructions of coupling
modifications that we consider.
The applications of the constraints
on the modifiers $c_i$ and ultimately
on $\mathrm{BR}_{\rm inv}$
obtained in this way to the parameter space of
UV-complete BSM theories,
in which both
$c_i$ and $\mathrm{BR}_{\rm inv}$ are
functions of the model
parameters, will be discussed
in \refse{sec:application}.

\subsection{\texorpdfstring{$h_{125}$}{h125}
with SM couplings and
a non-zero
\texorpdfstring{$\mathrm{BR}_{\rm inv}$}{BRinv}}
\label{sec:smlikecpls}
In the simplest hidden sector models,
the SM is augmented by
new fields which do not give rise
to modifications of the
couplings of $h_{125}$ compared to
a SM Higgs boson. 
As a result,
the cross sections at colliders for
the production of $h_{125}$
are unchanged compared to the SM predictions.
However, as discussed above,
the branching ratios
of $h_{125}$ can be modified due to additional
decay modes into the hidden sector if such
decay modes are kinematically allowed.
In this case,
the modifications of the properties of $h_{125}$
compared to the SM predictions can be quantified
exclusively in terms
of the invisible
branching ratio $\mathrm{BR}_{\rm inv}$.
In the following we will discuss how non-zero
values of $\mathrm{BR}_{\rm inv}$ give rise to
a suppression of the signal rates in the
conventional decay modes of $h_{125}$.
By comparing the predicted signal rates to
the experimental measurements, we will derive
an upper limit on $\mathrm{BR}_{\rm inv}$ at
the $95\%$~CL via the condition
shown in \refeqq{eq:1dimcond}. Finally, we will
compare the indirect limit resulting from the global
Higgs-boson measurements to the direct limit
on $\mathrm{BR}_{\rm inv}$ from
direct searches for the invisible decay
of $h_{125}$.

\begin{figure}[t]
\includegraphics[width=0.7\textwidth]{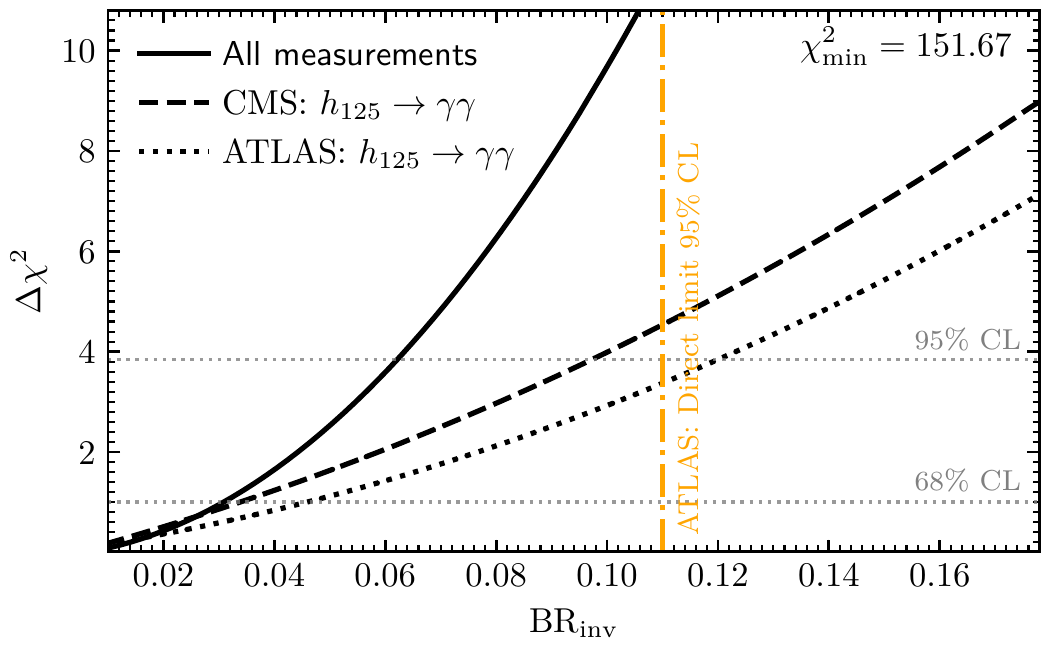}
\caption{\small
  Result of the global fit to the
  Higgs signal-rate measurements assuming
  a Higgs boson $h_{125}$ 
  that apart from
  the presence of an invisible decay mode
  behaves according to the predictions
  of the SM model, i.e.~the couplings of
  $h_{125}$ to the SM particles are
  unchanged. The black solid line indicates the value
  $\Delta \chi^2$ taking into account
  the full \texttt{HiggsSignals} data set.
  The black dashed and the black dotted lines indicate
  the value of $\Delta \chi^2$
  considering only the measurements
  in the diphoton final state
  utilizing the full Run~2 data
  by CMS~\cite{CMS:2021kom} and
  ATLAS~\cite{ATLAS:2020pvn}, respectively.
  The gray horizontal lines indicate the
  68\% and 95\% CL, respectively.
  The orange vertical line indicates
  the currently strongest upper limit
  on $\mathrm{BR}_{\rm inv}$ resulting
  from direct searches for
  $h_{125} \to \mathrm{inv}$ as published by
  ATLAS~\cite{ATLAS-CONF-2020-052}.}
\label{fig:SMinv}
\end{figure}

To this end, we show in \reffi{fig:SMinv}
the results of the $\chi^2$-analysis performed
with \texttt{HiggsSignals}, where we set
$c_i = 0$, and we varied
$\mathrm{BR}_{\rm inv}$ from 0\% to 18\%.
The black solid line indicates the value of
$\Delta \chi^2$, which is obtained including
the complete data set implemented
in \texttt{HiggsSignals}.
We find the best-fit
with $\Delta \chi^2 = 0$
at $\mathrm{BR}_{\rm inv} = 0$, with
$\chi^2_{\rm min} = \chi^2_{\rm SM} = 151.67$.
For increasing values of $\mathrm{BR}_{\rm inv}$
the $\chi^2$ function is monotonically increasing.
Consequently, under the assumption that the
couplings of $h_{125}$ are unchanged compared
to the SM predictions, the presence of an
invisible decay mode of $h_{125}$ with sizable
branching ratio worsens the fit result to
the Higgs-boson measurements for all possible
values of $\brinv$. Based on the \texttt{HiggsSignals}
analysis, we find an indirect upper limit of
\begin{equation}
\brinv < 6.22\%~(3.08\%) \qquad \textrm{at }95\%~(68\%)~\textrm{CL}\qquad \textrm{for } c_i = 1 \ .
\label{eq:ourlimitsminv}
\end{equation}
The 95\% CL limit is almost a factor
of two smaller
than the currently strongest
limits from direct searches
for the decay mode $h_{125} \to \mathrm{inv}$
(see \refta{tab:directlims}).
The strongest direct limit
$\brinv < 11\%$ is indicated by the
orange dotted-dashed vertical line in
\reffi{fig:SMinv}~\cite{ATLAS-CONF-2020-052}.
Accordingly, for phenomenological studies
of models with hidden sectors in which
the couplings of $h_{125}$ are not modified
compared to the SM predictions, but where
the decay of $h_{125}$ into the hidden sector
is kinematically open, one should include
the indirect constraints from the measurements
of the signal rates of $h_{125}$, currently
giving rise to the upper limit shown
in \refeqq{eq:ourlimitsminv}. On the contrary,
taking into account only
the limit on $\brinv$ from direct searches
allows also parameter space regions
that are already
excluded by the measurements regarding
the discovered Higgs boson at a CL of
more than $3\sigma$, i.e.~$\Delta \chi^2 > 9$,
as is visible in \reffi{fig:SMinv}.

In order to shed light on which particular
measurements are most relevant for the
increase of $\chi^2$ with increasing values
of $\brinv$, we show in \reffi{fig:SMinv}
the $\chi^2$-values that are obtained
only taking into account the measurement
with the largest and the second largest
individual contributions
to the total $\chi^2$.\footnote{The total
$\chi^2$-value including all measurements
(black solid line)
can be smaller than the sum of the two individual
contributions from the $h_{125} \to \gamma\gamma$
measurements (black dashed and dotted lines)
shown in \reffi{fig:SMinv}, because
for the computation of the total $\chi^2$-values
the correlations are taken into account.}
The most important
$\chi^2$-penalty has its origin in the
CMS measurement of Higgs-boson production
and subsequent decay into di-photon
pairs including the full Run~2 data
at $13\tev$ center of mass
energy~\cite{CMS:2021kom}. The fit to
the corresponding cross-section measurements
gives rise to the $\Delta \chi^2$ values
indicated by the black dashed line
in \reffi{fig:SMinv}.
One can see that based on this measurement
alone one obtains a stronger upper limit
on $\brinv$ than the one obtained from
direct searches for $h_{125} \to \mathrm{inv}$.
Also the second largest $\chi^2$-penalty
is caused by signal rates of $h_{125}$
utilizing the di-photon decay mode, but
here as a result of the corresponding
ATLAS measurements~\cite{ATLAS:2020pvn},
indicated by the black dotted line
in \reffi{fig:SMinv}.
The fact that the two most relevant
measurements are both related to the
$h_{125} \to \gamma\gamma$ decay mode can
be understood by realizing that, from the
experimental side, it is the most
precisely measured decay mode, and
that, from the theory side, the
di-photon branching ratio is
very sensitive to modifications of
the total width of $h_{125}$.

We finally compare the
indirect limit on $\brinv$ to similar
results that have been obtained in
the past and to projections for future runs of the LHC and for the high-luminosity LHC (HL-LHC).
An early analysis including Tevatron
and LHC first-year Run~1 data
found a limit of
$\brinv \leq 23\%$ via a global fit to
the Higss-boson data assuming SM-like
couplings~\cite{Belanger:2013kya}.
Shortly after, in \citere{Bechtle:2014ewa}
a slightly stronger
limit of $\brinv \leq 17\%$ was found
making use of \texttt{HiggsSignals} using
the same fit strategy as applied here.
Taking into account the full
Run~1 LHC data, an upper limit of
$\brinv \leq 12\%$ has been determined
in \citere{Bernon:2014vta}.
Global fits including also 13~TeV LHC data
have been performed in
\citere{Kraml:2019sis} and
\citere{Bechtle:2020uwn},
finding upper limits on $\brinv$ under the
assumption of SM-like couplings of
5\% and 10\%, respectively.
These numbers are of comparable size
as the limit $\brinv \leq 6.2\%$
found in this analysis.

It is interesting to note that the
current indirect limits 
are also comparable to the HL-LHC
projections assuming that
3000\,fb$^{-1}$ will be
collected at
14~TeV~\cite{Cepeda:2019klc}.
The reason for this
is mainly that the projected limits on $\brinv$
were obtained under the assumptions that the
experimental data will be in agreement with the
SM predictions, whereas the current limits
on $\brinv$ have been determined
from actual data in which the central
values of the various measurements
naturally fluctuate within statistical
and systematic uncertainties.
If differences exist between the central
values in the current data
and the SM predictions (which are expected at least at the level of statistical fluctuations),
the extracted
upper limit on $\brinv$ derived today can be
smaller than the projected HL-LHC limit even
though the precision of the individual cross-section
measurements is larger now compared to what
is expected in the future for the HL-LHC.
This also means that,
even though the uncertainties of
the cross-section measurements
of~$h_{125}$ will improve
in the future,
the indirect limit on $\brinv$
might not necessarily become substantially
stronger, because the limit will depend on
how the measured central values of the
cross sections will evolve within the current
uncertainty bands.

\subsection{\texorpdfstring{$h_{125}$}{h125} with universally
modified couplings
and a non-zero \texorpdfstring{$\mathrm{BR}_{\rm inv}$}{BRinv}}
\label{sec:uni}
In the previous section we discussed the case
in which the Higgs sector remained
unchanged compared to the SM except for the
presence of an additional invisible decay mode into
a hidden sector. However, in many BSM scenarios additional scalar fields can mix with the SM-like Higgs boson
at $125\gev$. In this case, the properties of
$h_{125}$ can be modified not only by additional
decay modes into invisible final states, but also
the couplings of $h_{125}$ to the fermions and
gauge bosons of the SM can be modified compared
to the SM. In this section we will analyze the most simple possibility, in which
the couplings of $h_{125}$ to
the SM particles are modified by a universal
coupling modifier
$c_{\rm uni} = c_V = c_u = c_d = c_\ell$.
The simplest
UV-complete model in which this pattern
of coupling modifiers arises
is a model in which a gauge singlet scalar field
mixes with $h_{125}$, giving rise to a suppression
of the couplings of $h_{125}$ by a universal
factor $c_{\rm uni}  < 1$ which can be identified as $c_{\rm uni} = \cos \theta$ where  $\theta$ is the singlet mixing angle with $h_{125}$.
Models in which also a coefficient $c_{\rm uni} > 1$
can be realized comprise, for instance,
extensions of the SM by a real or
complex SU(2) triplet Higgs field.

\begin{figure}[t]
\includegraphics[width=0.7\textwidth]{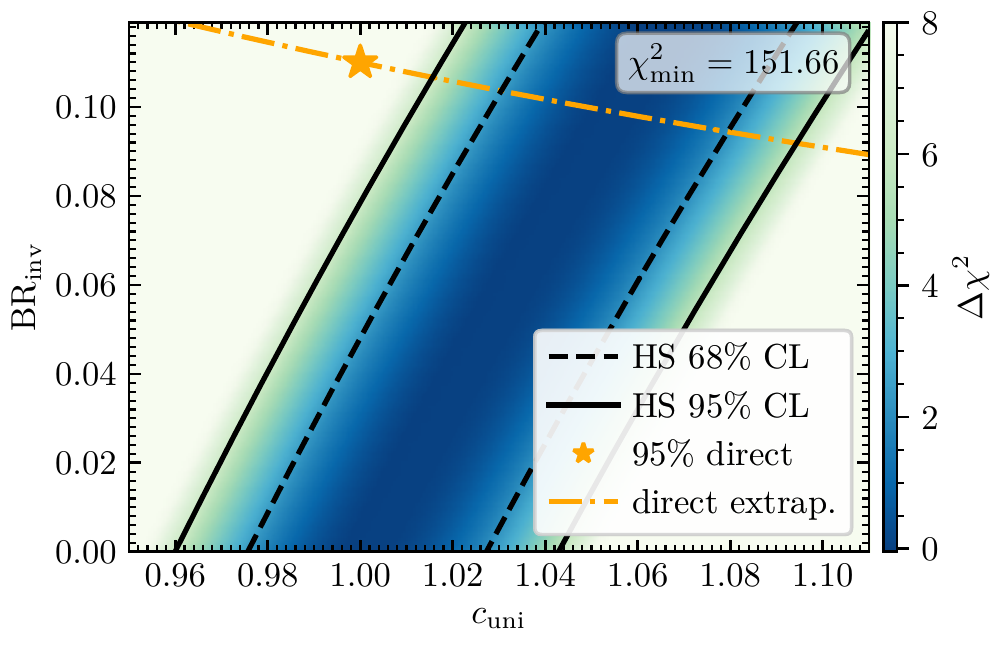}
\caption{\small
  Result of the global fit to the Higgs
  signal-rate measurements assuming a
  Higgs boson $h_{125}$ that has an
  invisible decay mode and whose couplings
  to the SM particles are modified
  w.r.t.~the SM predictions by a universal
  coupling coefficient $c_{\rm uni}$.
  The color coding indicates the value
  of $\Delta \chi^2$ resulting from
  the \texttt{HiggsSignals} analysis.
  The black solid and dashed lines
  indicate the allowed regions at
  the $95\%$ and the $68\%$ confidence
  level, respectively. The orange star
  indicates the currently strongest
  upper limit on $\mathrm{BR}_{\rm inv.}$
  resulting from direct searches
  for $h_{125} \to \mathrm{inv}$ obtained
  according to the assumption that
  $h_{125}$ is produced as predicted
  in the SM, i.e.~$c_{\rm uni}
  = 1$~\cite{ATLAS-CONF-2020-052}.
  The orange dashed line
  indicates an approximate extrapolation
  of this direct limit for the case
  $c_{\rm uni} \neq 1$ (see text for
  details).
}
\label{fig:BRinvCeff}
\end{figure}

Under the assumption of the universal
coupling coefficient $c_{\rm uni}$ and
the presence of an invisible decay mode
$h_{125} \to \text{inv}$, we can
perform a $\chi^2$-fit to the measurements
of the properties of $h_{125}$ in dependence
of the two free parameters $c_{\rm uni}$
and $\mathrm{BR}_{\rm inv}$
with the help of 
\texttt{HiggsSignals}.
By doing so,
we obtain the two-dimensional $\chi^2$-distribution
that is shown in \reffi{fig:BRinvCeff}.
In this plot the color grading indicates
the value of $\Delta \chi^2$ in the
parameter plane $\{c_{\rm uni},\brinv\}$,
and the solid and dashed lines indicate
the $95\%$ and the $68\%$ CL limits,
respectively. Also indicated with an
orange star is the 95\%
CL upper limit $\brinv < 11\%$ which was obtained
assuming a VBF production cross sections of
$h_{125}$ according to the SM
prediction~\cite{ATLAS-CONF-2020-052}.
As such, the direct limit can
in principle only be
applied for $c_{\rm uni} = 1$, in which
case the production cross sections of the
state at $125\gev$ considered here agree
with the SM predictions.
However, for $c_{\rm uni} \ne 1$
the direct limit on $\brinv$ can still be
applied in an approximate form. Taking into
account that the VBF production cross sections
is given by $\sigma_{\rm VBF} = c_{\rm uni}^2
\times \sigma_{\rm VBF}^{\rm SM}$
at leading order, with
$\sigma_{\rm VBF}^{\rm SM}$ being the SM
cross section, the number of signal events
would be enhanced or reduced by the factor
$c_{\rm uni}^2$ depending on whether
$c_{\rm uni}$ is larger or smaller than one,
respectively. Under the assumption that
the kinematical shape of the signal events
is not modified substantially, which can
be expected to be the case for small
deviations of $c_{\rm uni}$ from unity,
one can therefore
apply the upper limit $\brinv < 11\% / c_{\rm uni}^2$
from direct searches for $h_{125} \to \mathrm{inv}$
in order
to account for the modification of the
production cross section $\sigma_{\rm VBF}$
if $c_{\rm uni} \ne 1$. We will call this limit
an extrapolation of the direct limit on $\brinv$,
and this extrapolated limit is shown as the
orange dotted-dashed line in \reffi{fig:BRinvCeff}.

One can see in \reffi{fig:BRinvCeff} that there
is a flat direction with $\Delta \chi^2 \simeq 0$
in the $\{\cuni,\brinv\}$ plane. Along this flat direction, values of 
$\cuni > 1$ give rise to an enhancement of
the cross sections for the production of
$h_{125}$, but at the same time values of
$\brinv > 0$ suppress the decay modes of
$h_{125}$ decaying into SM particles
(see also \citere{Bechtle:2014ewa}).
As a result,
for each value of $\cuni > 1$ there is a value
of $\brinv > 0$ for which production cross sections times branching
ratios remain equal to the
SM predictions. Accordingly, we find
that in the range $c_{\rm uni} \gtrsim 1.015$
the direct upper limit on
$\brinv$ (orange lines)
is stronger than the indirect
limit on $\brinv$ resulting from the fit to the
signal-rate measurements of $h_{125}$.
For $\cuni \gtrsim 1.085$ the values of $\brinv$
that are required in order to cancel the
enhancement of the production
cross sections of $h_{125}$ are larger than
the direct limit, such that the combination
of the indirect constraints from signal-rate
measurements and the direct constraint on
$\brinv$ is able to exclude this part of
the parameter plane entirely.

In contrast to the case with $\cuni > 1$,
in models in which the couplings of $h_{125}$
are suppressed compared to the SM, i.e.~$\cuni < 1$,
one can observe that the indirect limit on
$\brinv$ from the fit to the signal-rate
measurements (black solid line) is always
substantially stronger than the direct
limit from searches for $h_{125} \to \mathrm{inv}$
(orange dotted-dashed line).
In this region of the parameter plane,
both the values of $\cuni < 1$ and
$\brinv > 0$ give rise to a suppression
of the signal rates of $h_{125}$. This is why
we observe that, the smaller the value of
$\cuni$, the smaller is the indirect upper limit on
$\brinv$ resulting from the signal-rate
measurements of $h_{125}$.
Finally we note that,
according to the discussion of
\refse{sec:smlikecpls},
also for $\cuni = 1$ the indirect
limit is stronger than the direct limit,
as indicated with the orange star.

An approximate estimate of the allowed parameter space, corresponding to the region delimited by the two black lines in Fig.~\ref{fig:BRinvCeff}, is given at 95\% CL by the condition
\begin{equation}
0.520 \, \brinv+0.960\,<\, \cuni \,<\,
0.565 \,  \brinv+1.043 \,.
\label{eq:constraintsBRinv_cuni}
\end{equation}
The fact that for $\cuni < 1$ the indirect constraints
provide the strongest constraints on $\brinv$,
whereas the direct limits are (in the considered
scenario) essentially irrelevant, is important
for phenomenological studies in models
with extended Higgs sectors featuring
a gauge-singlet scalar.
Then the deviations
of the universal coupling modifier
$c_\text{uni}$ to unity is generated
via the mixing of the SM-like
Higgs boson with the singlet scalar, and
the coupling coefficient can be written
as $c_\text{uni}=\cos \theta \simeq 1- \theta^2/2$
for small $\theta$, where $\theta$
is the mixing angle. 
In such case, the constraints from 
\refeqq{eq:constraintsBRinv_cuni} translate
in very good approximation into
\begin{equation}
\brinv < 0.078 \left( 1 -  \left(
\dfrac{\theta}{0.285} \right)^2  \right)\ ,
\label{eq:upperboundtheta}
\end{equation}
which implies that in the
limit $\brinv \rightarrow 0$ values
$\theta < 0.285$ are excluded and reciprocally
if $\theta \rightarrow 0$, $\brinv$ is constrained
to be $\brinv < 0.078$, which is stronger than all
current direct bounds listed in
\refta{tab:directlims}.\footnote{If the invisible
decay of $h_{125}$ is absent, and modifications of
the properties of $h_{125}$ arise only by means
of a universal coupling modification, we find
in a one-dimensional parameter
estimation 95\%~CL limits of
$\cuni > 0.968$, or equivalently
$\theta < 0.253$.}
We will discuss the parameter constraints resulting from the analysis of this section in two
concrete BSM theories, a hidden-sector composed
of a fermionic DM candidate as well as a
singlet scalar in \refse{sec:singletportal},
and a construction featuring (pseudo)
Nambu-Goldstone bosons in \refse{sec:PNGB}.

\subsection{\texorpdfstring{$h_{125}$}{h125} with non-universally
modified couplings
and a non-zero \texorpdfstring{$\mathrm{BR}_{\rm inv}$}{BRinv}}
\label{sec:nonunicoupling}

In the previous section we investigated
the case in which the couplings of the
Higgs boson $h_{125}$ are modified via
a universal coupling coefficient $\cuni$.
Although this procedure captures the
modifications that arise in many BSM constructions featuring additional scalars, there
is also a wide range of models in which
the modifications to the couplings of
$h_{125}$ compared to the SM prediction
cannot be captured in terms of a single
coefficient. Since it is not feasible to use independent
coupling modifiers for each of the
couplings of $h_{125}$, we have to make
assumptions on the total number of
coupling modifiers that we include, and on
how the couplings of $h_{125}$ depend on
these modifiers.
As already discussed in
\refse{sec:limitsfromhs}, for the
case of non-universal coupling modifiers
we will consider the four independent coefficients
$c_V$, $c_u$, $c_d$ and $c_\ell$ defined
in \refeq{eq:defcouplingmodifiers}, and for the
loop-induced couplings to gluons and photons
we assume that no additional BSM loop-contributions
play a role.
Thus, including $\mathrm{BR}_{\rm inv}$
as an additional free parameter,
\texttt{HiggsSignals} can be used to
determine $\Delta \chi^2$
as a function of up to five independent parameters
(see also
\refta{tab:my_table} for a summary of the combinations
of coupling modifiers for each case
considered in this work).
In order to be able to present
the results in a clearer
fashion, we will restrict our analysis
in this section
to parameter scans with only two
independent coupling modifiers.
The corresponding parameter space can
be understood as a subspace of the
more complicated coupling configurations
allowed by the effective Lagrangian
shown in \refeq{eq:defcouplingmodifiers},
requiring further relations amongst the four
individual coupling modifiers
$c_V$, $c_u$, $c_d$ and $c_\ell$.
This approach allows us to show the allowed regions
of the two varied parameters in two-dimensional
plots, and we can investigate how these
regions change with increasing values
of~$\mathrm{BR}_{\rm inv}$.

As a first example, we show constraints in a
benchmark model commonly utilized by the CMS and
the ATLAS collaborations (see, for instance,
\citere{ATLAS:2016neq,
CMS:2018uag,ATLAS:2019nkf,
CMS:2022dwd,ATLAS:2022vkf}),
in which it is assumed
that the couplings of $h_{125}$ to fermions
are modified by a common factor $c_f = c_u =
c_d = c_\ell$, and, in addition, the coupling
to the massive vector bosons is allowed to vary,
i.e.~$c_V \neq 1$.
Such coupling modifications can be present
in various different UV-complete BSM theories.
For example, the presence of a gauge-singlet scalar
that mixes with $h_{125}$ gives rise to
$c_V = c_f < 1$, as discussed in
\refse{sec:uni}. Other examples comprise the 2HDM
with Yukawa structure of type~I and extensions
thereof, where $c_V \leq 1$, and where $c_f$ can be
both smaller and larger than 1.
Also extensions of the SM containing SU(2)
triplet scalar fields that mix with the
SM Higgs field can give rise to coupling
modifications in terms of $c_V$ and $c_f$
as discussed above, where in contrast to
2HDMs also $c_V > 1$ is possible (at tree level).
All these models can be extended
by a hidden sector in order to
feature one or more valid DM
particles, and where the Higgs sector can act
as the portal between the hidden and the SM
sector (see, for instance,
\citeres{FileviezPerez:2008bj,No:2015xqa,
Arina:2019tib,Biekotter:2021ovi}). Thus,
if the decay of $h_{125}$ into states of
the hidden sector
is kinematically allowed, it
is interesting to analyze how the experimental
limits on the coupling modifiers are modified
for different values of the invisible branching
ratio of $h_{125}$.

\begin{figure}[t]
  \centering
  \includegraphics[width=0.98\textwidth]{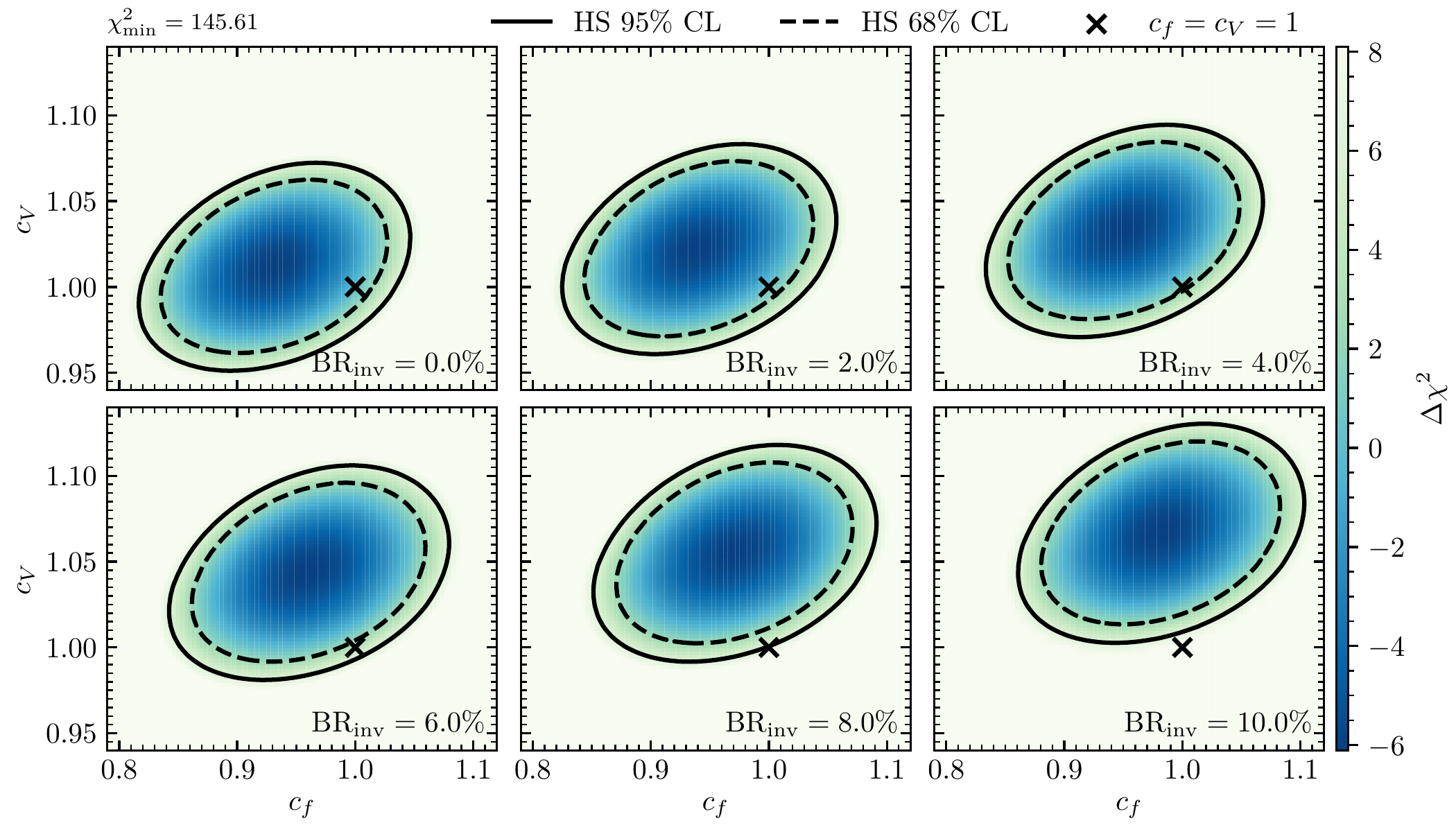}
  \caption{\small
  Result of the global fit to the Higgs-boson
  signal-rate measurements
  in the plane $\{c_f,c_V\}$ for
  different values of $\brinv$.
  The color coding indicates the value
  of $\Delta \chi^2$ resulting from
  the \texttt{HiggsSignals} analysis.
  The black solid and dashed lines
  indicate the allowed regions at
  the $95\%$ and the $68\%$ confidence
  level, respectively.
  The black crosses  indicate the point
  at which the couplings of $h_{125}$
  are identical to the SM predictions,
  i.e.~$c_f = c_V = 1$.}
  \label{fig:kapvkapf}
\end{figure}

In \reffi{fig:kapvkapf} we show the values of
$\Delta \chi^2$ as a function of $c_V$
and $c_f$ for different choices of
$\mathrm{BR}_{\rm inv}$. The upper left plot
is obtained assuming that there is no invisible
decay mode of $h_{125}$, and the other plots
are obtained by increasing $\mathrm{BR}_{\rm inv}$
in steps of 2\% up to $\mathrm{BR}_{\rm inv} =
10\%$ (lower right plot), covering the
allowed range of $\brinv$ according to the
direct limits. The solid and the dashed
contours indicate the 95\% and 68\% CL
exclusion limits based on the \texttt{HiggsSignals}
$\chi^2$-fit to the cross-section measurements
of $h_{125}$. One can see that with increasing
value of $\mathrm{BR}_{\rm inv}$ the allowed
regions of the coupling modifiers move to larger
values. This observation is in line with the
discussion in \refse{sec:uni}, where
we demonstrated that
the suppression of the decay modes into SM
particles due to the presence of the invisible
decay mode can be compensated by en enhancement
of the production cross sections as a result
of coupling modifiers larger than one
(see also \reffi{fig:BRinvCeff}).
In contrast to the location of the allowed regions,
the shape and the size of the allowed
regions are practically unchanged in all plots.
Furthermore, also the values of $\Delta \chi^2$
featured by the best-fit points for the
different values of $\mathrm{BR}_{\rm inv}$
are effectively the same.
Based on the results shown in \reffi{fig:kapvkapf}, one
can conclude that for BSM scenarios in which
the restriction $c_V \leq 1$ applies, the presence
of the invisible decay mode gives rise to a
decrease of the allowed ranges of 
both $c_V$ and $c_f$.
On the other hand, for models like the Higgs-triplet
extension in which one can find
values of $c_V > 1$, one can see that the
presence of an invisible decay mode of
$h_{125}$ with sizable branching ratios can
open up parameter space regions that would
be excluded if such a novel decay mode is not present.
It should be noted also that the best-fit
points do not lie at the point at which
the couplings of $h_{125}$ are identical
to the SM predictions (black crosses) even
if the decay mode $h_{125} \to \mathrm{inv}$
is absent. Instead, for $\brinv = 0\%$ the
parameter points with the smallest values
of $\chi^2$ are found for $c_V > 1$ and
$c_f < 1$, featuring values of $\Delta \chi^2
\approx -6$.\footnote{The values of $\Delta \chi^2 < 0$
are driven here by the following
measurements, where for simplicity
we only quote inclusive signal
strength measurements here, and we
remind the reader that $\mu = 1$
according to the SM predictions: (i)
$\mu(\mathrm{ttH},H \to b \bar b) =
0.35^{+0.36}_{-0.34}$ reported by
ATLAS~\cite{ATLAS:2021qou},
where it should be taken into account that
the measurement uncertainty is dominated by
systematic uncertainties regarding the
background estimation, (ii)
$\mu(\mathrm{ggH}, H \to \tau^+ \tau^-) =
0.59^{+0.28}_{-0.32}$ and
$\mu(\mathrm{VBF}, H \to \tau^+ \tau^-) =
1.39^{+0.56}_{-0.47}$ reported by
CMS~\cite{CMS:2021sdq}, (iii) $\mu(\mathrm{ttH}) =
0.58^{+0.36}_{-0.33}$ reported by
ATLAS~\cite{ATLAS-CONF-2019-045},
where the signal
extraction is affected by the fact that
the ttW background was found to be larger
than the SM prediction.}
Hence, if one would define the exclusion regions
based on $\Delta \chi^2 = \chi^2 - \chi^2_{\rm min}$,
instead of $\Delta \chi^2 = \chi^2 - \chi^2_{\rm SM}$
as applied throughout this paper, one would
obtain even stronger exclusion limits in
the $\{c_f,c_V\}$ plane (see also
the related discussion in \refse{sec:limitsfromhs}).
The location of the best-fit points
as a function of $\brinv$ moves to larger values
of both coupling modifiers investigated here
for increasing values of $\brinv$.

\begin{figure}[t]
  \centering
  \includegraphics[width=0.98\textwidth]{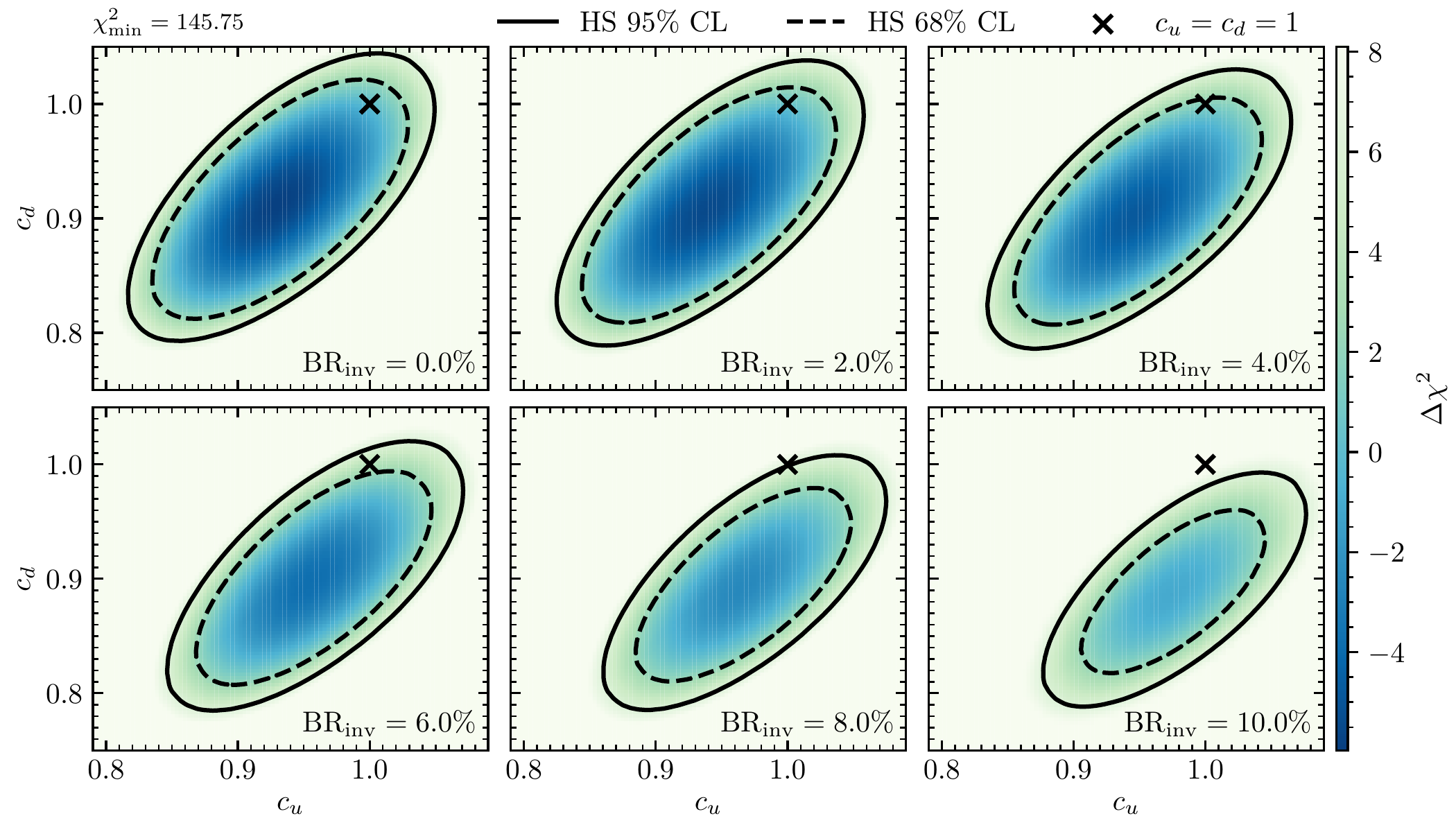}
  \caption{\footnotesize
  As \reffi{fig:kapvkapf}, but in the plane
  $\{c_u,c_d\}$ and assuming that
  $c_\ell = c_d$ and $c_V = 1$.}
  \label{fig:kapukapd}
\end{figure}

A second BSM scenario that we consider is
the case in which the coupling of
$h_{125}$ to gauge bosons is not modified
compared to the SM predictions
($c_V = 1$), in contrast to all the other
scenarios discussed above, but in which
the couplings to the fermions
can deviate from the SM ($c_u,c_d,c_\ell \neq 1$).
We furthermore impose the condition
$c_\ell = c_d$ in order to stick to a
parameter scan with two free parameters for
a given value of $\brinv$.
UV-complete models in which coupling
modifications of this form
arise comprise generically extensions of
the SM containing a second SU(2) doublet scalar
field in which the so-called Yukawa
structure of type~II (see \refse{sec:2HDM}
for details) is imposed, which is also the Yukawa
structure that arises in supersymmetric
extensions of the SM
(see e.g.~\citere{Aguilar-Saavedra:2021ijx}).
We show the $\chi^2$-distribution in the
$(c_u,c_d)$ for different values of
$\brinv$ as they were obtained with the
help of \texttt{HiggsSignals}
in \reffi{fig:kapukapd}, where the color
coding and the definitions of the CL exclusion
limits is as in \reffi{fig:kapvkapf}.
One can observe that increasing values
of $\brinv$ shifts the allowed parameter regions
to larger values of $c_u$ and smaller values
of $c_d = c_\ell$. The reason for the preference
of larger values of $c_u$ if $\brinv > 0$ lies
in the fact that the enhancement of the coupling
of $h_{125}$ to top quarks gives rise to an
enhancement of the gluon-fusion production
cross section that partially compensates the
suppression of the branching ratios for
the ordinary decay modes of $h_{125}$.
However, the VBF production mode, for instance,
is unchanged since $c_V = 1$, such that
the minimal values of $\chi^2$ 
that are found for each value
of $\brinv$ considered increase with
$\brinv$,
in contrast to the results
depicted in \reffi{fig:kapvkapf}.
The trend towards lower values of $c_d$
with increasing values of $\brinv$ has its
origin in the fact that a suppression of
the coupling of $h_{125}$ to bottom quarks
suppresses the total width of $h_{125}$.
As a consequence, sizable values of
the decay width $\Gamma_{\rm inv}$
(see \refeqq{eq:gammainv}) can be
be in agreement with the signal-rate
measurements of $h_{125}$, because
the measured branching ratios of the ordinary
decays of $h_{125}$
remain closer to the SM predictions.
However, since $c_V = 1$ and $c_\ell = c_d$
in this scenario, the suppression of the
couplings of $h_{125}$ to tau leptons if
$c_d < 1$, and the unchanged partial decay
with for the decay modes $h_{125} \to W^+W^-$
and~$ZZ$
still result in significant modifications
of the branching ratios of $h_{125}$ compared
to the SM predictions, such that overall the
fit result becomes worse with increasing
values of $\brinv$.
As a result, even though for all possible
values of $\brinv$ one can find ranges of
$c_u$ and $c_d$ that describe the experimental
data regarding $h_{125}$ as accurate as the
SM, the size of the allowed regions of
the coupling modifiers decreases and the
fit result deteriorates with increasing value
of $\brinv$.

The two examples discussed above demonstrate
that the presence of sizable values of
$\brinv$ can improve or worsen the fit result
to the Higgs-boson measurements, depending on
which ranges of the coupling modifiers are
considered, and depending on whether additional
relations between the different modifiers
are imposed.
Since in both cases viable parameter ranges
of the coupling modifiers survived even for
values of $\brinv \simeq 10\%$, it becomes
apparent that both the direct and
the indirect constraints
on $\brinv$ have to be considered in order
to be in agreement with the measurements
regarding $h_{125}$.
In \refse{sec:2HDM} we will analyze the
impact of the values of $\brinv$ on the
allowed parameter regions in UV-complete
models in which the couplings of the
state at $125\gev$ can be captured by
the full set of coupling modifiers considered
in our analysis, as defined in
\refeqq{eq:defcouplingmodifiers}.
Therein, as an illustrative example
of popular BSM theories for which our results
are relevant, we will focus on
models containing, in addition to the hidden
sector as the origin of the invisible decay
mode of~$h_{125}$, a second Higgs doublet.

\section{Application to hidden sector models}
\label{sec:application}
In this section we will apply the
indirect constraints on $\brinv$ obtained previously to concrete BSM scenarios featuring a
hidden sector. We will start by considering
Higgs-portal dark-matter models in
\refse{sec:Higgsportal}
in which no coupling modifications
of $h_{125}$ are present, i.e.~$c_i = 1$.
As demonstrated in \refse{sec:smlikecpls}, in this
case the indirect constraints on~$\brinv$ are
stronger than the direct constraints,
such that the latter do not have to be considered.
Afterwards we will discuss in
\refses{sec:singletportal}--\ref{sec:2HDM}
models featuring a hidden sector and
an extended Higgs sector, in which
the couplings of $h_{125}$ to ordinary
matter deviate from the SM predictions.
Following the discussions in \refse{sec:uni}
and \refse{sec:nonunicoupling},
here depending on the
values of the coupling modifieres
$c_i$ predicted in each model, both
the indirect or the direct constraints
on $\brinv$ can be stronger, such that
both constraints will have to be considered
and their complementarity can be studied.

\subsection{The Higgs portal dark matter model}
\label{sec:Higgsportal}

\begin{figure}[t]
    \centering
  \includegraphics[width=0.32\textwidth]{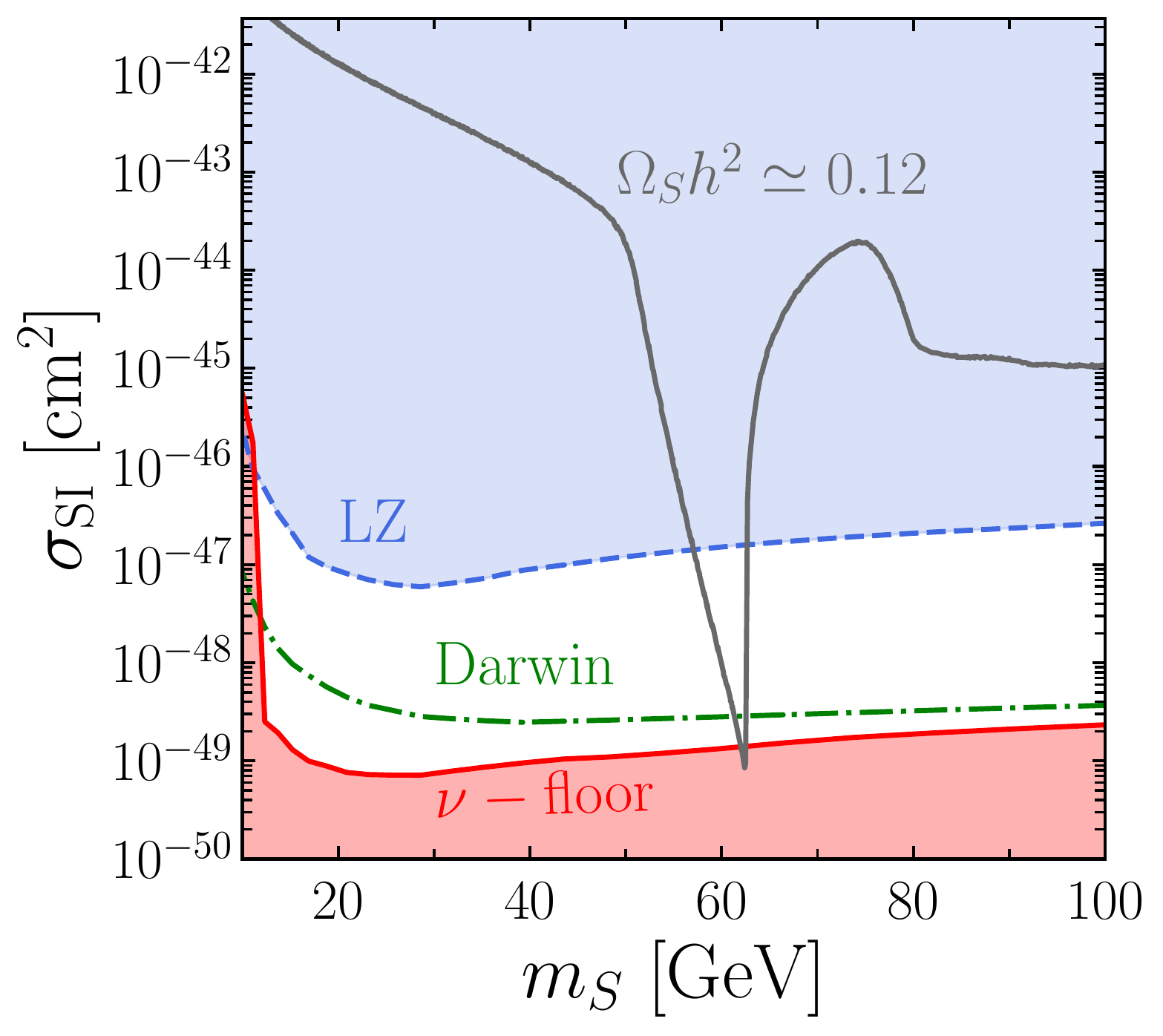}
  \includegraphics[width=0.32\textwidth]{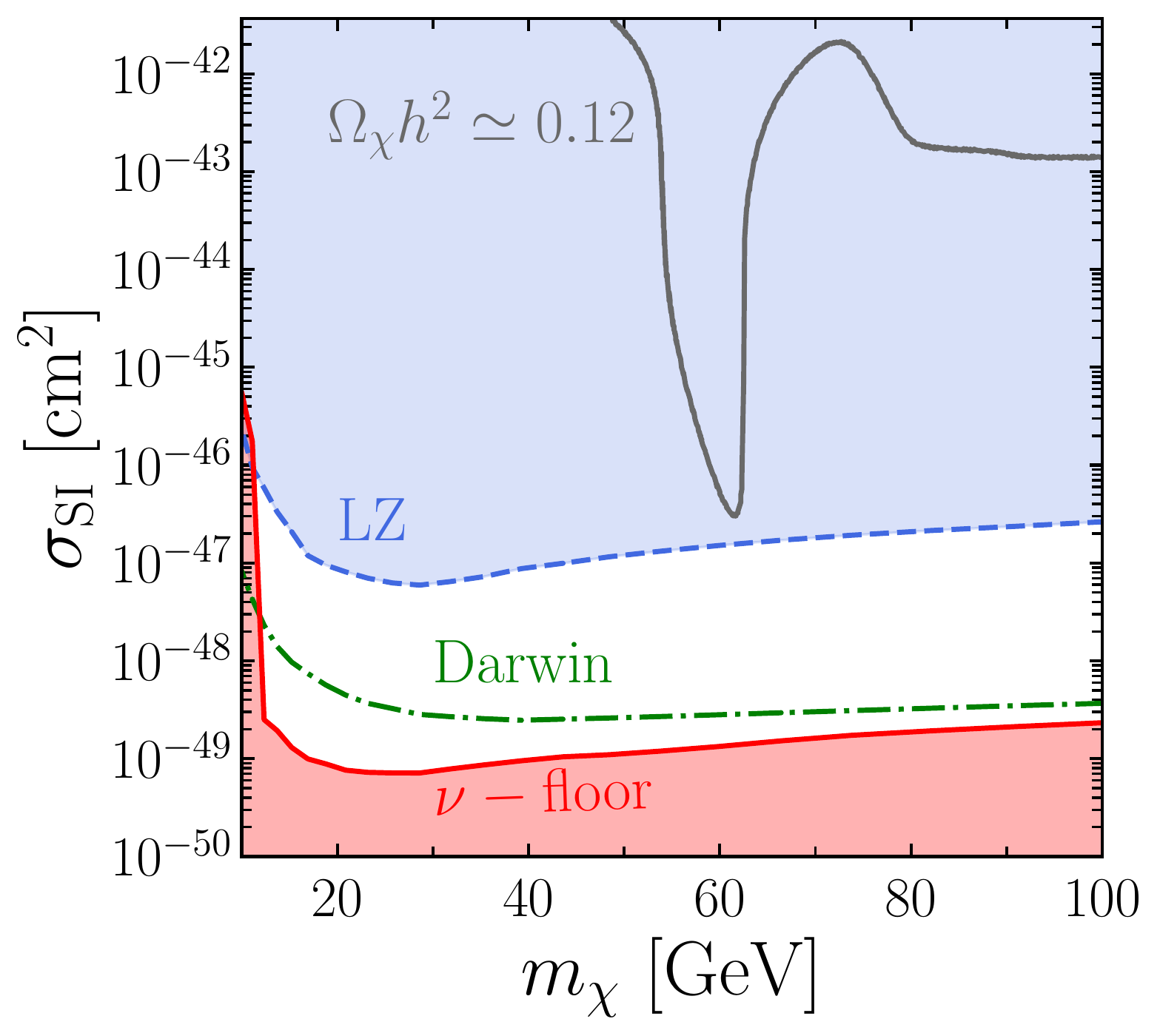}
  \includegraphics[width=0.32\textwidth]{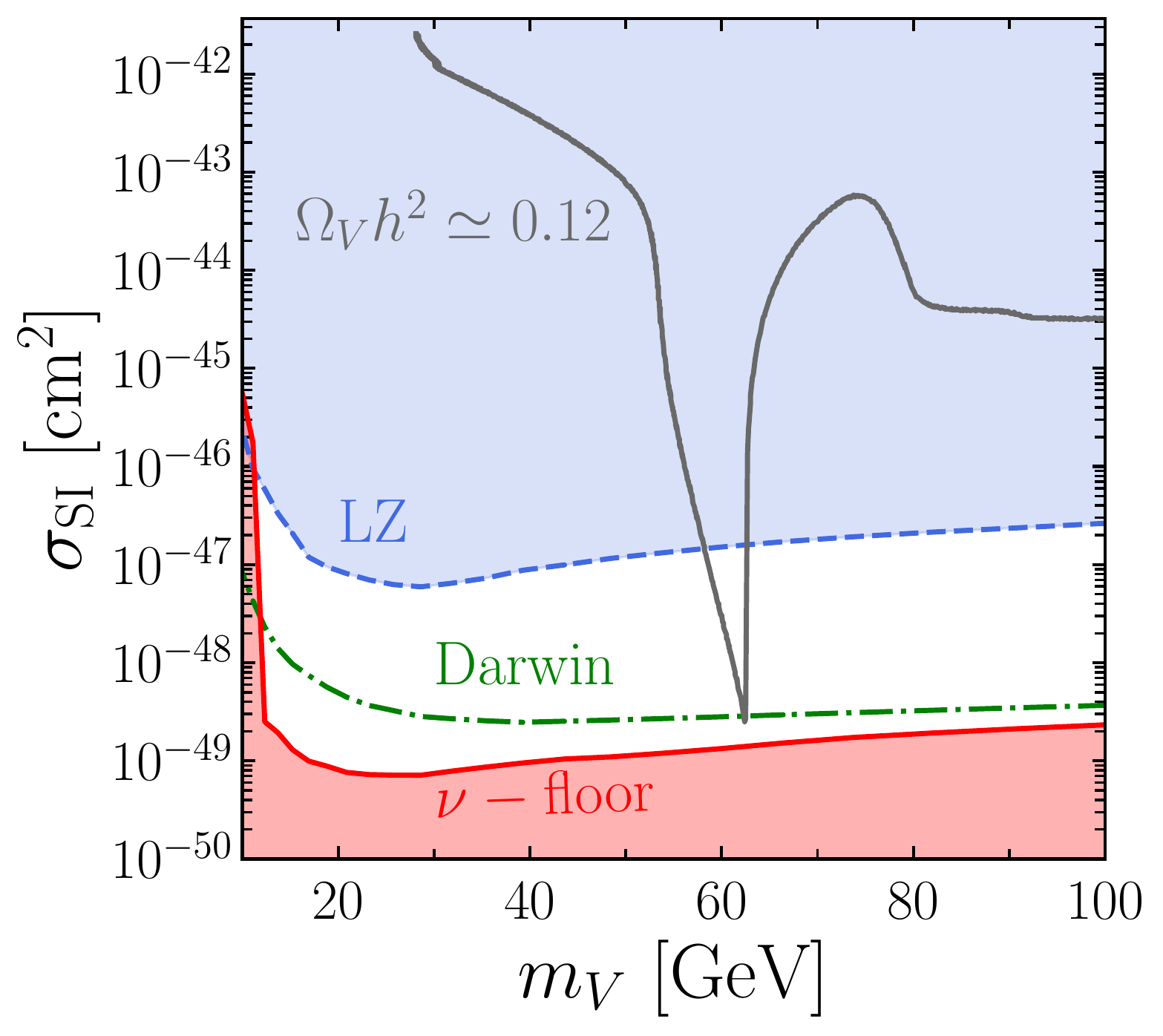}
    \caption{\small
    $\sigma_{\rm SI}$ as a function of the DM mass
    for the Higgs-portal
    parameter space in which the relic
    abundance as observed by Planck is
    reproduced (dark grey), for scalar
    DM (left), 
    Dirac fermion DM (center) and 
    vector DM (right).
  Constraints from LZ, projection for Darwin and the neutrino floor are represented in blue (dashed), green (dotted-dashed) and red (solid) respectively.}
    \label{fig:HiggsportalDD}
\end{figure}

One of the simplest approaches to accommodate
a valid DM candidate in a BSM scenario consists
of the so-called Higgs portal scenario. Here it is assumed
that the discovered Higgs boson $h_{125}$ is the
only portal to the dark sector, such that no
direct (gauge) interactions with the SM fermions
and the gauge bosons exist.
We will consider here three possible kind of DM
candidates coupled to the Higgs boson via $\mathcal{L}=\mathcal{L}_\text{SM}+\mathcal{L}_\text{DM}$ where $\mathcal{L}_\text{SM}$ is the SM Lagrangian and $\mathcal{L}_\text{DM}=\mathcal{L}_S$ for real scalar DM ($\equiv S$), $\mathcal{L}_\text{DM}=\mathcal{L}_\chi$ for Dirac fermion DM ($\equiv  \chi$) and $\mathcal{L}_\text{DM}=\mathcal{L}_V$ for real vector DM ($\equiv  V^\mu$) with
\begin{equation}
\mathcal{L}_S \, = \, -\dfrac{1}{4} \lambda_S |H|^2 S^2 \,, \quad
\mathcal{L}_\chi \, = \, -\dfrac{1}{4} \dfrac{\lambda_\chi}{\Lambda}|H|^2 \bar \chi \chi \, , \quad
\mathcal{L}_V \, = \, -\dfrac{1}{4} \lambda_V |H|^2V^\mu V_\mu \, ,
\end{equation}
where $\lambda_{S,\chi,V}$ and $\Lambda$  are respectively dimensionless and dimension-one couplings. We parameterize the SM Higgs doublet in unitary gauge $H=(0, ~v+h)^T/\sqrt{2}$ with $v\simeq 246$ GeV and $h=h_{125}$ is the SM Higgs boson. In addition, in each of the three cases we consider
bare mass terms for our DM candidates.
We will denote
the physical scalar, fermion and vector
dark-matter masses after
electroweak symmetry breaking
by $m_{S,\chi,V}$, respectively.
The partial widths of the Higgs-boson
decays into a pair
of DM particles are given by~\footnote{The expression for the scalar DM is in
agreement with \citere{Casas:2017jjg} and
\citere{Cline:2013gha}. The expression for
the vector DM is in
agreement with Ref.~\cite{Lebedev:2011iq}.}
\begin{align}
    \Gamma_{h\rightarrow SS} \, = & \, \dfrac{\lambda_S^2 v^2}{128\pi m_h }   \left(1-\frac{4 m_{S }^2}{m_h^2}\right)^{1/2}    \nonumber \,, \\
    \Gamma_{h\rightarrow \bar \chi \chi} \, = & \, \dfrac{\lambda_\chi^2 v^2 m_h}{128\pi \Lambda^2 }   \left(1-\frac{4 m_{\chi }^2}{m_h^2}\right)^{3/2}     ~  \nonumber \,, \\
    \Gamma_{h\rightarrow VV} \, = & \, \dfrac{\lambda_V^2 v^2 m_h^3}{512\pi m_V^4 }  \left( 1- 4\dfrac{m_V^2}{m_h^2} +12 \dfrac{m_V^4}{m_h^4} \right)  \left(1-\frac{4 m_{V }^2}{m_h^2}\right)^{1/2}     ~.
\end{align}
Higgs-portal models are strongly constrained by the
null-results of direct-detection experiments. Currently, the most stringent experimental constraints
on the DM-nucleon spin-independent scattering cross section $ \sigma_\text{SI}$ were reported by the Xenon1T~\cite{Aprile:2018dbl},
PandaX-4T~\cite{PandaX-4T:2021bab} and the
LUX-ZEPLIN (LZ) experiments~\cite{LZnew},
whose most recent results exclude $\sigma_\text{SI}\gtrsim
10^{-47}\,\text{cm}^2$ for a $50$ GeV DM mass and
up to  $\sigma_\text{SI}\gtrsim 10^{-45}\,\text{cm}^2$ for $10$ TeV DM mass.
The sensitivity of the upcoming Darwin experiment~\cite{Aalbers:2016jon} should improve the current bounds from LZ by more than an order of magnitude, and will almost reach the so-called neutrino
floor~\cite{Billard:2013qya}.
The direct detection cross
section with a nucleon $N$, mediated by the Higgs boson,
can be expressed for the three scenarios
considered here as
\begin{equation}
   \sigma^{S}_\text{SI}\,= \,\dfrac{\lambda_S^2 \mu_{S N}^2 m_N^2}{16\pi m_S^2 m_h^4}  f_N^2  ~,  \quad
    \sigma^{\chi}_\text{SI}\,= \,\dfrac{\lambda_\chi^2 \mu_{\chi N}^2 m_N^2}{16\pi \Lambda^2 m_h^4}  f_N^2 ~,  \quad
   \sigma^{V}_\text{SI}\,= \,\dfrac{\lambda_V^2 \mu_{V N}^2 m_N^2}{16\pi m_V^2 m_h^4}  f_N^2 ~,
   \label{eq:directdetection_Higgsportal}
\end{equation}
with $\mu_{AB}=m_A m_B/(m_A+m_B)$.\footnote{The expression for scalar DM is in agreement with Ref.~\cite{Casas:2017jjg,Arcadi:2021mag,Cline:2013gha}, and the one for vector DM is in agreement with \citeres{Lebedev:2011iq,Arcadi:2021mag}}
The nucleon form factor $f_N\sim 0.3$ is defined
as $f_N \equiv f_{Tu}^{(N)}+f_{Td}^{(N)}+
f_{Ts}^{(N)}+(6/27)f_{TG}^{(N)}$ where
$f_{Tq}^{(N)}\equiv \langle N | m_q \bar q q | N \rangle/m_N$ is the contribution from a quark $q$ to the nucleon mass 
and $f_{TG}^{(N)}\equiv 1 - \sum_q f_{Tq}^{(N)}$ is the gluon contribution.\footnote{Numerical values for these form factor can be found in Ref.~\cite{DelNobile:2013sia}.} 

\begin{figure}[t!]
    \centering
  \includegraphics[width=0.48\textwidth]{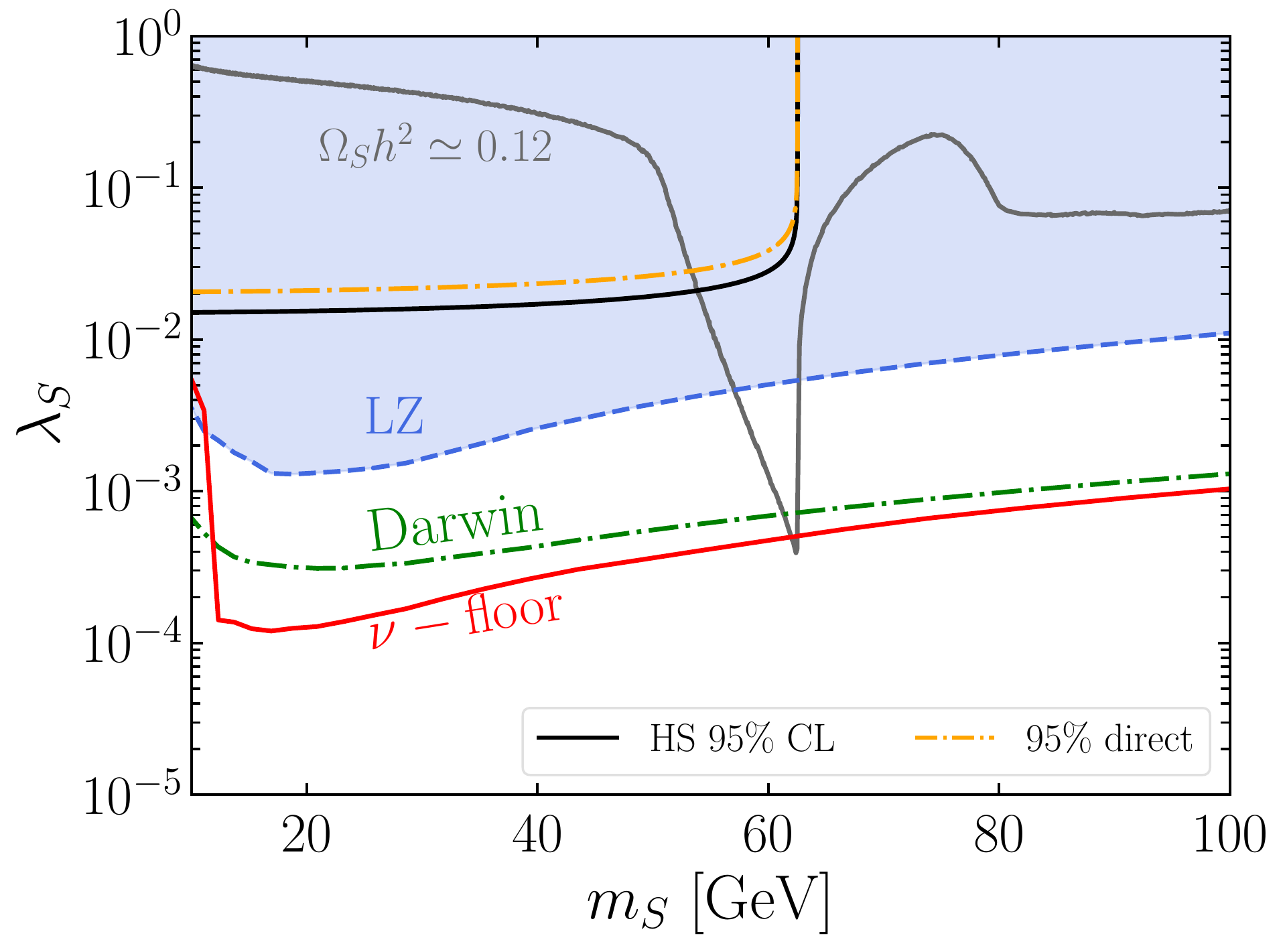}
   \includegraphics[width=0.48\textwidth]{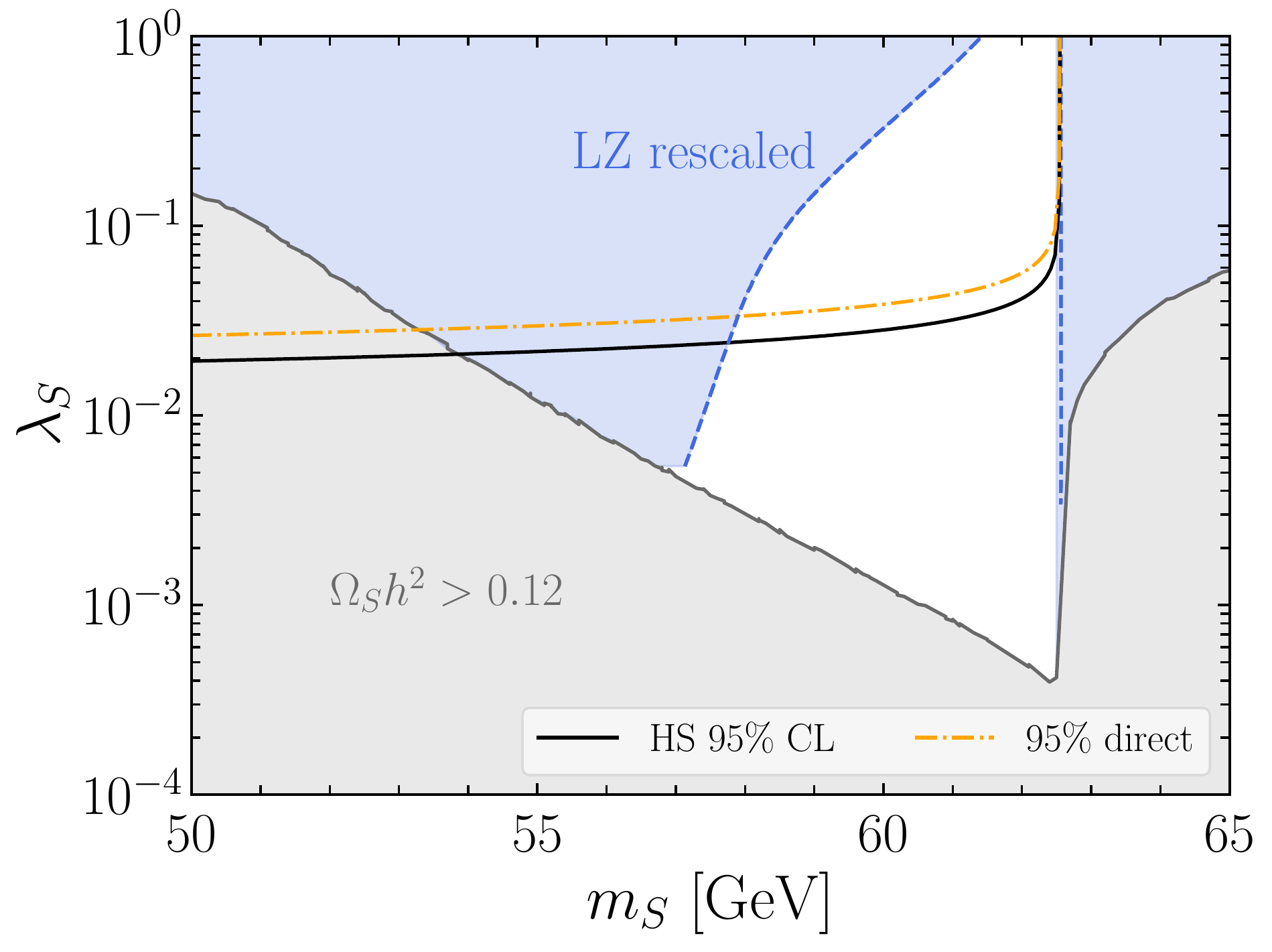}
  \includegraphics[width=0.48\textwidth]{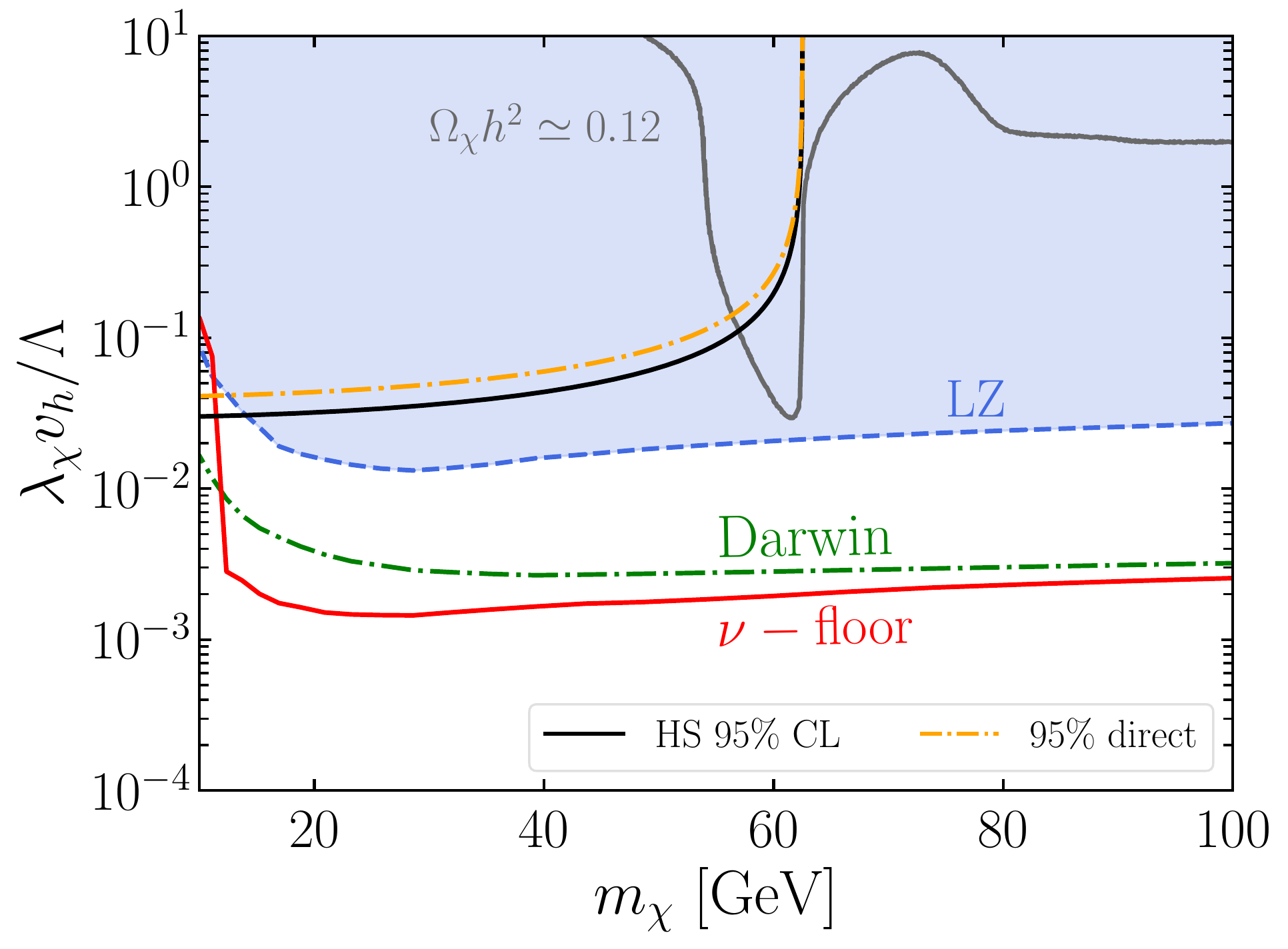}
   \includegraphics[width=0.48\textwidth]{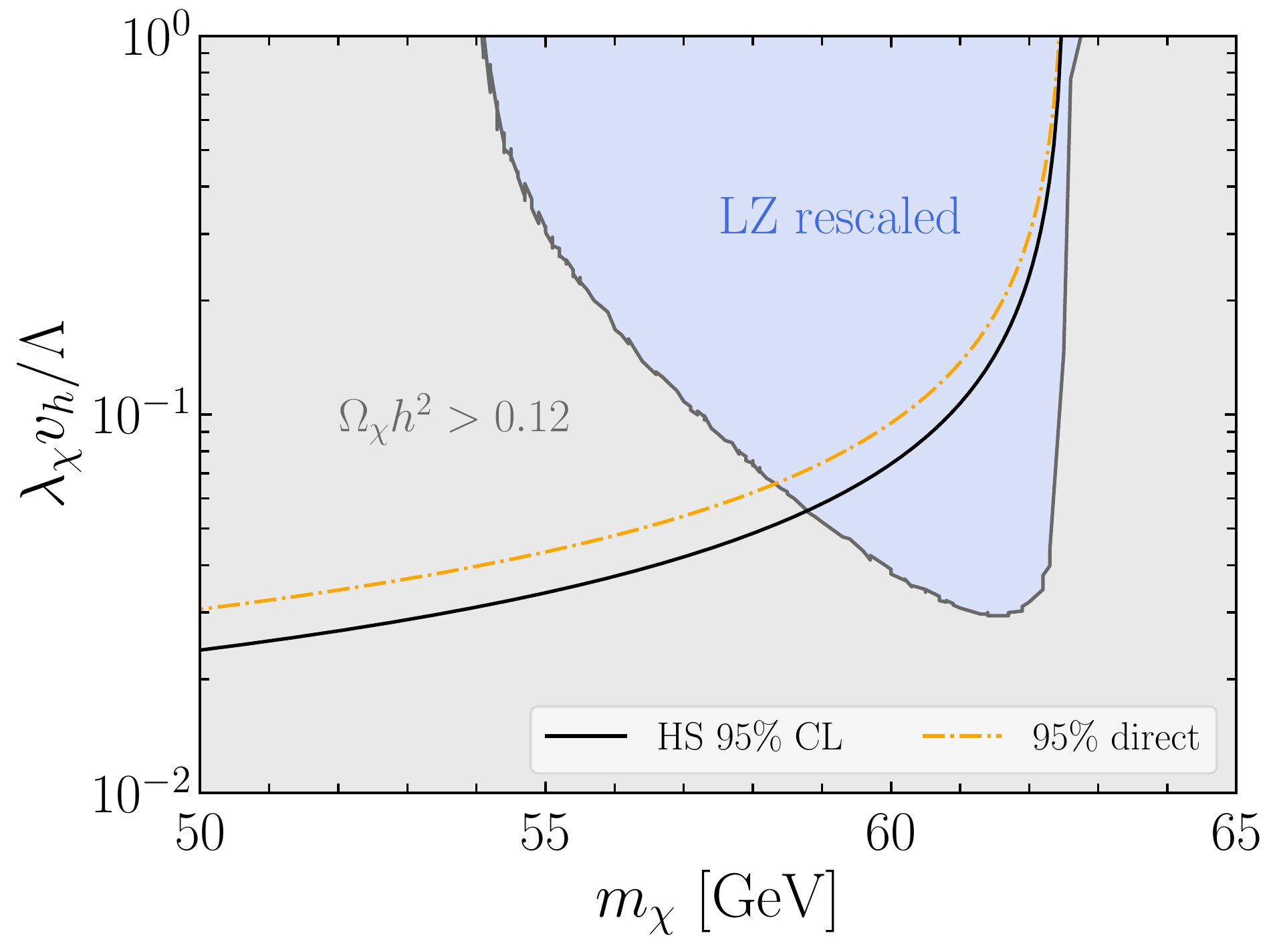}
   \includegraphics[width=0.48\textwidth]{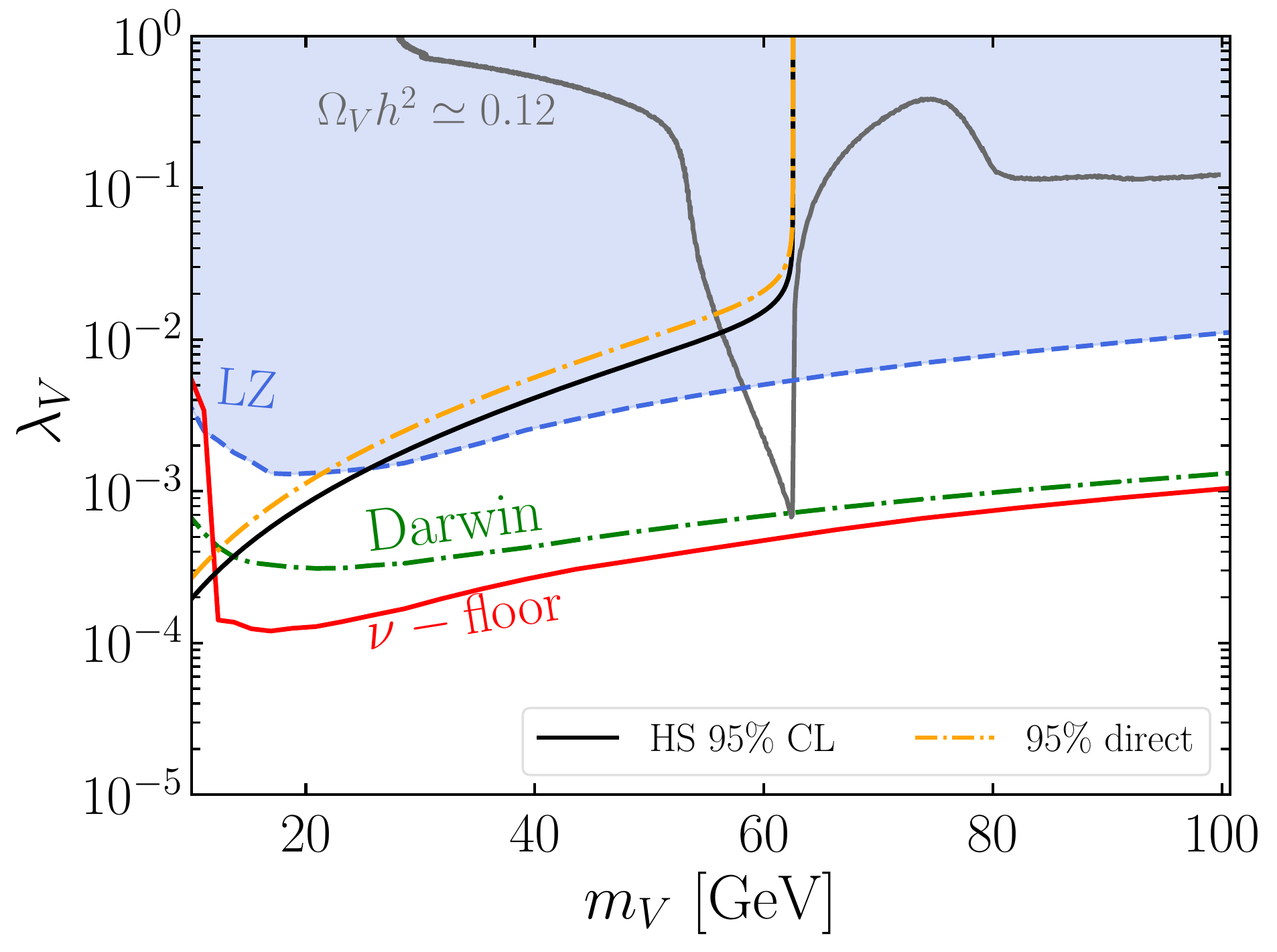}
   \includegraphics[width=0.48\textwidth]{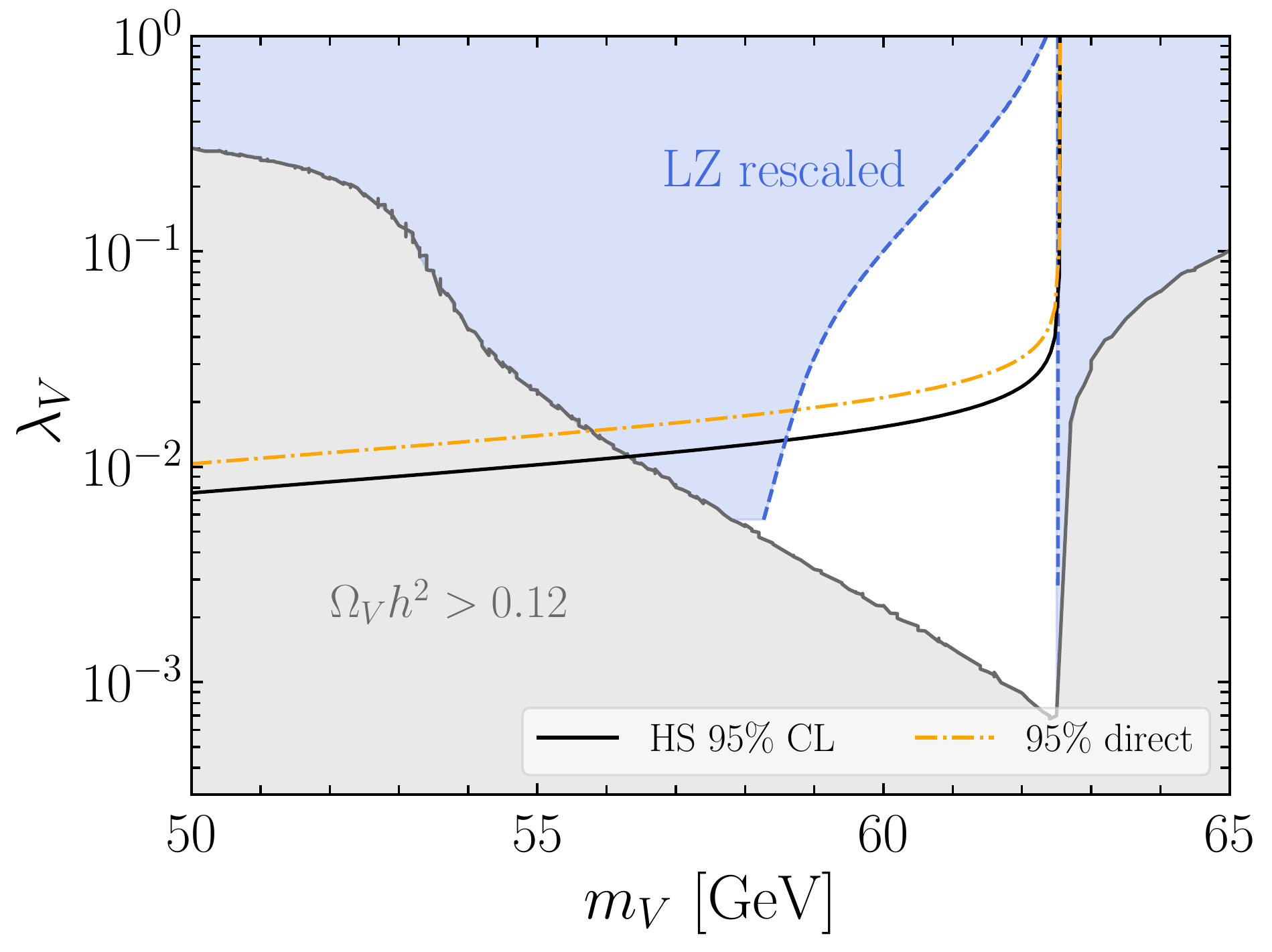}
    \caption{\small
    Higgs-portal
    parameter plane of dimensionless
    couplings 
    and the DM mass for
    scalar DM $S$ (first row), 
    Dirac fermion DM $\chi$ (second row) and
    vector DM $V$ (third row). The dark-grey lines
    indicates the parameter space
    reproducing the relic abundance as
    observed by Planck.
    Constraints from LZ, projection for Darwin
    and the neutrino floor are represented
    in dashed blue, dotted-dashed green
    and solid red respectively.  
    In the plots on the right, the LZ constraint
    is rescaled according to the predicted
    DM abundance via \refeqq{eq:rescaling_LZ}, and
    the light-grey areas represent regions where
    the DM relic abundance exceeds the Planck value.
    The solid black and dotted-dashed orange lines
    represent respectively our constraints derived
    using \texttt{HiggsSignal} and constraints
    from direct searches for the invisible
    decay of $h_{125}$.}
    \label{fig:Higgsportal}
\end{figure}

After implementing the models in \texttt{Feynrules}~\cite{Alloul:2013bka}, we used \texttt{micrOMEGAs}~\cite{Belanger:2018ccd,Belanger:2013oya} to compute numerically both relic abundance and direct detection cross sections.\footnote{With \texttt{micrOMEGAs}, we obtained numerical results for the direct detection cross section within a $10\%$ agreement with respect to our analytical expression of Eq.~(\ref{eq:directdetection_Higgsportal}).}
In \reffi{fig:HiggsportalDD} we represented with
a dark gray line
in the plane $\{\sigma_{\rm SI},m_{S,\chi,V}\}$
the parameter points that satisfy the relic
density as determined by the Planck
collaboration~\cite{Aghanim:2018eyx} within
a $3\sigma$ interval around the best-fit value,
$\Omega_{ \chi} h^2  \in [0.11933 - 3 \times 0.00091, 0.11933 + 3 \times 0.00091]$ for the three kind of dark matter candidates of the Higgs portal scenarios considered in this work. We also indicate the constraints from LZ in dashed blue,
sensitivity prediction for Darwin in dotted-dashed
green and the neutrino floor in solid red.
From Fig.~\ref{fig:HiggsportalDD}, one can see that the Higgs portal models considered in this section are essentially only allowed for bosonic dark matter candidates and for a very narrow region of DM masses close to the resonance $ m_\chi \simeq m_{h}/2 \simeq 62.5~\text{GeV}$, if one assumes that the dark matter abundance saturates the Planck best fit value. The fermionic dark matter candidate considered in this section is entirely excluded by the most recent results from the LZ collaboration. \par \medskip

We represented the relic density and direction detection constraints in the left panels of Fig.~\ref{fig:Higgsportal} in terms of a dimensionless coupling as a function of the DM mass. In addition, we represented constraints on Higgs physics from~\texttt{HiggsSignal} and direct limits from ATLAS respectively in black and dotted-dashed orange line. For a given set of parameters, one can rescale the LZ constraints according to the predicted value for the relic abundance via
\begin{equation}
    \sigma_\text{SI}^\text{LZ}\,=\,    \sigma_\text{SI}^\text{LZ}|_{\Omega_\text{DM} h^2=0.11933} \left(\dfrac{0.11933}{\Omega_\text{DM} h^2} \right)\,,
    \label{eq:rescaling_LZ}
\end{equation}
where $\sigma_\text{SI}^\text{LZ}|_{\Omega_\text{DM} h^2=0.11933}$ is the LZ constraint assuming a DM abundance $\Omega_\text{DM} h^2=0.11933$ and $\Omega_\text{DM} h^2$ is the DM abundance predicted for a given set of parameters. The rescaled LZ constraint is represented on the plots in the right panels of Fig.~\ref{fig:Higgsportal} where the blue areas represent the excluded regions. One can see from this figure that such constraint weaken precisely around the peak value for bosonic DM candidates, as a smaller DM abundance $\Omega_\text{DM} h^2<0.11933$ is generated in this region. However, the fermionic dark matter candidate is completely excluded, even if it constitutes just a fraction of the total dark
matter.\footnote{Notice that in this work we
considered a scalar operator $\bar \chi \chi$
connected to the Higgs field.
Our statement here depends on the choice of
operators and could differ, for example,
for a pseudoscalar operator $\bar \chi \gamma_5 \chi$.
It remains to be investigated whether
one can evade the LZ constraints
upon inclusion of the pseudoscalar
operator, or whether the presence of
new mediators between the hidden sector and
the visible sector at or below the electroweak
scale have to be introduced in order to
predict a viable fermionic Higgs-portal DM
scenario~\cite{Freitas:2015hsa}
(see also \refse{sec:singletportal}).}

\par \medskip

One can relax the condition of achieving the correct relic abundance via a single Higgs-portal parameter by assuming the presence of additional DM annihilation channels as typically expected in constructions featuring a more complex hidden sector. In this case, the correct relic abundance could be achieved for a smaller value of the Higgs-portal dimensionless coupling. As can be seen in the left panels of Fig.~\ref{fig:Higgsportal}, constraints from Higgs physics would become the strongest, for fermion and vector DM, in the region of the parameter space corresponding to masses $\lesssim  m_h/2$. In addition, if one assumes a multi-component dark matter setup, direct detection constraints would have to be rescaled according to the local density of  the relevant dark matter component. The parameter space at small masses $\lesssim  m_h/2$ could open up and the constraints derived in this work would become the strongest. In such cases, for DM masses typically $ \lesssim m_h/2$, our constraints derived using \texttt{HiggsSignal}, independent of the local dark matter density, could be stronger than both direct detection bounds and from an extrapolation of direct searches for a invisible decay of the Higgs boson.

\subsection{Singlet portal dark matter}
\label{sec:singletportal}

The simplest possibility to extend the Higgs portal DM
scenario is to assume the presence
of an additional real singlet scalar field $\Phi$
that can act as a portal between the dark sector
and the visible sector. This can be achieved by considering a discrete $\mathbb{Z}_4$ symmetry
acting on $\Phi \rightarrow - \Phi$~\cite{Yaguna:2021rds} and by introducing components of a fermionic Dirac
DM candidate $\chi_{L,R}$ with opposite chiralities $\chi\equiv \chi_L + \chi_R$ that couples to $\Phi$ via
\begin{equation}
\mathcal L \supset - y_\chi \Phi \bar \chi_L \chi_R+\text{h.c.} \,,
\label{dmyuk}
\end{equation}
where $y_\chi$ is a Yukawa coupling.\footnote{Under the $\mathbb{Z}_4$ symmetry, the two chirality components of $\chi $ would transform as $\chi_R \rightarrow i \chi_R $ and $\chi_L \rightarrow - i \chi_L$.}
The most general scalar potential respecting
the $\mathbb{Z}_4$ symmetry is given by
\begin{equation}
V = \mu_H^2 H^\dagger H +
\frac{1}{2} \mu_\Phi^2 \Phi^2 +
\lambda_H \left( H^\dagger H \right)^2 +
\frac{1}{4} \lambda_\Phi \Phi^4 +
\frac{1}{2} \lambda_{\Phi H} \Phi^2 H^\dagger H \ ,
\label{eq:potentialsingletportal}
\end{equation}
where $\mu_{H,\Phi}$ and $\lambda_{H,\Phi,\Phi H}$ are respectively dimension-one and dimensionless parameters. $H$ is a $SU(2)$ Higgs doublet, following the same notation as in the previous section. The quartic coupling $\lambda_{\Phi H}$ in
combination with the Yukawa interaction of \refeqq{dmyuk} allow for the interactions between
the DM candidate and the SM. The $\mathbb{Z}_4$ symmetry is spontaneously broken down to a remaining $\mathbb{Z}_2$ symmetry by the vacuum expectation value (vev) of the singlet field $\langle \Phi \rangle = v_\phi$. We parametrize the scalar field as $\Phi = v_\phi + \phi$ where $\phi$ denotes a real scalar degree of freedom. The singlet vev $v_\phi$ generates a mass term $m_\chi = y_\chi v_\phi $ for the DM fermion, whose stability is ensured by the remaining $\mathbb{Z}_2$ symmetry. In the following, we will assume that no bare mass term for the DM
fermion is present (or equivalently that it can be neglected), such that the physical mass of $\chi$ is given by $m_\chi$. In order to obtain the physical mass eigenstates
$h_{1,2}$ 
an orthogonal field transformation
can be performed, parameterized by an angle $\theta$
\begin{equation}
\begin{pmatrix}
h_1 \\
h_2
\end{pmatrix}
=
\begin{pmatrix}
c_\theta & -s_\theta \\
s_\theta & {\color{white}-}c_\theta
\end{pmatrix}
\begin{pmatrix}
h \\ \phi
\end{pmatrix} \ ,
\label{eqscarot}
\end{equation}
with $s_\theta\equiv \sin \theta$ and  $c_\theta\equiv \cos \theta$. The expression for the mixing angle $\theta$ in terms of the vevs and the quartic
scalar couplings and details about the minimization of the scalar potential are given in \refap{app:singletportal}. In the following,
the physical state $h_1\simeq h_{125}$ denotes the neutral scalar whose mass is identical to
the 
discovered Higgs boson, i.e.~$m_{h_1}\simeq 125$ GeV,
but with couplings modified by factors of
$c_{\rm uni} = \cos \theta$ with respect to the SM prediction.
In the following
the mixing angle $\theta$ is taken as free parameter
in combination with the
physical masses $m_{h_1}$ and $m_{h_2}$
and the vev of the singlet $v_\phi$.
Taking into account that the Yukawa
coupling $y_\chi$ is fixed by the
DM mass $m_\chi =  y_\chi v_\phi$ if
$v_\phi$ is used as free parameter, we are
left with $(m_{h_1}, m_{h_2}, m_\chi,
v_\phi, \theta)$
as set of independent free parameters.\par \medskip

\begin{figure}[t]
  \centering
   \includegraphics[width=0.32\textwidth]{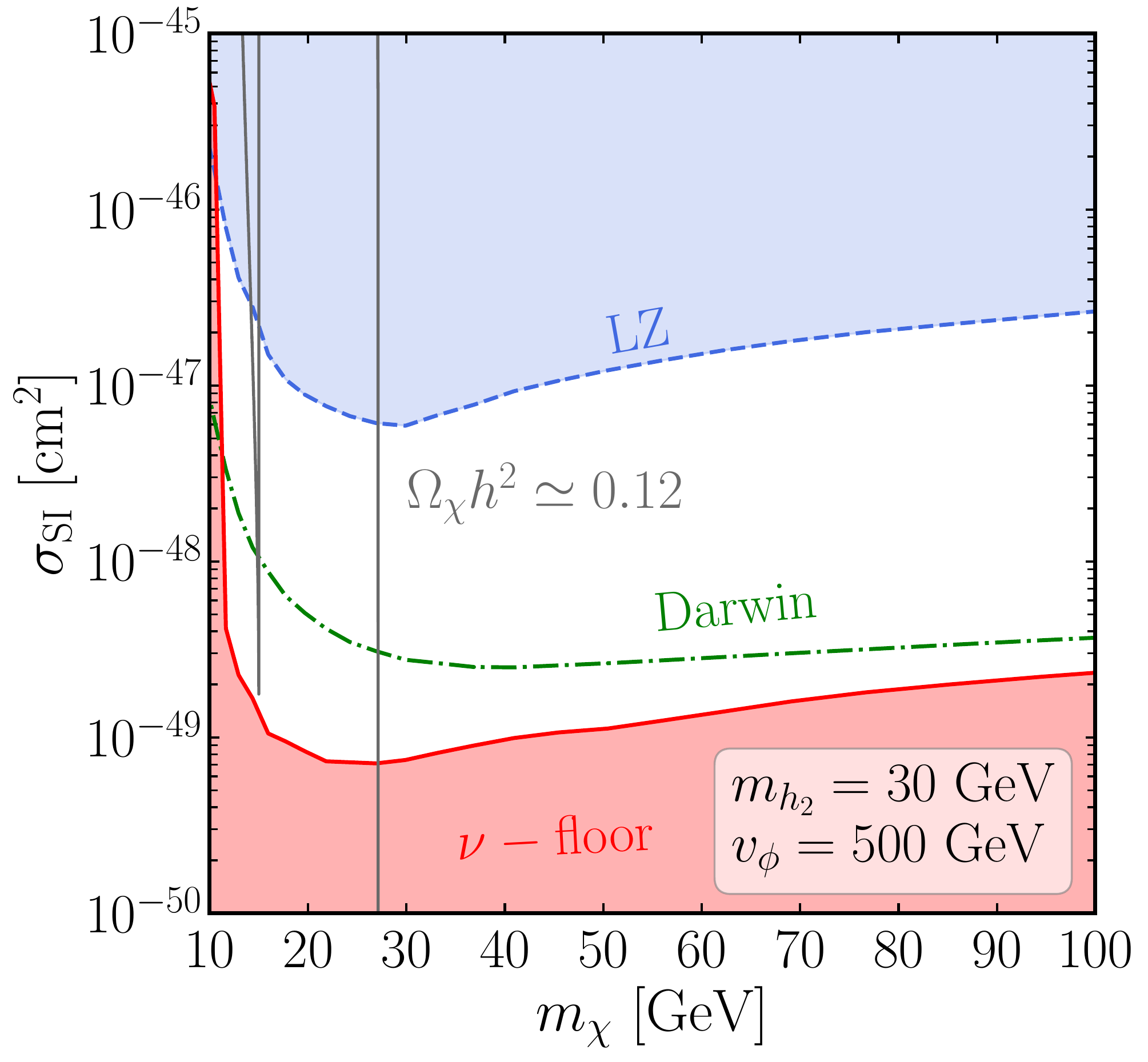}
      \includegraphics[width=0.32\textwidth]{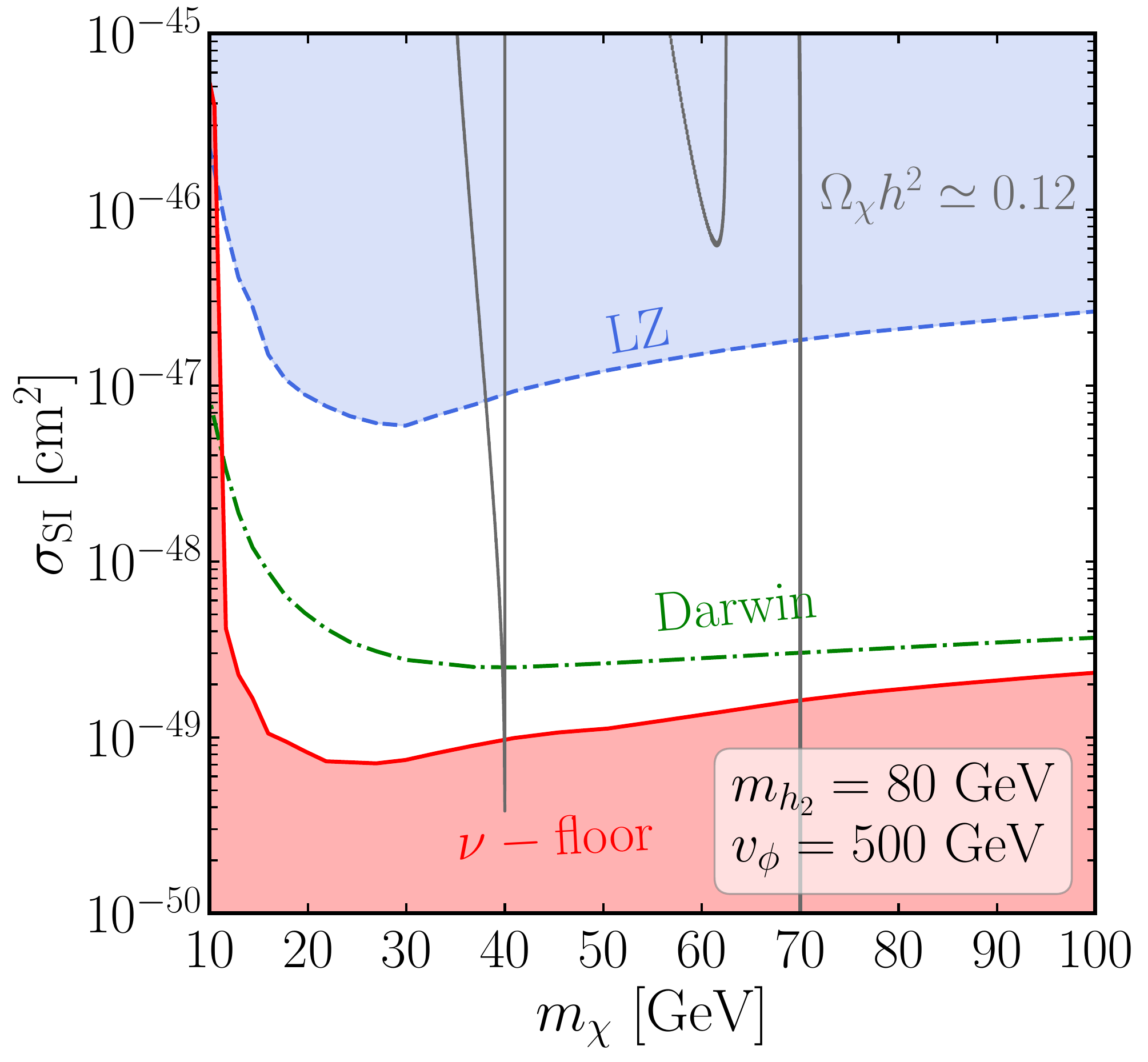}
   \includegraphics[width=0.32\textwidth]{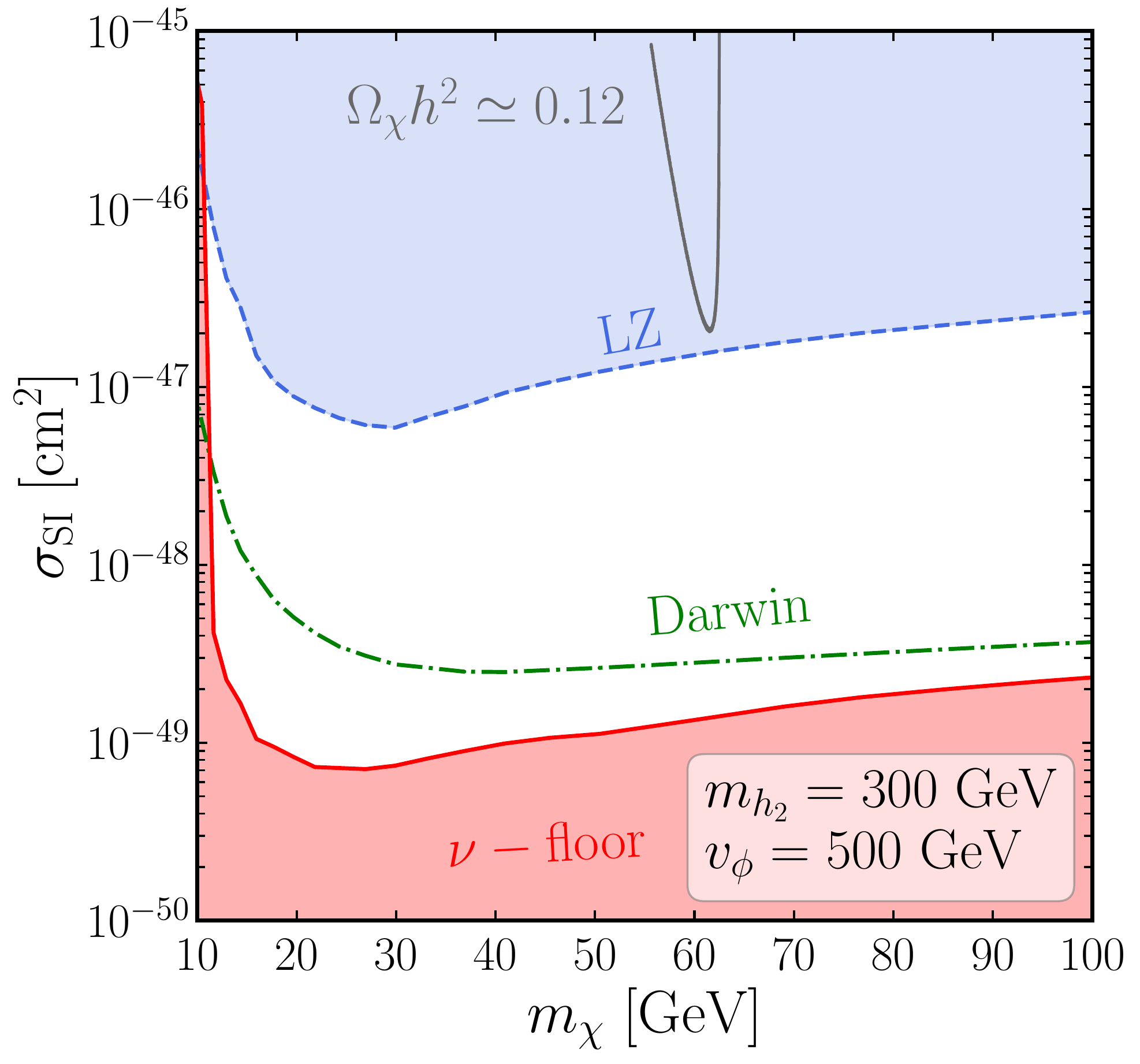}
  \caption{\small As in \reffi{fig:HiggsportalDD} for the
  singlet portal DM scenario for
  different values of the mass of
  the additional scalar $m_{h_2}$
  and the vev $v_\phi$.}
  \label{fig:singletDMplusscalar_sigmaSI}
\end{figure}

Both physical scalars mediate DM-nucleon scatterings whose corresponding cross section can be expressed as
\begin{equation}
    \sigma_\text{SI}\,=  \,\dfrac{\mu_{\chi N}^2 }{\pi} \left[ \dfrac{m_N m_\chi}{v v_\phi}  c_\theta s_\theta \left( \dfrac{1}{m_{h_2}^2}-\dfrac{1}{m_{h_1}^2} \right)\right]^2  f_N^2 ~. 
   \label{eq:directdetection_singletDMplusscalar}
\end{equation}
where notations are identical to the ones used in
\refse{sec:Higgsportal}. In \reffi{fig:singletDMplusscalar_sigmaSI} we represented the value of the direct detection cross section as a function of the dark matter mass for the parameter space allowing to reproduce the correct relic abundance for selected value of $v_\phi$ and $m_{h_2}$ as dark-grey lines.\footnote{The DM-nucleon scattering cross sections
obtained with the analytical expression shown in
\refeqq{eq:directdetection_singletDMplusscalar}
are in agreement within $10\%$ with the
results obtained using the public code
\texttt{micrOMEGAs}~\cite{Belanger:2018ccd,Belanger:2013oya}.} In \reffi{fig:singletDMplusscalar_sigmaSI} the resonances correspond to DM annihilating via (quasi) on-shell mediators ($h_1$ and $h_2$) and therefore are peaked around  $ m_\chi \simeq m_{h_{1}}/2 \simeq 62.5~\text{GeV}$
and $m_\chi \simeq m_{h_{2}}/ 2$.
The vertical lines at $m_\chi \simeq
30\gev$ in the left plot and
$m_\chi \simeq 80\gev$ in the middle plot, respectively,
have their origin in the channel
$\bar \chi \chi \rightarrow h_2 h_2$ opening, which is
kinematically forbidden at zero temperature for
$m_\chi<m_{h_2}$ and exponentially suppressed at
finite temperature but yet still efficient enough
to yield the correct relic abundance.
For this annihilation channel the DM abundance is set by the value of the Yukawa coupling between the fermionic DM and the singlet scalar, which also sets the DM mass resulting in $\theta-$independent generated relic abundance. Larger DM masses correspond to larger annihilation cross sections and a resulting dark matter under-abundance.

We also indicate
in \reffi{fig:singletDMplusscalar_sigmaSI}
the currently strongest upper limit on
$\sigma_{\rm SI}$ from the LZ collaboration
with the blue shaded region.
One can see that the LZ constraints exclude
most parts of the parameter space that
predicts the experimentally determined
DM relic abundance.
In the scenario with $m_{h_2} < m_{h_1} / 2$
(left plot) and
$m_{h_1} / 2 < m_{h_2} < m_{h_1}$
(middle plot)
only DM masses of $m_\chi \simeq m_{h_2} / 2$
and $m_\chi \lesssim m_{h_2}$ remain viable,
whereas the scenario with $m_{h_2} > m_{h_1}$
(right plot) is entirely ruled out
in the low-mass regime investigated here.
Thus, we will not discuss the latter scenario
any further in the following.

In the singlet portal dark matter model
the presence of a second scalar particle
gives rise to additional experimental constraints
that have to be applied.
The 
Higgs boson $h_1 \simeq h_{125}$ can decay invisibly
into a dark matter pair, but $h_1$ can also decay
into pairs of light scalars $h_2$ 
if $m_{h_2} < m_{h_1} / 2$. 
Depending on whether the singlet-like state
$h_2$ decays predominantly into SM particles
or into DM pairs, the decay mode $h_1 \to h_2 h_2$
either gives rise to exotic visible decay modes
of $h_1$, mainly resulting in
$b \bar b b \bar b$, $b \bar b \tau^+ \tau^-$
and $b \bar b \mu^+ \mu^-$ final states, or the
decay mode $h_1 \to h_2 h_2$ gives rise to
an additional invisible decay mode of $h_1$.
The 
decay rates for the decays of $h_1$
into BSM states are given in \refap{app:decayratesingletscalarDM}. Accordingly, the additional
Higgs boson $h_2$ can be searched for
via the decays $h_1 \to h_2 h_2$
if kinematically allowed in the final states mentioned above.
In addition, $h_2$
can be directly searched for
at $pp$ colliders via its production in the
ggH or VBF production modes and
at lepton colliders via Higgsstrahlung production.
Therefore, we here also include
constraints from collider searches
for additional Higgs bosons by using
the public code 
\texttt{HiggsBounds}~\cite{Bechtle:2008jh,
Bechtle:2011sb,Bechtle:2013wla,Bechtle:2020pkv,hsnew},
which are complementary to the constrains resulting
from the signal-rate measurements of $h_1$.

\begin{figure}[t!]
  \centering
   \includegraphics[width=0.45\textwidth]{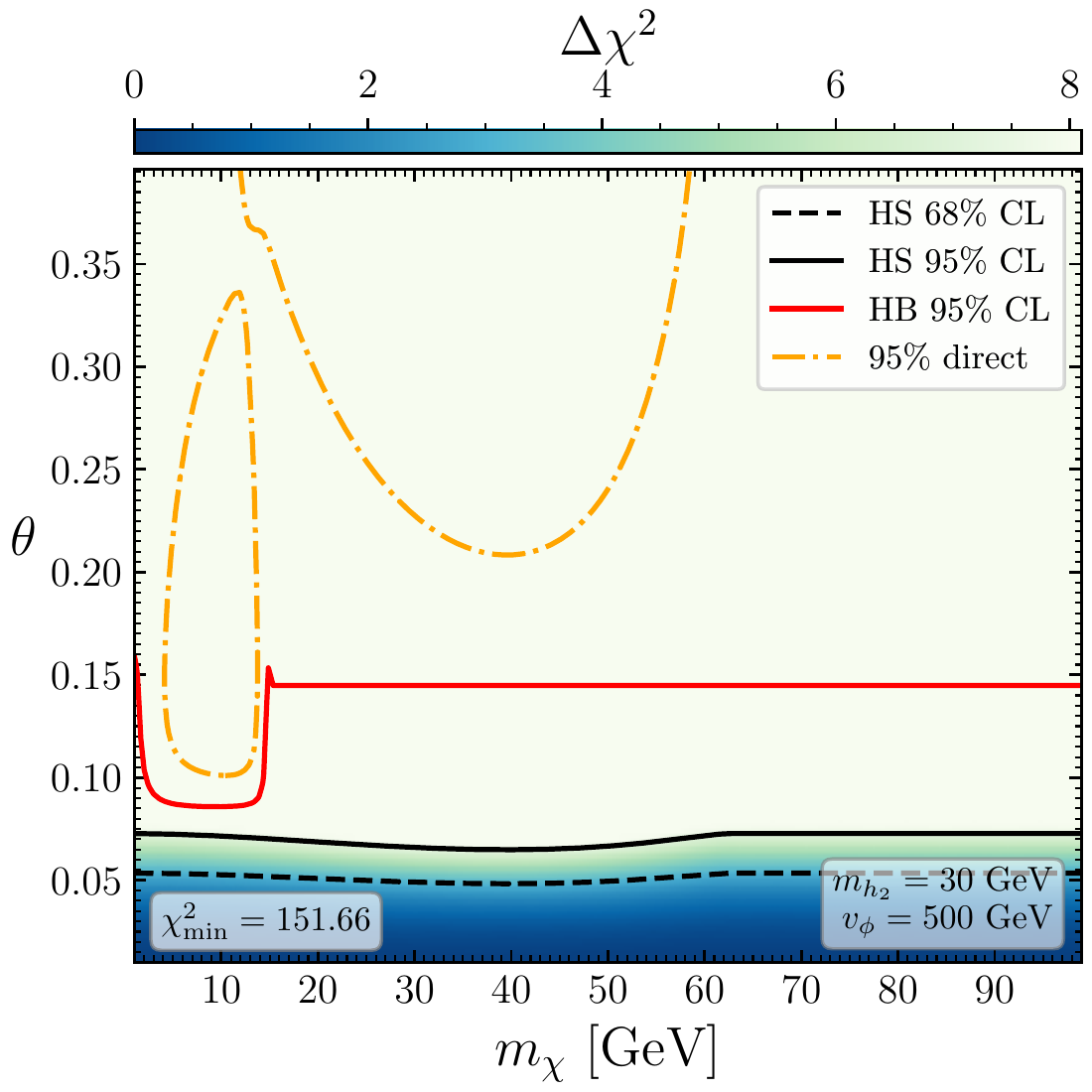}
      \includegraphics[width=0.45\textwidth]{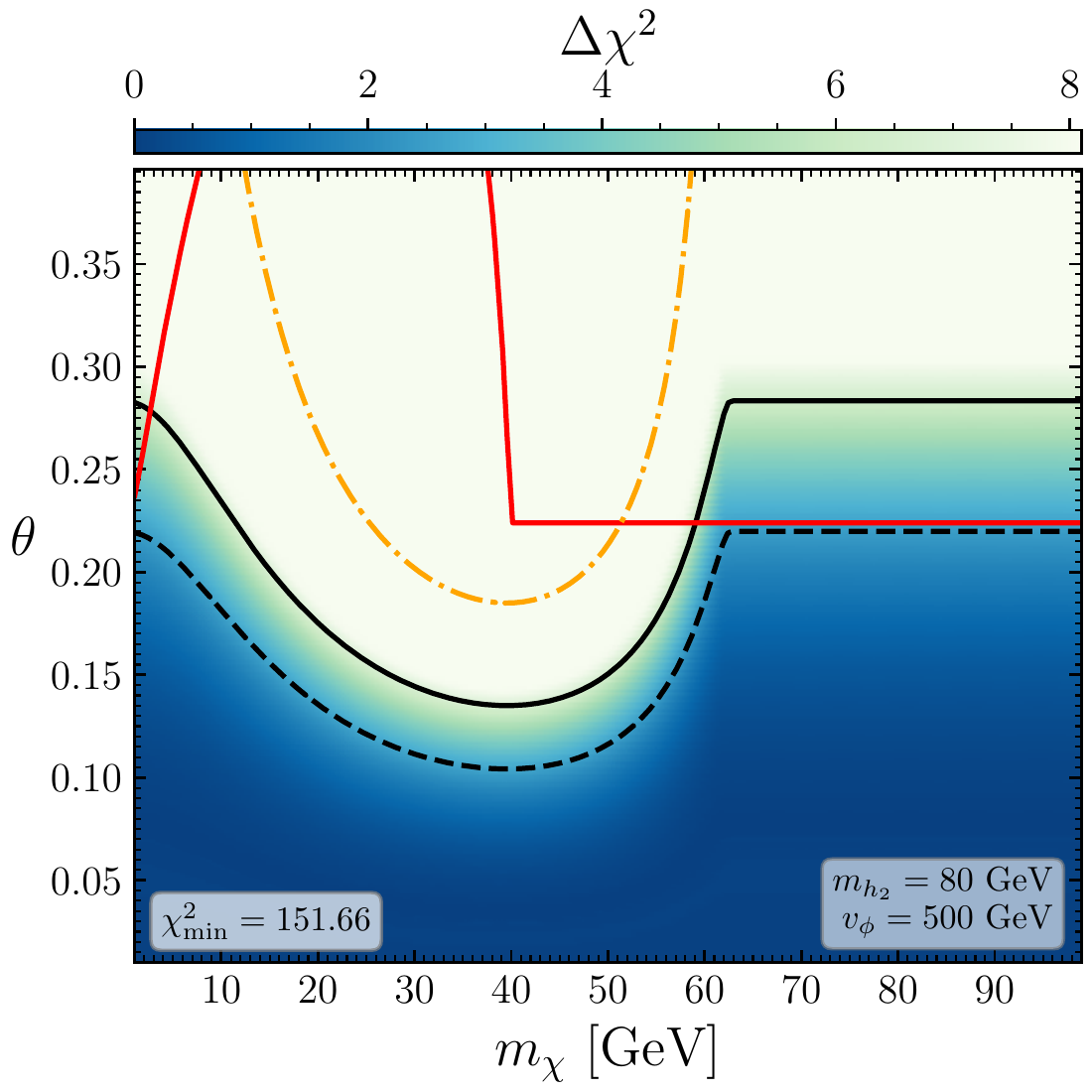}\\[0.4em]
   \includegraphics[width=0.45\textwidth]{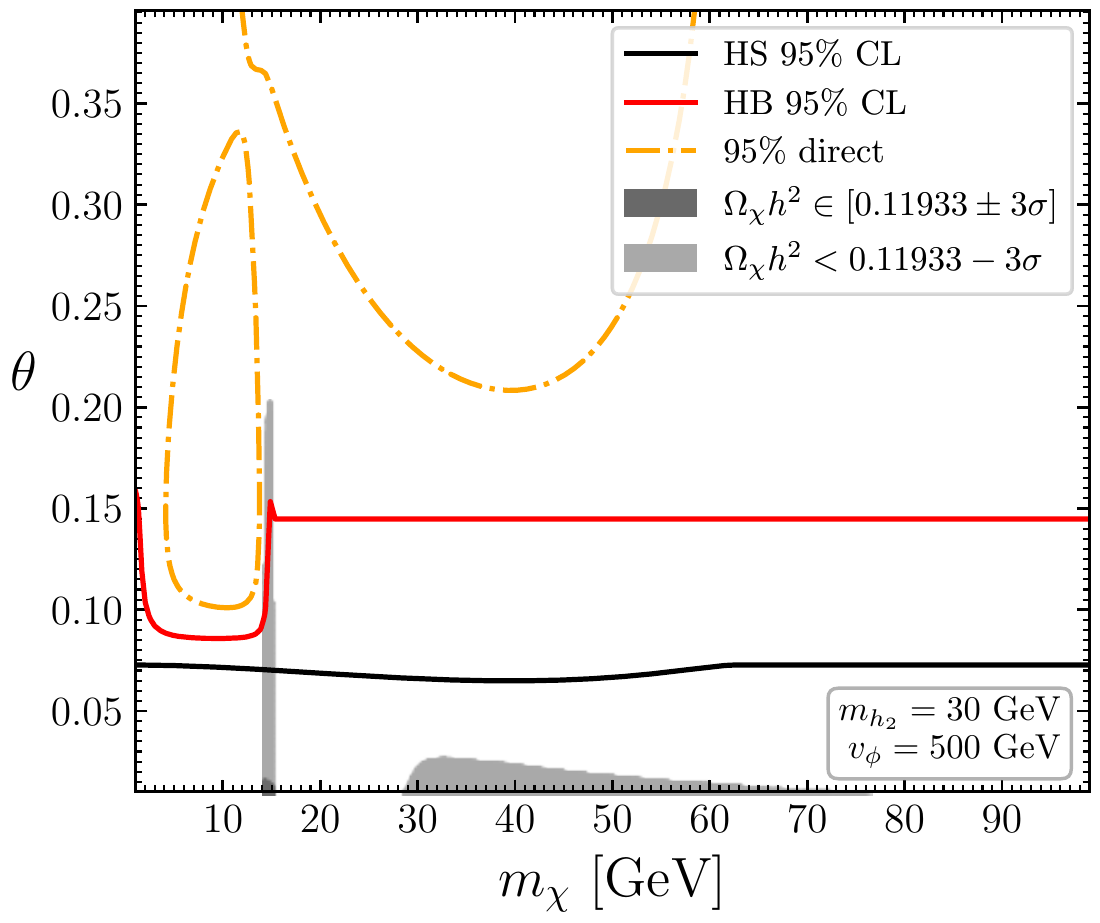}
      \includegraphics[width=0.45\textwidth]{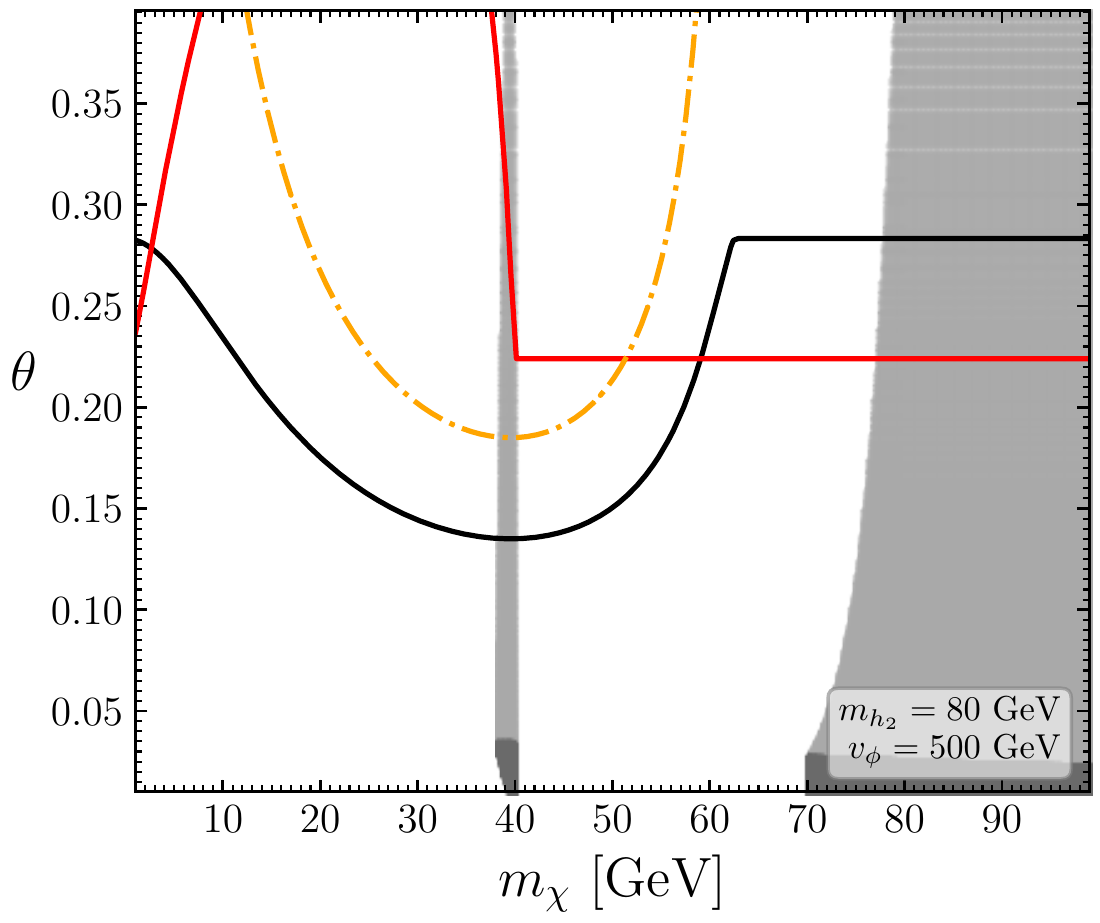}
  \caption{\small In the upper plots we show
  $\Delta \chi^2$ in the
  $\{m_\chi,\theta\}$ plane for two
  different choices of $m_{h_2}$ and $v_\phi$.
  The resulting exclusion limits
  at 95\% and 68\% CL are shown
  with solid and dashed black lines, respectively,
  excluding the areas above the lines.
  The 95\%~CL exclusion region from searches
  for additional scalars is shown in red,
  excluding the areas above the lines.
  The parameter points for which
  $\brinv$ is equal to the upper limit
  from direct searches for the invisible
  decay of $h_{125}$ are indicated with
  the orange lines.
  In the lower plots we additionally indicate the
  parameter space allowing to reproduce all
  (dark grey) or part of (light grey) the
  measured relic abundance.}
      \label{fig:singletDMplusscalar_Higgsconstraints}
\end{figure}

We 
present in
\reffi{fig:singletDMplusscalar_Higgsconstraints}
the constraints from the 
Higgs-boson measurements using \texttt{HiggsSignal}
and from searches for additional Higgs bosons
using \texttt{HiggsBounds}.
The top-panels show 
$\Delta \chi^2$ obtained 
with \texttt{HiggsSignal}, and the corresponding
upper bounds on the mixing angle as a function of
the dark matter mass for selected value
of $v_\phi$ and $m_{h_2}$ at the 68\% and
the 95\% CL indicated by the black dashed and
solid lines, respectively.
We also indicate the region which is excluded
by the observed 95\%~CL cross-section limits with
regards to $h_2$ with the red line,
where the regions above
the red lines are excluded (further details
are given below). Finally, the orange
dotted-dashed lines indicate the regions at which
the predicted invisible branching ratio of
the discovered Higgs boson at $125\gev$ are equal
to the upper limit obtained from direct searches
for the invisible decay mode,
i.e.~$\brinv = 11\%$. As mentioned above, here $\brinv$
is determined by adding the contributions
from the decay modes $h_1 \to \bar \chi \chi$
and $h_1 \to h_2 h_2 \to \bar \chi \chi \bar \chi \chi$
if kinematically allowed.
One can see that in both scenarions
the indirect constraints on $\brinv$
from the cross-section measurements of $h_{125}$
(black lines)
are stronger than the direct limit on $\brinv$
for all DM masses considered.

The lower panels show
the same exclusion lines superimposed to
the parameter space allowed by the LZ bounds,
and in which 
the correct relic abundance (dark grey)
or a lower value (light grey) is achieved.
As discussed in \refse{sec:Higgsportal},
the LZ constraints
are rescaled according to \refeqq{eq:rescaling_LZ}.
According to the previous discussion,
the parameter space allowed by the LZ 
constraints corresponds to two different regimes:
a narrow region around $m_\chi \simeq m_{h_2}/2$
where DM annihilation is resonantly enhanced by
$s-$channel diagrams and the region $m_\chi \simeq m_{h_2}$
where $\bar \chi \chi \rightarrow h_2 h_2$
DM annihilation processes
are efficient enough
in order to not overclose the universe.
In the following, we discuss in more detail the two
benchmark scenarios with $m_{h_2} < m_{h_1} / 2$
(left) and $m_{h_1}/2 < m_{h_2} < m_{h_1}$ (right)
that are depicted
in \reffi{fig:singletDMplusscalar_Higgsconstraints}.

\paragraph{A second Higgs boson below
$\mathbf{125/2}$~GeV:}
In the scenario depicted in
the left panels of Fig.~\reffi{fig:singletDMplusscalar_Higgsconstraints}, the decay of the
SM-like Higgs boson $h_1$ into a pair of
scalars $h_2$ is kinematically allowed,
which mainly determines the exclusions obtained
from the experimental data regarding $h_1 \simeq h_{125}$.
For DM masses of $2m_\chi>m_{h_2}=30\gev$,
the scalar $h_2$ decays to visible
final state (mostly $\bar  b b$).
If 
$2 m_\chi>m_{h_1}\simeq 125\gev$
the \texttt{HiggsSignals} constraint 
($\theta<0.073$ at $95\%$ CL)
is then insensitive to the DM mass.
For $m_\chi\lesssim m_{h_2}/2$,
the invisible decay channel
$h_2 \rightarrow \bar \chi \chi$ becomes relevant,
such that $\brinv > 0$ and one finds
slightly 
stronger constraints
on $\theta$ from the cross-section measurements
of $h_{125} \simeq h_1$.
One can compare the 
indirect constraints from the
cross-section measurements
of~$h_{125}$,
indicated by the black line, to the exclusions
from the direct constraint
$\brinv < 11\%$ obtained from searches for the
invisible decay of $h_{125}$, indicated by the
orange line. The direct limit on $\brinv$
gives rise to
two disconnected exclusion regions.
The parameter space above the right parabola-shaped
orange line is excluded because the decay mode
$h_1 \to \chi \bar \chi$ has a branching ratio
larger than~11\%. The second exclusion region
is the parameter space that lies within the
closed orange line at $m_\chi < 15\gev$, in which
additionally the decay mode $h_1 \to h_2 h_2 \to
\chi \bar \chi \chi \bar \chi$ contributes to $\brinv$.
However, both exclusion regions resulting from the
direct limit on $\brinv$ lie above the black line
and are therefore already excluded based on the
signal-rate measurements of~$h_{125}$.
Finally, the exclusions as a consequence of the
searches for additional Higgs bosons, indicated by
the red line, have their origin in two different
collider searches. For $m_\chi > 15\gev$ we find
an upper limit of $\theta \lesssim 0.15$ due to
constraints from searches for Higgs-boson decays
into a pair of two lighter scalars in the
$b \bar b \mu^+ \mu^-$ final state performed
by the ATLAS collaboration at
$13\tev$~\cite{ATLAS:2021hbr}.
For $m_\chi < 15\gev$ we find a substantially
stronger limit of $\theta \lesssim 0.10$ from
searches for invisibly decaying scalars
performed by the OPAL collaboration at the
LEP collider at up to
$209\gev$~\cite{OPAL:2007qwz}.
However, for all values of $m_\chi$ considered
the constraints from
the searches for additional Higgs bosons are
weaker than the constraints determined with
\texttt{HiggsSignals}.

\paragraph{A second Higgs boson below
$\mathbf{125}$~GeV:}
Whereas in the previous case the
indirect constraints from the $h_{125}$
measurements were dominantly determined by
the presence of the decay mode $h_2 \to h_1 h_1$,
in the benchmark scenario depicted in the
right plot of \reffi{fig:singletDMplusscalar_Higgsconstraints}
this decay is kinematically not possible.
Thus, the constraints depend,
on the one hand, on the presence of
the mixing between $h_1$ and $h_2$, and,
on the other hand,
also on the presence
of the invisible decay $h_1 \to \chi \bar \chi$
if $2 m_\chi < 125\gev$.
We find that the \texttt{HiggsSignal}
constraints (black line) are weaker 
at large DM masses $m_\chi>m_{h_1}/2$,
yielding an upper limit of
$\theta<0.29$.
This limit
is only set by modifications of the SM
prediction of the Higgs-boson universal
couplings and matches our results from
\refse{sec:uni} of \refeqq{eq:upperboundtheta}.
The invisible decay of the SM-like Higgs boson
opens up for $m_\chi<m_{h_1}/2$,
and the exclusion line from the
\texttt{HiggsSignals} analysis drops below the
one from \texttt{HiggsBounds} (red line) at
DM masses below about~$60\gev$.
With regards to the direct limit on
$\brinv$ (orange line),
the island of an excluded region
at $m_\chi < 15\gev$ disappears in the
right plot, 
since only the invisible decay
$h_1 \rightarrow \bar \chi \chi$ is
kinematically allowed, whereas the decay
$h_1 \to h_2 h_2$ is not possible,
leaving only the parabolic-shaped upper
bound set by the invisible decay of the Higgs boson
$h_{125}$. As before, the indirect constraints from
the cross-section measurements of $h_{125}$
are always stronger than the constraints from the
direct limit on $\brinv$.
In contrast to the previous case, we find
for $m_{h_2} > m_{h_1} / 2$ that the \texttt{HiggsBounds}
analysis excluded parameter regions that otherwise
would be allowed. In the right plot,
with $m_{h_2} = 80\gev$, we find an upper
limit of $\theta \lesssim 0.225$ as a result of the
cross section limits from searches for scalar
particles produced via Higgsstrahlung production
and decaying into pairs of bottom quarks
performed at
LEP~\cite{LEPWorkingGroupforHiggsbosonsearches:2003ing}.\footnote{In
addition, the \texttt{HiggsBounds} analysis
provides constraints that are stronger than
the ones from \texttt{HiggsSignals}
at $m_\chi \lesssim 3\gev$, where
however the dark-matter direct-detection
constraints already rule out the relevant
part of the parameter space.
The most sensitive search as determined
by \texttt{HiggsBounds} here
is the search for invisibly
decaying scalars at LEP performed by the
L3 collaboration~\cite{L3:2004svb}.} The exclusion power of this search
becomes much weaker for $m_\chi < 40\gev$,
where the scalar $h_2$ is able to decay
into pairs of~$\chi$ and the branching ratio
for the decay $h_2 \to b \bar b$ is smaller.

\par \medskip

In this section we have considered a discrete $\mathbb{Z}_4$ symmetry as a specific example. However this construction could be extended by considering instead a global continuous $U(1)$ symmetry, broken spontaneously. In this case the imaginary part of the complex scalar field responsible for the symmetry breaking would be a massless Goldstone boson. The analysis performed in this section would not be drastically affected by the presence of this new state but would need to be accounted for in the possible DM annihilation final states. Constraints from Higgs physics would have to be modified as the SM-like Higgs boson could decay into a pair of Goldstone bosons. This precise point is discussed in the following subsection.

\subsection{(Pseudo) Nambu-Goldstone bosons}
\label{sec:PNGB}

In this section we consider a model with
an additional (approximate) global $U(1)$ symmetry and
a complex scalar field parameterized in the
exponential form by $\Phi = (v_\phi+\phi
)e^{ia/v_\phi}/\sqrt{2}$, singlet under the SM
gauge group but charged under the new global $U(1)$. All SM fields are singlets under the
additional $U(1)$ symmetry.
The scalar potential including all terms that
respect this symmetry has the same form as
the one of \refeqq{eq:potentialsingletportal},
and we adopt the same notation as in
\refse{sec:singletportal}.
Depending on the specificity of the model,
the global symmetry could be spontaneously
broken ($v_\phi\neq0$) or not ($v_\phi=0$).
If the symmetry is spontaneously broken,
a mixing by an 
angle $\theta$ between the real part
of the complex field and the Higgs field is generated,
while the 
state $a$ is a Nambu-Goldstone Boson (NGB) or
Pseudo Nambu-Goldstone Boson (PNGB)
depending on whether the broken symmetry
was 
exact or approximate before spontaneous
symmetry breaking.
In the latter case, we assume that any explicit
symmetry breaking term is small compared to the
electroweak scale such that the mass of the
(P)NGB can be  
set to zero in the following analysis.
Using the exponential parametrization, the
field $a$ disappears from the scalar
potential and the only coupling between the
Higgs boson at $125\gev$ and the
(P)NGB is generated via mixing from the kinetic
terms of the scalar $\Phi$ after spontaneous symmetry
breaking~\cite{Weinberg:2013kea, Fernandez-Martinez:2021ypo},
\begin{equation}
    \mathcal{L}\supset |\partial_\mu \Phi|^2 = \dfrac{1}{2} \left( \partial_\mu \phi \partial^\mu \phi + \partial_\mu a \partial^\mu a \left( \dfrac{\phi^2}{v_\phi^2}+2 \dfrac{\phi}{v_\phi}+1  \right) \right)\,,
\end{equation}
before performing a rotation to the mass eigenstate basis.
We obtain  
canonically normalized kinetic terms for the massless pseudoscalar field $a$ 
in addition to higher dimensional operators
involving derivative couplings.
One can perform a rotation to the physical
CP-even neutral scalar mass eigenstate
basis $\{h_1,h_2\}$ by considering the
transformation described in the previous section
(see \refeqq{eqscarot}).
By selecting only terms relevant for the
decay of the SM-like Higgs boson into
pairs of (P)NGBs, one obtains
\begin{equation}
       \mathcal{L}\supset  \dfrac{1}{2} \partial_\mu a \partial^\mu a \Big(1 -2 \dfrac{s_\theta h_1}{v_\phi}  \Big)  \supset - \partial_\mu a \partial^\mu a  \dfrac{s_\theta h_1}{v_\phi}  \,.
       \label{eq:HiggsPNGBcoupling}
\end{equation}
This parametrization makes obvious that the decay into pairs of (P)NGBs
of the SM-like Higgs state is triggered by the
mixing angle $\theta$.
The coupling term of \refeqq{eq:HiggsPNGBcoupling}
generates a partial decay width  
\begin{equation}
    \Gamma_{h_1 \rightarrow a a }\,=\,\dfrac{s_\theta^2}{32 \pi} \dfrac{m_{h_1}^3}{v_\phi^2}\,.
    \label{eq:pnginv}
\end{equation}
One can recover this result by choosing the linear parametrization $\Phi=(v_\phi+\phi+ia)/\sqrt{2}$
and by expressing the scalar potential of
\refeqq{eq:potentialsingletportal} in terms of
the mass eigenstates,
as performed in Ref.~\cite{Arina:2019tib}. \par \medskip

\begin{figure}[t]
\includegraphics[width=0.6\textwidth]{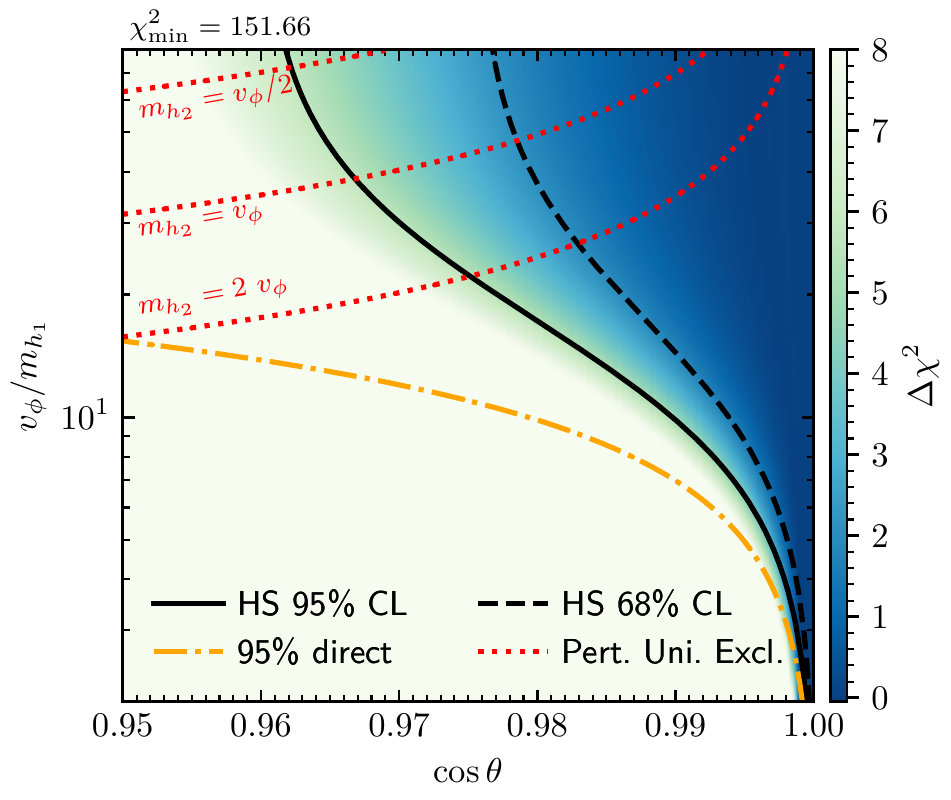}
\caption{\small
$\Delta \chi^2$,
indicated by the colour coding, in the
$\{\cos\theta,v_\phi/m_{h_1}\}$ plane
for the (Pseudo) Nambu-Goldstone boson
scenario. The resulting exclusion limits
at 95\% and 68\% CL are shown respectively
with solid and dashed black lines, excluding the
areas below the lines. Areas below the orange dashed line
are excluded based on
the experimental limit on $\brinv$ from direct searches
for $h_{125} \to \mathrm{inv}$.
The areas above the dotted red lines are excluded
by constraints from perturbative unitarity
assuming
$m_{h_2} = v_\phi / 2, v_\phi,2 v_\phi$
(see discussion in
\refse{sec:PNGB} for details).}
\label{fig:PNGB}
\end{figure}

As the couplings between the SM field content and
the  
Higgs boson $h_1 \simeq h_{125}$ are modified
by a universal factor
$c_\mathrm{uni}=\cos\theta$,
we can derive constraints in the plane
$\{\cos \theta,(v_\phi/m_{h_1})\}$, where we divide the vev $v_\phi$
by $m_{h_1} = 125\gev$ in order to
have a dimensionless quantity.
We depict
in \reffi{fig:PNGB}
with the color coding
the $\Delta \chi^2$-values from \texttt{HiggsSignals}
in this plane.
In addition, we 
show  
with the black dashed and solid
lines the resulting exclusion limits
at the 68\% and 95\%~CL,
respectively, and we also show
exclusion lines based on
perturbative unitarity constraints on the
dimensionless coupling of the scalar potential
for a selection of masses
$m_{h_2}=v_\phi/2,v_\phi,2v_\phi$ (explicit
expressions for these constraints can
be found, for instance,
in \citere{Dawson:2017jja}).\footnote{The
$m_{h_2}$ dependence of these limits enters
via the relations shown in
\refeqs{eq:rotationsingletscalar1}--(\ref{eq:rotationsingletscalar3}).}
In agreement with the discussions
of \refse{sec:uni},
we find that the indirect constraints
from the cross-section measurements of~$h_{125}$ are
considerably stronger than constraints
from the experimental
limit on $\brinv$ from direct searches
for the invisible decay of $h_{125}$,
where the latter are indicated with the
orange dashed-dotted line in \reffi{fig:PNGB}.
For instance,
for $v_\phi/m_{h_1}=10$ 
we find that
the mixing angle is constrained to be
$\cos \theta> 0.99$ or
equivalently $\theta< 0.14$ at
$95\%$~CL based on the values of
$\Delta\chi^2$, 
whereas the bounds from direct searches
for the decay $h_{125} \to \mathrm{inv}$
allow for values
as small as $\cos \theta> 0.98$ or
equivalently $\theta< 0.20$.
As can be seen in \refeqq{eq:pnginv},
in the limit where the ratio $v_\phi/m_{h_1}$ becomes
large, the partial width for the invisible
decay of $h_1 \simeq h_{125}$ becomes negligible.
Accordingly, in this limit the bound
on the mixing angle saturates at a
constant value of
$\cos \theta > 0.959$ which matches our results
derived in \refse{sec:uni} of \refeqq{eq:constraintsBRinv_cuni}
in the limit $\brinv\rightarrow 0$.
The limits on $\cos\theta$ can be even stronger
when one combines the limits from the
cross-section measurements of~$h_{125}$
with the perturbative unitarity constraints.
For instance, assuming that
$m_{h_2}=v_\phi$ imposes
$\cos \theta> 0.966$,
corresponding to $\theta< 0.26$.

It is tempting to demand independent limits
on $\brinv$ and $\theta$ in a phenomenological
analysis in order to account
in an approximate form
for the experimental
constraints, on the one hand,
from the invisible decay mode of~$h_{125}$
and, on the other hand, from
the mixing between $h_{125}$ and the 
additional scalar state.
In many studies this amounts to applying
the direct limit on $\brinv$ (orange line)
and a constant lower limit on the
coupling modifier $c_{\rm uni} = \cos\theta
(> 0.959)$,
where the value in the brackets is
the limit we found to be valid
in the limit $\brinv \to 0$ according
to the discussion above.
The results depicted in
\reffi{fig:PNGB} illustrate that this approach
would allow substantial parts of the
parameter space that are actually excluded by
the cross-section measurements of $h_{125}$
(black line).
This demonstrates that in order to
fully exploit the
experimental data with regards to
the discovered Higgs boson in models
with light hidden sectors it is vital to
take into account \textit{simultaneously}
the presence of a non-zero $\brinv$ and
the modifications of the couplings
of~$h_{125}$ to SM particles, as was
done in our global $\chi^2$-analysis
using \texttt{HiggsSignals}.
\par \medskip

The results derived in this section
go beyond the global symmetry case
and can also be applied to the case
of a local $U(1)$ gauge symmetry to
some extent. In the regime where the mass of the new light vector state  $m_{Z^\prime} = g_{Z^\prime} q_\phi v_\phi \ll m_{h_{125}}$,
with $g_{Z^\prime}$ being the extra $U(1)$
gauge coupling and $q_\phi$ the charge of
the complex scalar breaking this symmetry, by virtue of the
Goldstone boson equivalence theorem the decay rate $\Gamma_{h_{125} \rightarrow aa}$ into a pair
of Goldstone bosons $a$
is  
identical
to the rate $\Gamma_{h_{125} \rightarrow Z^\prime Z^\prime}$ into a pair of massive gauge
fields $Z^\prime$ in the small mass limit.
Therefore, our constraints also apply to this
case, provided that the produced $Z^\prime$
is sufficiently weakly coupled in order to
escape the detector without interacting.

\subsection{Two Higgs doublet models}
\label{sec:2HDM}

One of the most prominent example where
observables related to the Higgs boson
depart from their SM predicted values
are models containing two Higgs doublet
fields $\Phi_1$ and $\Phi_2$,
called 2 Higgs doublet models (2HDM)~\cite{Lee:1973iz,Kim:1979if}
(see also \citere{Branco:2011iw} for a review).
In the CP-conserving 2HDM, the physical
Higgs spectrum
consists of two CP-even states $h$ and $H$,
where $h$ in the following plays the role of
$h_{125}$, a CP-odd state $A$, and a pair of
charged Higgs bosons $H^\pm$.
In the decoupling limit of the 2HDM, in which
the BSM particle states have masses considerably larger
than the electroweak scale, the couplings of the
state $h = h_{125}$ to the fermions
and gauge bosons are determined by only
two parameters: the ratio of the vacuum expectation
values of the neutral CP-even components
of the Higgs doublets $v_1$ and $v_2$ written in terms of
the parameter
\begin{equation}
    \tan \beta \equiv  \dfrac{v_2}{v_1}\, ,
    \textrm{ where } v=\sqrt{v_1^2+v_2^2}\simeq246\gev \ ,
\end{equation}
and the rotation angle $\alpha$ that determines
the mixing of the two CP-even states $h$ and $H$.
In the so-called alignment limit, defined by
the condition $\cos(\alpha - \beta) = 0$,
the couplings of $h$ to the SM particles
are identical to the predictions of the SM.

\begin{table}
    \centering
    \begin{tabular}{l||c|c|c}
     \text{Model}  & $u_R$   & $d_R$ & $\ell_R$ \\
     \hline
     \hline
        \text{Type I}  & $\Phi_2$ & $\Phi_2$ & $\Phi_2$  \\
        \text{Type II} & $\Phi_2$ & $\Phi_1$ & $\Phi_1$ \\
        \text{Type III (lepton-specific)} & $\Phi_2$ & $\Phi_2$ & $\Phi_1$ \\
        \text{Type IV (flipped)}  & $\Phi_2$ & $\Phi_1$ & $\Phi_2$ \\
    \end{tabular}
    \caption{Summary of which of the two Higgs doublet fields
    $\Phi_{1,2}$ is coupled to up-type fermions ($u_R$),
    down-type fermions ($d_R$) and charged leptons ($\ell_R$)
    in the four Yukawa types of the 2HDM.}
    \label{tab:my_table2}
\end{table}

In order to eliminate sources of
flavour-changing neutral currents
at the classical level, one can
introduce a softly broken $\mathbb{Z}_2$ symmetry,
under which one of the Higgs doublets changes sign, whereas the second Higgs doublet
transforms trivially, and the $\mathbb{Z}_2$ charges
of the fermions depend on the so-called
Yukawa type that is assumed.
In total, there are four different possibilities to
assign the fermion charges. Depending on the
fermion charge, either $\Phi_1$ or $\Phi_2$
can be coupled to the corresponding fermion.
In \refta{tab:my_table2} we explicit
which Higgs doublet field is coupled to
which kind of fermion in each of the
four types of the 2HDM considered in this work.
The resulting structure of the coupling modifiers
that is realized in each Yukawa type
is given in \refta{tab:my_table}.
The explicit dependence of the modifiers
$c_{V,u,d,\ell}$ on the parameters $\alpha$
and $\beta$ can be found, for instance,
in \citere{Branco:2011iw}.

In the following we
perform a $\chi^2$-scan in the plane
$\{\cos(\alpha-\beta),\tan \beta\}$ for the four
types of the 2HDM in the decoupling limit,
and we analyze how the allowed
regions of parameter space are modified by the
existence of an invisible decay mode of $h$.
The resulting constraints in terms of exclusion
regions can be applied to a variety of 2HDMs
that are extended by a hidden sector. Such
models comprise, for instance,  scalar
gauge-singlet extensions of the 2HDM
such as the N2HDM~\cite{Grzadkowski:2009iz},
the S2HDM~\cite{Biekotter:2021ovi,Biekotter:2022bxp} or
the 2HDM+a~\cite{Bauer:2017ota}.
Here it should be taken into account that
for a concrete model realization of a
particular configuration of 2HDM-like
coupling modifications also other
experimental and theoretical constraints
would have to be considered. We refrain
from doing such a model-specific analysis
here, as we want to focus on the constraints
that arise from the properties of~$h_{125}$
under the presence of an additional invisible
decay mode.

\begin{figure}
\centering
\includegraphics[width=0.98\textwidth]{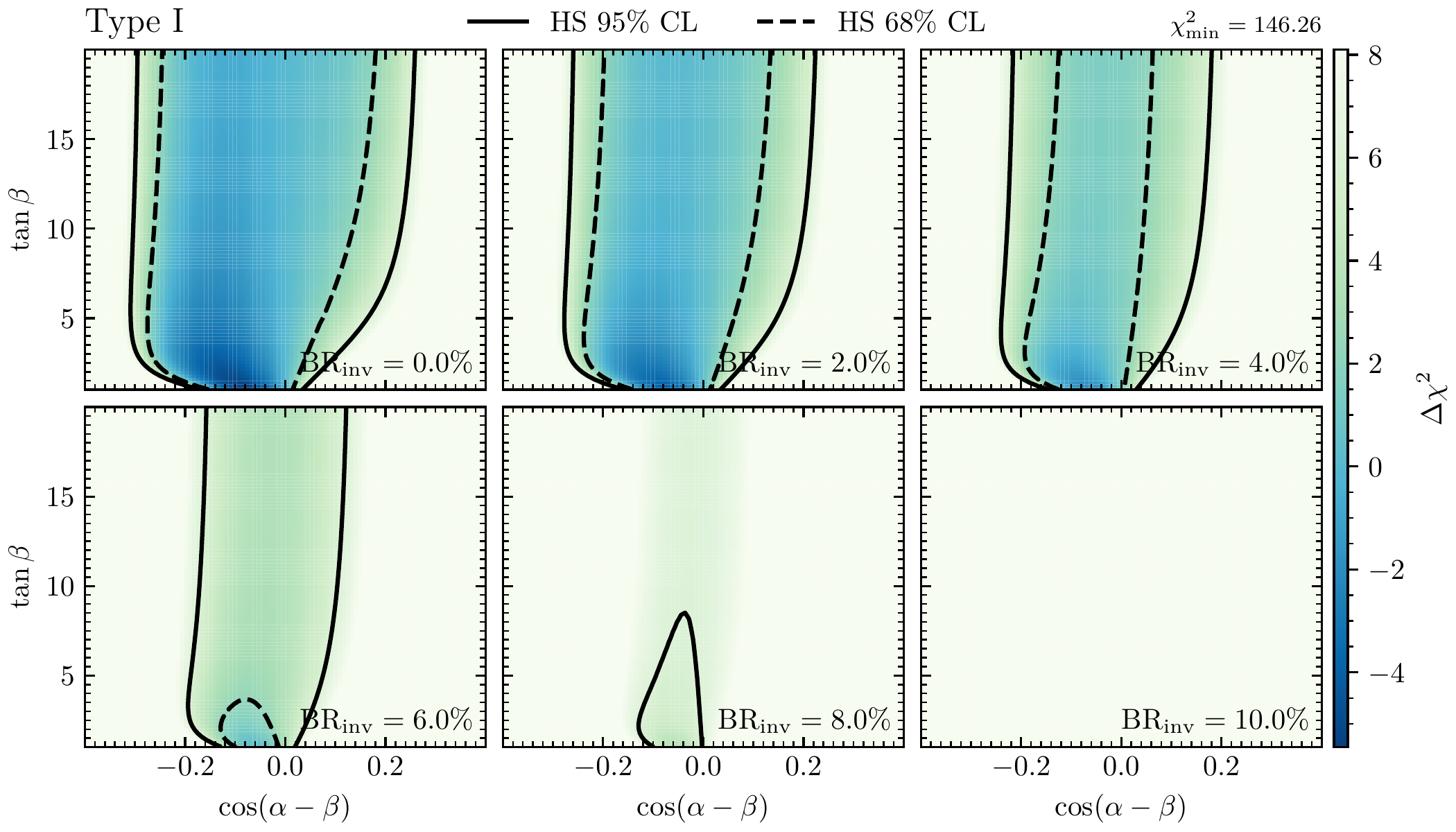}\\[1.0em]
\includegraphics[width=0.98\textwidth]{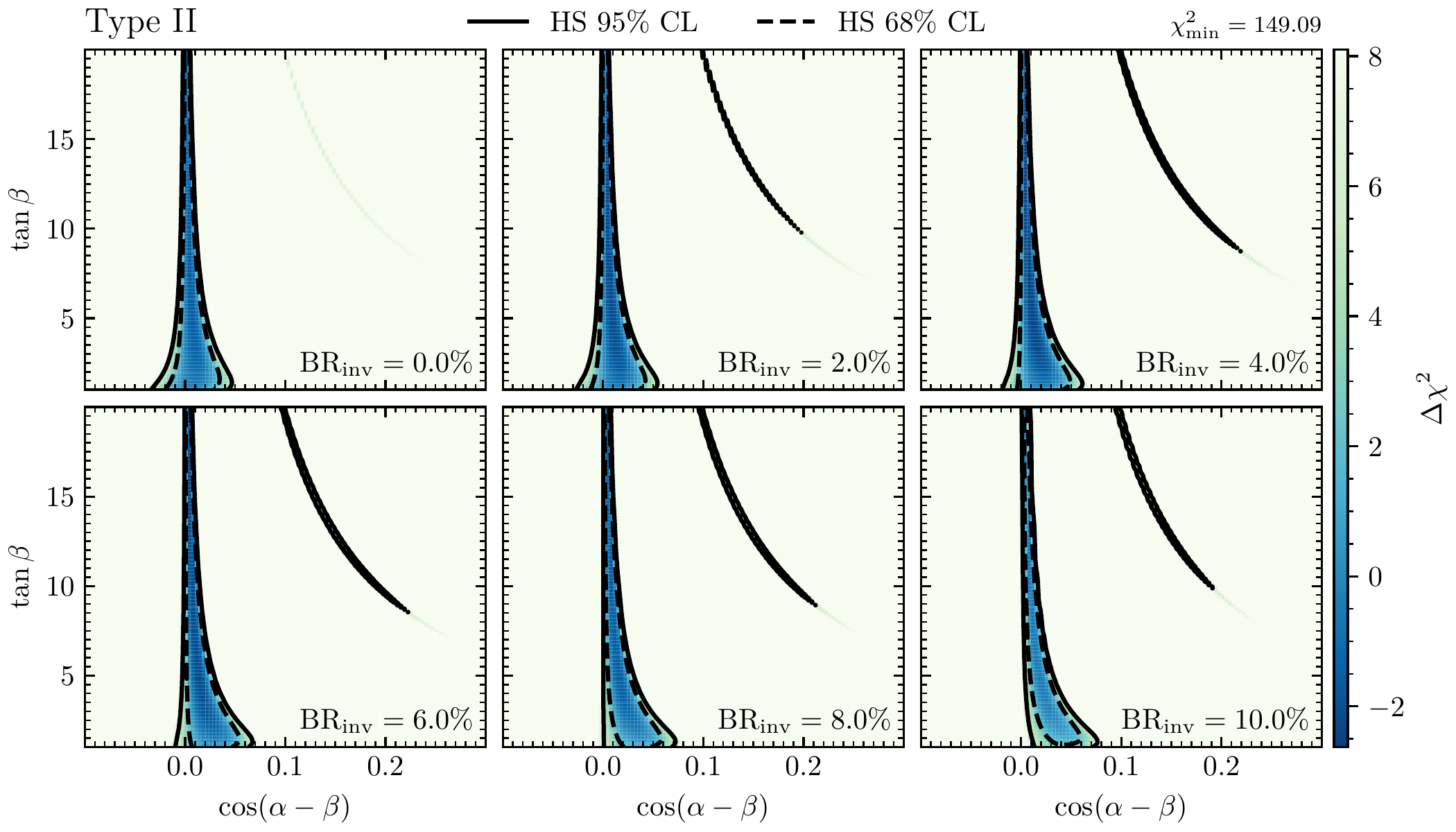}
\caption{$\Delta \chi^2$-distribution in the
$\{\cos(\alpha-\beta),\tan\beta\}$ plane for
different values of $\brinv$ in the type~I (top)
and the type~II (bottom). The dashed black and the solid
black lines indicate the exclusion limits at
the 68\% and the 95\% confidence level, respectively.}
\label{fig:2HDMTypeIandII}
\end{figure}

\paragraph{Type I} We show in the
upper plots of \reffi{fig:2HDMTypeIandII}
the $\chi^2$-distribution in the
$\{\cos(\alpha-\beta),\tan\beta\}$ plane
for the Yukawa type~I for different
values of $\brinv$ (varied in steps
of 2\%). The black dashed and
solid lines indicate the excluded regions of the
model parameters at the 68\% and the 95\%~CL,
respectively.
One can see that, as expected, the best agreement
with the experimental data regarding $h_{125}$ is
found at or about the alignment limit
$\cos(\alpha - \beta) = 0$, in which the state
$h$ resembles a SM Higgs boson. 
By comparing the plots for different values
of $\brinv$ it becomes apparent that the values
of $\Delta \chi^2$ increase with increasing
values of $\brinv$ in the whole parameter plane.
Consequently, in the Yukawa type~I the presence
of an invisible decay mode of $h_{125}$ deteriorates
the fit result to the experimental Higgs-boson data
independently of the values of the coupling
modifiers. For a value of $\brinv = 6\%$ only
a small region with $\tan\beta \lesssim 5$ and
$\cos(\alpha - \beta) \simeq 0$ remains allowed
at the $1\sigma$ level, whereas for $\brinv = 8\%$
even the alignment limit is almost excluded at
the 95\%~CL. The type~I Yukawa structure
is excluded at more than 95\%~CL
for $\brinv = 10\%$, such that the experimental
limits from direct searches for $h_{125} \to \mathrm{inv}$
(currently at the level of $\brinv < 11\%$ or larger)
do not provide additional constraints.
The observations made here are
in agreement with the results
depicted in \reffi{fig:kapvkapf}
for the scan in the benchmark
model featuring the coupling modifiers
$c_V$ and $c_f = c_u = c_d = c_\ell$, which is
also the pattern of coupling modifications that
is realized in the type~I, where here the additional
restriction $c_V \leq 1$ applies.
Accordingly, we also observe here values
of $\Delta \chi^2$ substantially smaller
than zero. The origin of
these values was discussed already in
\refse{sec:nonunicoupling}.

\begin{figure}
\includegraphics[width=0.98\textwidth]{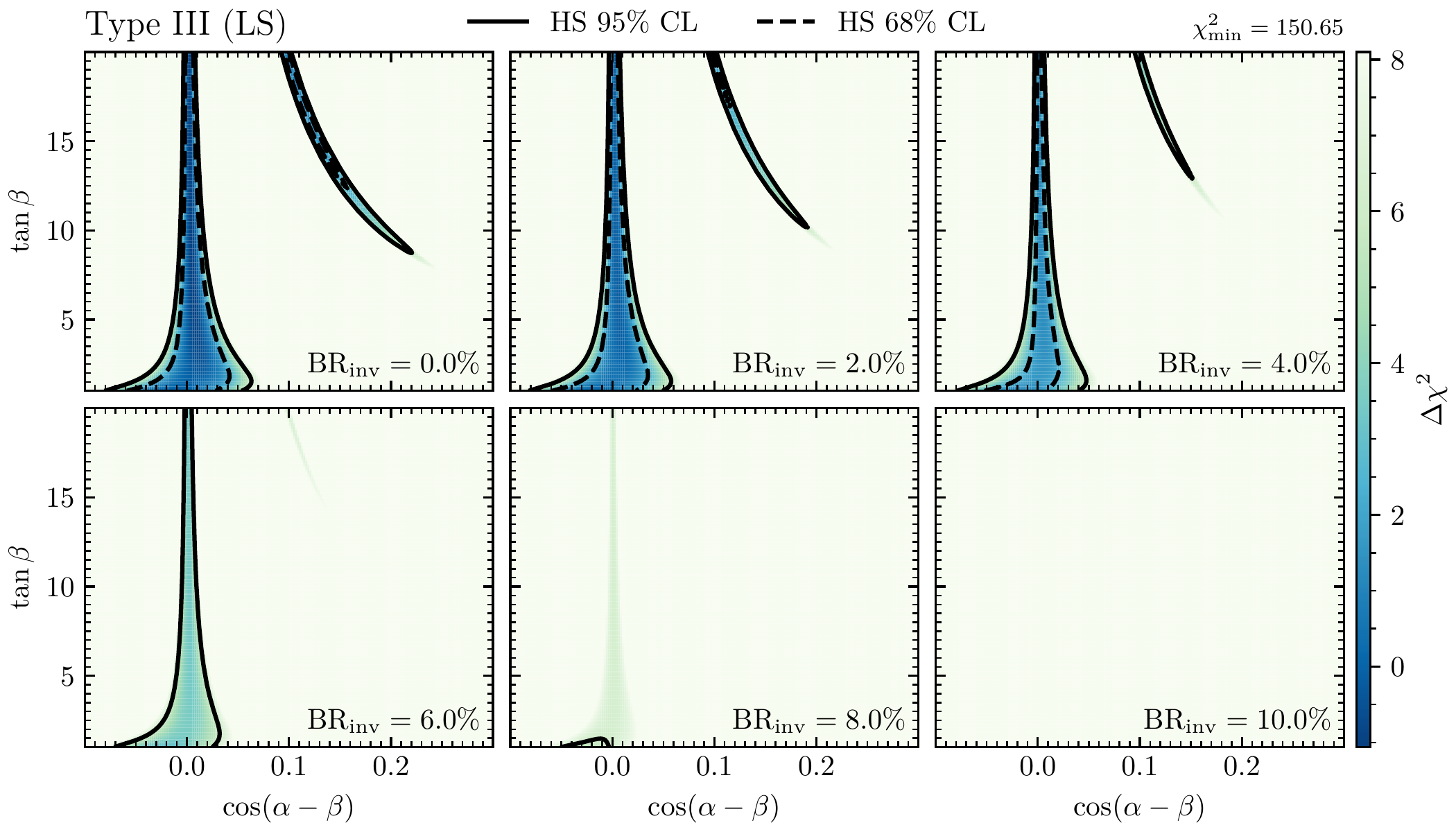}\\[1.0em]
\includegraphics[width=0.98\textwidth]{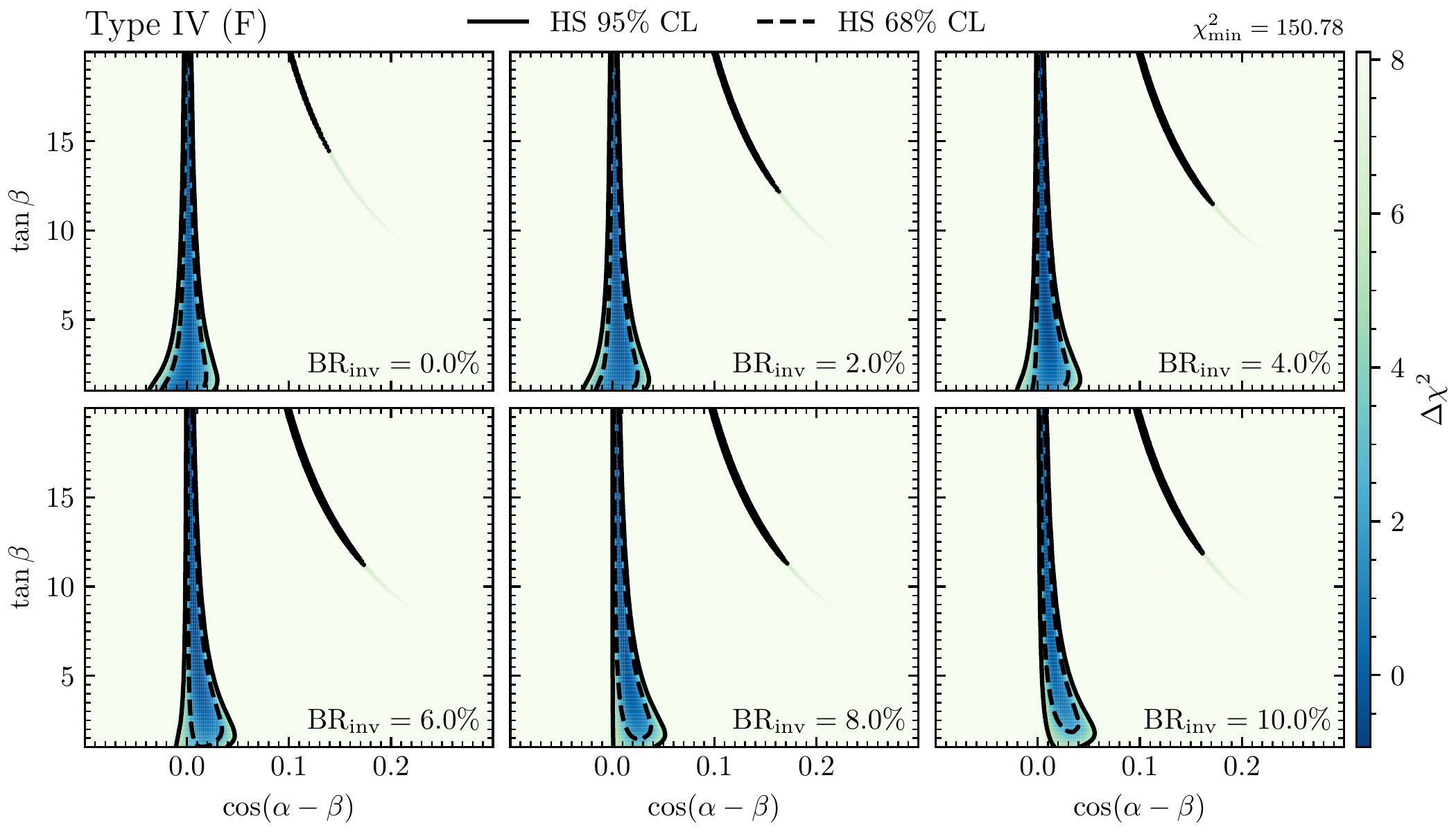}
\caption{Same as in \reffi{fig:2HDMTypeIandII},
but for the type~III (top)
and the type~IV (bottom).}
\label{fig:2HDMTypeIIIandIV}
\end{figure}

\paragraph{Type II} The corresponding results for
the Yukawa type~II are depicted in the bottom
panels of \reffi{fig:2HDMTypeIandII}.
Regarding the fit result in the vicinity of the
alignment limit of the Yukawa type~II, one can observe
that with increasing value of $\brinv$ the
parameter region that is in agreement with
the measurements of $h_{125}$ moves towards
values of $\cos(\alpha - \beta) > 0$.
This is related to the fact that one
finds $|c_b| < 1$ and $|c_V|\simeq |c_u| \simeq 1$ for
$\cos(\alpha - \beta) > 0$, such that
the additional contribution $\Gamma_{\rm inv}$
to the total decay width of $h_{125}$ is
compensated by a reduced partial decay
width $\Gamma_{b \bar b}$, while at
the same time the
ggH and the VBF production
cross sections of $h_{125}$ are not
significantly modified compared to the
SM predictions.
On the other hand,
for $\cos(\alpha - \beta) < 0$ one finds
$|c_d| > 1$, such that the branching ratio
of the decay $h_{125} \to \gamma\gamma$ is
suppressed by both the non-zero value of
$\Gamma_{\rm inv}$ and the enhancement
of $\Gamma_{b \bar b}$, and consequently
the fit result deteriorates in this case
with increasing values of $\brinv$.

In the type~II, in addition to the alignment limit,
there is a tight branch in the parameter plane
with values of $\cos(\alpha - \beta) > 0.08$ and
$\tan\beta \gtrsim 7$ that predicts a state $h$ that
can be in agreement with the experimental measurements
and hence features small values of $\Delta \chi^2$.
This branch is a 2HDM realization
of the so-called \textit{wrong-sign Yukawa
coupling regime}, in which the absolute values of the
coupling modifiers are approximately equal to one,
i.e.~$|c_{V,u,d,\ell}| \simeq 1$, but in which
the couplings to down-type quarks have the opposite
sign as compared to the couplings to gauge bosons
and up-type
quarks~\cite{Espinosa:2012im,Falkowski:2013dza} (see also
\citere{Ferreira:2014naa} for a discussion of
this limit in the 2HDM
in light of recent LHC measurements).
In the absence of a decay mode $h_{125} \to \mathrm{inv}$
the wrong-sign Yukawa coupling regime is in tension
with the signal-rate measurements of $h_{125}$,
as can be seen in the upper left plot. This
tension is mainly
driven by the enhancement of the gluon-fusion
production
cross section of $h_{125}$ compared to the SM
prediction as a result of the modified interference
effects between the contributions from
the top-quark loop and the bottom-quark
loop~\cite{Ferreira:2014naa}.
We find here that the agreement with the experimental
data can be improved significantly in the
wrong-sign Yukawa coupling regime if a sizable value
of $\brinv$ is present. Even for values as large
as $\brinv = 10\%$ we find parameter points in this
limit that fit the signal-rate measurements of
$h_{125}$ just as well as the SM, whereas the same
parameter points would be excluded at a CL
substantially larger than $2\sigma$
for $\brinv = 0$. Here, the additional partial
decay with $\Gamma_{\rm inv}$ leads to a
suppression of the ordinary decay modes
of $h_{125}$ which compensates the
enhancements of the gluon-fusion production
cross sections. The
relative sizes of the signal rates
of $h_{125}$ can then remain close
to the SM predictions because the type~II
Yukawa structure allows for the individual
modifcation of both $c_u$ and $c_d$, such
that one can find $|c_u| \simeq |c_V| \simeq 1$ and
$|c_d| < 1$ (compare also to \reffi{fig:kapukapd} and
the related discussion).

As a consequence of the fact that we find both
in the vicinity of the alignment limit and
in the wrong-sign Yukawa coupling regime parameter
points that are in agreement with the indirect
constraints from the signal-rate
measurements of $h_{125}$ even for values of
$\brinv = 10\%$, whereas the same parameter points
would be excluded in the absence of the
invisible decay mode, it becomes clear
that in the Yukawa type~II
the upper limits on $\brinv$ from the direct
searches for $h_{125} \to \mathrm{inv}$
are (and will be) relevant and have to be taken
into account. Finally, we emphasize that in order
to find a valid parameter point in the
allowed parameter regions for $\brinv > 0$ in
a concrete BSM theory also other constraints,
for instance, from LHC searches for the BSM
Higgs bosons $H$, $A$ and $H^\pm$ would have to
be checked against. We will leave such a
model-specific analysis for future investigations.

\paragraph{Type III} The results for the
type~III (lepton-specific) Yukwa
structure are depicted in the top
panels of \reffi{fig:2HDMTypeIIIandIV}.
The couplings of $h_{125}$ to vector
bosons and to up- and
down-type quarks in type~III are identical 
to the ones in type~I. Thus, the predictions
for the production
cross sections of $h_{125}$ are very similar
to type~I. Accordingly, we find for type~III
that the agreement with the indirect constraints
becomes worse with increasing value
of $\brinv$, similarily to what we observed
for type~I (see \reffi{fig:2HDMTypeIandII}).
This is also true for the wrong-sign Yukawa
coupling regime, which cannot be realized
in type~I, whereas, as can be seen in the
top row of plots in
\reffi{fig:2HDMTypeIIIandIV}, in type~III
the wrong-sign Yukawa coupling regime is
still allowed with regards to the constraints
from the signal-rate measurements of
$h_{125}$. However, in contrast to what
we observed in type~II, in type~III
the fit-result in the
wrong-sign Yukawa coupling regime
becomes worse with increasing value
of $\brinv$, and it is effectively absent
for values of $\brinv > 6\%$.
We find that for $\brinv = 8\%$ only
a small region around the alignment limit
remains allowed if exclusion limits
at the 95\%~CL are applied.
For $\brinv = 10\%$ the whole parameter
plane is excluded in type~III. Hence, in this
case the upper limits on $\brinv$
resulting from the direct searches for
$h_{125} \to \mathrm{inv}$ do not provide
additional constraints, and one should instead
apply the indirect constraints resulting from
the signal-rate measurements of $h_{125}$.

\paragraph{Type IV} The results for the Yukawa
type~IV (flipped) are depicted in the
bottom panels of \reffi{fig:2HDMTypeIIIandIV}.
In this type, the coupling modifiers $c_V$, $c_u$
and $c_d$ are identical to the ones of the
Yukawa type~II. As a result, the cross-section
predictions are also (practically) identical,
and also the main branching ratios of $h_{125}$
remain effectively unchanged compared to
type~II, with the exception of the decays
of $h_{125}$ into pairs of charged leptons.
Accordingly, we find that the allowed region
in the $\{\cos(\alpha-\beta),\tan\beta\}$ plane
resemble the allowed regions that we found
for type~II (see \reffi{fig:2HDMTypeIandII}).
Both in the vicinity of the alignment limit,
but also in the wrong-sign Yukawa coupling regime,
we find allowed regions that accommodate the
experimental data regarding $h_{125}$ within
the $1\sigma$ level. Hence, as in type~II,
the direct limit $\brinv < 11\%$ from the searches
for the $h_{125} \to \mathrm{inv}$ decay mode
have to be taken into account in type~IV, since
they exclude parameter space points with
larger values of $\brinv$ that otherwise
would be allowed.

\section{Summary and conclusions}
\label{sec:conclu}

We explored the possibility of
connecting a hidden sector to the
discovered Higgs boson at~$125\gev$.
We parameterized deviations of the Higgs-boson couplings
to SM states by introducing coupling
modifiers $c_i$ for $i={V,u,d,\ell},$
and by introducing a possible invisible branching
ratio $\brinv$ as a free parameter.
We performed $\chi^2$-analyses
taking into account a large set of cross-section
and signal-rate measurements of the discovered Higgs boson
by using the public code \texttt{HiggsSignals},
and we identified the 
ranges of the coupling modifiers that are
allowed or excluded at $68\%$ and $95\%$
confidence level depending on the
value of $\brinv$.
We furthermore derived constraints
on generic constructions of modifications
of the properties of $h_{125}$ compared
to the SM, by making general
assumptions about the structure of the couplings,
in order for the corresponding constraints to
be applicable to a large category of models.
Finally, we derived constraints on the
model parameters for a
variety of more specific BSM constructions 
featuring hidden
sectors such as Higgs-portal dark matter
models, scenarios featuring Nambu-Goldstone bosons
and extended Higgs sector in the context
of the~2HDM. \par \medskip

The first general conclusion of this paper is that currently in a wide
class of models 
the LHC measurements of the signal rates
of conventional visible signals
of $h_{125}$, in particular in the \textit{golden}
di-photon channel,
impose \textit{indirect}
constraints on the invisible branching ratio of the
discovered Higgs boson 
that can be substantially stronger than
the \textit{direct} constraints resulting from
searches for the invisible Higgs-boson decay.
Hence, for a phenomenological analysis of a model
in which decays of $h_{125}$ into hidden sectors
are kinematically open it is crucial
to not only
consider the direct limits on $\brinv$,
but to also take into account
the indirect constraints on $\brinv$
in order to not include parameter regions
that are in disagreement with the experimental
data from the LHC.

Secondly, we also found that the presence
of a sizable value of $\brinv$ can render
scenarios beyond the SM that predict modifications to the
couplings of $h_{125}$ compared to the SM viable which would
otherwise be excluded by the Higgs-boson signal-rate
measurements. This is trivially true in case
of a universal enhancement of all Higgs-boson
couplings by a common factor, where the
enhancements of the couplings cancel the
suppression of ordinary decay modes of
$h_{125}$ as a result of the additional
decay mode $h_{125} \to \mathrm{inv}$.
However, also more intricate coupling modifications
can be rendered viable via the presence
of an invisible decay mode of $h_{125}$.
The most striking example of our
analysis is the wrong-sign
Yukawa coupling regime realized  
in two Higgs doublet extensions of the SM
with Yukawa type~II and~IV, which we showed
to be in tension with the
cross-section  
measurements of $h_{125}$
assuming $\brinv = 0$, whereas it is in very
good agreement with these  
measurements
assuming values of $\brinv$ as large
as~$10\%$.\par \medskip

For the variety of structures of
coupling-modifiers and BSM model considered
in this work, our main results are summarized in the following.
\par \medskip

\paragraph{SM-like couplings $c_i=1$ and  $\mathbf{\brinv \neq 0}$.} If the Higgs boson possesses couplings to ordinary matter
as predicted by the SM, we found that the
invisible Higgs-boson branching ratio is constrained
to be $\brinv < 6.2\%$ at $95\%$ CL
as a result of the indirect constraints
from $h_{125}$ signal-rate measurements, which is
substantially smaller than the currently strongest limit
$\brinv < 11\%$ resulting from the direct searches
for the invisible decay of $h_{125}$.

\paragraph{Universal couplings $c_i=c_\text{uni}$ and $\text{BR}_\text{inv} \neq 0$.}
We found a flat direction with $\Delta \chi^2 = 0$ in the
$\{\cuni,\brinv\}$ plane where the universal enhancements
of the Higgs-boson couplings for $c_\mathrm{uni} > 1$
are compesated by a non-vanishing $\brinv$.
Based on the indirect constraints,
universal couplings are allowed at 95\% CL in the
interval
$0.520 \, \brinv+0.960\,<\,
\cuni \,<\, 0.565 \, \brinv+1.043 \,$,
and are ultimately only bound by the
direct limit on $\brinv$. In the case where the origin of the coupling
modification is the mixing of an additional
gauge-singlet scalar with the Higgs boson $h_{125}$,
i.e.~${\cuni=\cos\theta}$ with $\theta$
being the mixing angle,
we provided indirect limits on $\brinv$
as a function of $\theta$ in an approximate
form as given in \refeqq{eq:upperboundtheta}.

\paragraph{Non-universal couplings and  $\text{BR}_\text{inv} \neq 0$.}
We considered two examples for
non-universal coupling modifications
of the discovered Higgs boson.
The first example 
consists in a common coupling
modifier for the couplings to fermions,
i.e.~$c_f = c_u = c_d = c_\ell$,
and a second independent coupling
modifier 
$c_V$ 
for the couplings to the gauge bosons.
We found that if $c_V \leq 1$ applies,
invisible $h_{125}$ decay modes gives rise to a
decrease of the allowed ranges of both $c_V$ and $c_f$.
However, if also $c_V > 1$ is considered,
the $\chi^2$~values do not generically degrade as
$\mathrm{BR}_{\rm inv} $ is increased to 10\%.
Hence, the presence of an invisible decay mode of
$h_{125}$ with sizable branching ratios
opens up parameter space regions
in the $\{c_f,c_V\}$ plane that would
be excluded if such a novel decay mode is not present.
The second example  
considered is a SM-like Higgs-boson coupling to
vector bosons $c_V = 1$ but modified couplings
to up and down-type quarks with the additional condition
$c_\ell = c_d$. We found that increasing values
of $\brinv$ shifts the allowed parameter space
to larger values of $c_u$ and smaller values
of $c_d = c_\ell$, but overall yielding a degraded 
fit result.

\paragraph{Higgs portal dark matter.}
We considered dark matter candidates
(scalar, fermion and vector) coupled to the
Higgs boson via a single operator and computed
the relic abundance and direct detection cross
section in terms of the dimensionless
Higgs-portal coupling parameter
for each case.  
Direct detection experiments
tightly constrain the parameter space suitable 
to achieve the measured  
relic abundance, leaving for the bosonic
candidates only the mass
region at  
the Higgs-boson resonance,
i.e.~if the DM mass is
$\mathcal{O}(m_{h_{125}}/2)$.
We showed that new results
from the LZ collaboration rule
out the fermionic DM candidate within the
simplest Higgs-portal scenario under standard assumptions.\footnote{Given the fact that the LZ limit is close to the parameter space corresponding to the correct relic abundance, several uncertainties can affect this statement. To derive bounds on the DM nucleon scattering cross section, a local DM density of $\rho_\odot=0.3 \,\text{GeV}\, \text{cm}^{-3}$ and the standard halo model for the distribution of dark matter in the Milky Way are standard assumptions from direct detection collaborations. Both quantities suffer from astrophysical uncertainties that could affect constraints from direct searches by a $\mathcal{O}(1-3)$ factor~\cite{deSalas:2020hbh}. Such uncertainties could allow the parameter space around the resonance to still be viable but on the edge of being excluded.}. Around the resonance, if we relax the hypothesis
of the DM  
candidates to account for
all the dark matter, the indirect 
constraints resulting from the cross-section
measurements of $h_{125}$
derived using \texttt{HiggsSignal}
are the strongest bounds in the narrow
range of DM masses that is not excluded
by direct detection experiments.

\paragraph{Singlet portal dark matter.}

We studied
an extension of the Higgs-portal scenario comprising an
additional singlet-like scalar $h_2$ acting as the
portal between a dark matter fermion and the
visible sector.
As a consequence of the
presence of $h_2$, we also took into account
constraints from collider searches for
additional Higgs bosons using the
public code
\texttt{HiggsBounds}.
Firstly, we investigated a benchmark scenario in which
both the decays of $h_{125}$ into the dark-matter
state and into a pair of the additional scalar
are kinematically allowed, which both can
contribute to $\brinv$ depending on whether
$h_2$ decays invisibly.
Secondly, we investigated
a benchmark scenario in which the second scalar
has a mass of~$80\gev$, allowing for a larger
mixing between discovered Higgs boson and $h_2$.
We identified the regions of the parameter space
that are still
allowed by direct detection constraints
and in which part or all of the measured relic
abundance of dark matter is predicted.
We demonstrated that in such regions of the
parameter space the indirect constraints
from the cross-section measurements of~$h_{125}$
provide the strongest constraints, whereas
the direct limit on $\brinv$ from searches
for the invisible decay of~$h_{125}$ do not
give rise to any additional constraints.

\paragraph{(Pseudo) Nambu-Goldstone bosons ((P)NGB).}
We considered the possibility of the 
Higgs boson $h_{125}$ decaying into
pairs of (P)NGBs,  
where the (P)NGB arises as
the angular mode of a new complex scalar  
after the spontaneous breaking of an additional
(approximate) continuous symmetry
once the radial mode acquires a vev $v_\phi$.
Using the cross-section measurements
of~$h_{125}$, we determined constraints
on the mixing between~$h_{125}$ 
and the radial mode and on the
invisible branching ratio $\brinv$.
We highlighted that
if one would apply independent constraints on
the mixing angle $\theta$ and the invisible
branching ratio $\brinv$, one would consider
parameter regions as allowed
which are actually excluded by the
cross-section measurements of $h_{125}$,
as we showed by simultaneously
taking into account both
the mixing in the Higgs sector
and the presence of a non-vanishing
$\brinv$ in a global fit
to the experimental Higgs-physics observables.

\paragraph{Two Higgs doublet models.}
We derived constraints in the plane
$\{\tan \beta, \cos (\alpha-\beta)\}$ for
$\tan \beta\in [0,20 ]$ in the four Yukawa
types of the 2HDM.
We found that the $\chi^2$-fit result deteriorates in the
type~I and~III as $\brinv$ is increased,
leaving both types as excluded at the
95\%~CL for values of $\brinv > 8\%$.
On the other hand,
for type~II and~IV
a narrow region
with small but positive values of
$\cos(\alpha-\beta)$
remains viable even for $\brinv = 10 \%$.
We also found that in these two types the
wrong-sign Yukawa coupling regime, which
is barely in agreement with the measurements
of $h_{125}$ for $\brinv = 0$,
can be in very good agreement with the
experimental data if the presence of
an invisible decay mode of $h_{125}$
with $\brinv > 2\%$ is assumed to be present.


\section*{Acknowledgements}
The authors thank Anton Sokolov for useful discussions.
The authors acknowledge support by the Deutsche Forschungsgemeinschaft (DFG, German Research Foundation) under Germany‘s Excellence Strategy – EXC 2121 “Quantum Universe” – 390833306. This work has been partially funded by the Deutsche Forschungsgemeinschaft (DFG, German Research Foundation) - 491245950. This work was made possible by with the support of the Institut Pascal at Université Paris-Saclay during the Paris-Saclay Astroparticle Symposium 2021, with the support of the P2IO Laboratory of Excellence (program “Investissements d’avenir” ANR-11-IDEX-0003-01 Paris-Saclay and ANR-10-LABX-0038), the P2I axis of the Graduate School Physics of Université Paris-Saclay, as well as IJCLab, CEA, IPhT, APPEC, the IN2P3 master projet UCMN and EuCAPT ANR-11-IDEX-0003-01 Paris-Saclay and ANR-10-LABX-0038.

\appendix

\section{Singlet portal dark matter: additional content}
\label{app:singletportal}

\subsection{Minimization of the potential}

The vevs are related to the bilinear mass
parameters via the tadpole equations
\begin{equation}
\mu_H^2 = -\frac{1}{2} \left(
  2 \lambda_H v^2 + \lambda_{\Phi H} v_\phi^2 \right) \ , \quad
\mu_\Phi^2 = -\frac{1}{2} \left(
  \lambda_{\Phi H} v^2 + 2 \lambda_\Phi v_\phi^2 \right) \ .
\end{equation}
Using these relations, the mass matrix of
the CP-even fields $h$ and $\phi$ can be written as
\begin{equation}
\mathcal M^2 = \begin{pmatrix}
2 \lambda_H v^2 & \lambda_{\Phi H} v v_\phi \\
\lambda_{\Phi H} v v_\phi & 2 \lambda_\Phi v_\phi^2
\end{pmatrix} \ .
\end{equation}
This matrix can be diagonalized by a rotation of angle $\theta$, expressed in terms of the various parameters as
\begin{equation}
    \tan (2\theta)\,=\,\dfrac{\lambda_{\Phi H} v v_\phi}{\lambda_\Phi v_\phi^2-\lambda_H v^2}\,.
\end{equation}
The quartic scalar couplings are then
dependent parameters that can be computed
via the relations
\begin{align}
\lambda_H &= \frac{1}{2 v^2} \left(
  m_{h_1}^2 \cos^2\theta + m_{h_2}^2 \sin^2\theta \right) \ , \\
  \label{eq:rotationsingletscalar1}
\lambda_\Phi &= \frac{1}{2 v_\phi^2} \left(
  m_{h_1}^2 \sin^2\theta + m_{h_2}^2 \cos^2\theta \right) \ , \\
\lambda_{\Phi H} &= \frac{1}{v v_\phi} \left(
  m_{h_2}^2 - m_{h_1}^2 \right) \cos\theta \sin\theta \ .
  \label{eq:rotationsingletscalar3}
\end{align}

\subsection{Decay rates}
\label{app:decayratesingletscalarDM}
 The relevant partial decay width for the SM-like Higgs $h_1\simeq h_{125} $ are
\begin{equation}
  \Gamma_{h_1 \rightarrow h_2 h_2} \, = \, \frac{ \left(c_\theta^3 \lambda_{\Phi H} v +2 c_\theta^2 s_\theta v_\phi (\lambda_{\Phi H}-3 \lambda_{S})+2 c_\theta s_\theta^2 v (3 \lambda_{H}-\lambda_{\Phi H})-\lambda_{\Phi H} s_\theta^3 v_\phi\right)^2}{32 \pi  m_{h_1}} \sqrt{1-\dfrac{4m_{h_2}^2}{m_{h_1}^2 }} \,,
\end{equation}
Decay rates from the physical scalars to a DM pair are given by
\begin{equation}
    \Gamma_{h_1 \rightarrow \bar \chi \chi} \, = \,  \dfrac{y_\chi^2 s_\theta^2}{8\pi} m_{h_1} \left(1-\frac{4 m_{\chi }^2}{m_{h_1}^2}\right)^{3/2}\,=\,\frac{s_{\theta }^2  m_{h_1} m_{\chi }^2 }{8 \pi  v_\phi^2}\left(1-\frac{4 m_{\chi }^2}{m_{h_1}^2}\right)^{3/2}~,
\end{equation}
and
\begin{equation}
    \Gamma_{h_2 \rightarrow \bar \chi \chi} \, = \,  \dfrac{y_\chi^2 c_\theta^2}{8\pi} m_{h_2} \left(1-\frac{4 m_{\chi }^2}{m_{h_2}^2}\right)^{3/2}\,=\,\frac{c_{\theta }^2  m_{h_2} m_{\chi }^2 }{8 \pi  v_\phi^2}\left(1-\frac{4 m_{\chi }^2}{m_{h_2}^2}\right)^{3/2}~.
\end{equation}

\addcontentsline{toc}{section}{References}
\bibliography{references}

\providecommand{\href}[2]{#2}\begingroup\raggedright\begin{thebibliography}{10}

\bibitem{Aad:2012tfa}
{\scshape ATLAS} collaboration, \emph{{Observation of a new particle in the
  search for the Standard Model Higgs boson with the ATLAS detector at the
  LHC}}, \href{https://doi.org/10.1016/j.physletb.2012.08.020}{\emph{Phys.
  Lett.} {\bfseries B716} (2012) 1}
  [\href{https://arxiv.org/abs/1207.7214}{{\ttfamily 1207.7214}}].

\bibitem{Chatrchyan:2012xdj}
{\scshape CMS} collaboration, \emph{{Observation of a new boson at a mass of
  125 GeV with the CMS experiment at the LHC}},
  \href{https://doi.org/10.1016/j.physletb.2012.08.021}{\emph{Phys. Lett.}
  {\bfseries B716} (2012) 30}
  [\href{https://arxiv.org/abs/1207.7235}{{\ttfamily 1207.7235}}].

\bibitem{CMS:2022dwd}
{\scshape CMS} collaboration, \emph{{A portrait of the Higgs boson by the CMS
  experiment ten years after the discovery}},
  \href{https://doi.org/10.1038/s41586-022-04892-x}{\emph{Nature} {\bfseries
  607} (2022) 60} [\href{https://arxiv.org/abs/2207.00043}{{\ttfamily
  2207.00043}}].

\bibitem{ATLAS:2022vkf}
{\scshape ATLAS} collaboration, \emph{{A detailed map of Higgs boson
  interactions by the ATLAS experiment ten years after the discovery}},
  \href{https://doi.org/10.1038/s41586-022-04893-w}{\emph{Nature} {\bfseries
  607} (2022) 52} [\href{https://arxiv.org/abs/2207.00092}{{\ttfamily
  2207.00092}}].

\bibitem{LHCHiggsCrossSectionWorkingGroup:2013rie}
{\scshape LHC Higgs Cross Section Working Group} collaboration, \emph{{Handbook
  of LHC Higgs Cross Sections: 3. Higgs Properties}},
  \href{https://arxiv.org/abs/1307.1347}{{\ttfamily 1307.1347}}.

\bibitem{Lineros:2020eit}
R.A.~Lineros and M.~Pierre, \emph{{Dark matter candidates in a type-II
  radiative neutrino mass model}},
  \href{https://doi.org/10.1007/JHEP06(2021)072}{\emph{JHEP} {\bfseries 21}
  (2020) 072} [\href{https://arxiv.org/abs/2011.08195}{{\ttfamily
  2011.08195}}].

\bibitem{CMS:2020xrn}
{\scshape CMS} collaboration, \emph{{A measurement of the Higgs boson mass in
  the diphoton decay channel}},
  \href{https://doi.org/10.1016/j.physletb.2020.135425}{\emph{Phys. Lett. B}
  {\bfseries 805} (2020) 135425}
  [\href{https://arxiv.org/abs/2002.06398}{{\ttfamily 2002.06398}}].

\bibitem{ATLAS:2022net}
{\scshape ATLAS} collaboration, \emph{{Measurement of the Higgs boson mass in
  the $H \rightarrow ZZ^* \rightarrow 4\ell$ decay channel using 139 fb$^{-1}$
  of $\sqrt{s}=13$ TeV $pp$ collisions recorded by the ATLAS detector at the
  LHC}},  \href{https://arxiv.org/abs/2207.00320}{{\ttfamily 2207.00320}}.

\bibitem{ATLAS:2021wwb}
{\scshape ATLAS} collaboration, \emph{{Evidence for Higgs boson decays to a
  low-mass dilepton system and a photon in pp collisions at s=13 TeV with the
  ATLAS detector}},
  \href{https://doi.org/10.1016/j.physletb.2021.136412}{\emph{Phys. Lett. B}
  {\bfseries 819} (2021) 136412}
  [\href{https://arxiv.org/abs/2103.10322}{{\ttfamily 2103.10322}}].

\bibitem{ATLAS:2020qcv}
{\scshape ATLAS} collaboration, \emph{{A search for the $Z\gamma$ decay mode of
  the Higgs boson in $pp$ collisions at $\sqrt{s}$ = 13 TeV with the ATLAS
  detector}}, \href{https://doi.org/10.1016/j.physletb.2020.135754}{\emph{Phys.
  Lett. B} {\bfseries 809} (2020) 135754}
  [\href{https://arxiv.org/abs/2005.05382}{{\ttfamily 2005.05382}}].

\bibitem{CMS:2022ahq}
{\scshape CMS} collaboration, \emph{{Search for Higgs boson decays to a Z boson
  and a photon in proton-proton collisions at $\sqrt{s}$ = 13 TeV}},
  \href{https://arxiv.org/abs/2204.12945}{{\ttfamily 2204.12945}}.

\bibitem{Aghanim:2018eyx}
{\scshape Planck} collaboration, \emph{{Planck 2018 results. VI. Cosmological
  parameters}},
  \href{https://doi.org/10.1051/0004-6361/201833910}{\emph{Astron. Astrophys.}
  {\bfseries 641} (2020) A6}
  [\href{https://arxiv.org/abs/1807.06209}{{\ttfamily 1807.06209}}].

\bibitem{LUX:2016ggv}
{\scshape LUX} collaboration, \emph{{Results from a search for dark matter in
  the complete LUX exposure}},
  \href{https://doi.org/10.1103/PhysRevLett.118.021303}{\emph{Phys. Rev. Lett.}
  {\bfseries 118} (2017) 021303}
  [\href{https://arxiv.org/abs/1608.07648}{{\ttfamily 1608.07648}}].

\bibitem{Aprile:2018dbl}
{\scshape XENON} collaboration, \emph{{Dark Matter Search Results from a One
  Ton-Year Exposure of XENON1T}},
  \href{https://doi.org/10.1103/PhysRevLett.121.111302}{\emph{Phys. Rev. Lett.}
  {\bfseries 121} (2018) 111302}
  [\href{https://arxiv.org/abs/1805.12562}{{\ttfamily 1805.12562}}].

\bibitem{PandaX-4T:2021bab}
{\scshape PandaX-4T} collaboration, \emph{{Dark Matter Search Results from the
  PandaX-4T Commissioning Run}},
  \href{https://doi.org/10.1103/PhysRevLett.127.261802}{\emph{Phys. Rev. Lett.}
  {\bfseries 127} (2021) 261802}
  [\href{https://arxiv.org/abs/2107.13438}{{\ttfamily 2107.13438}}].

\bibitem{LZnew}
{\scshape LUX-ZEPLIN} collaboration, \emph{{First Dark Matter Search Results
  from the LUX-ZEPLIN (LZ) Experiment}},
  \href{https://arxiv.org/abs/2207.03764}{{\ttfamily 2207.03764}}.

\bibitem{Arcadi:2017kky}
G.~Arcadi, M.~Dutra, P.~Ghosh, M.~Lindner, Y.~Mambrini, M.~Pierre et~al.,
  \emph{{The waning of the WIMP? A review of models, searches, and
  constraints}},
  \href{https://doi.org/10.1140/epjc/s10052-018-5662-y}{\emph{Eur. Phys. J. C}
  {\bfseries 78} (2018) 203}
  [\href{https://arxiv.org/abs/1703.07364}{{\ttfamily 1703.07364}}].

\bibitem{Mambrini}
Y.~Mambrini, \emph{Particles in the dark Universe}, no.~ISBN 978-3-030-78139-2,
  Springer (2021).

\bibitem{Escudero:2016gzx}
M.~Escudero, A.~Berlin, D.~Hooper and M.-X.~Lin, \emph{{Toward (Finally!)
  Ruling Out Z and Higgs Mediated Dark Matter Models}},
  \href{https://doi.org/10.1088/1475-7516/2016/12/029}{\emph{JCAP} {\bfseries
  12} (2016) 029} [\href{https://arxiv.org/abs/1609.09079}{{\ttfamily
  1609.09079}}].

\bibitem{Casas:2017jjg}
J.A.~Casas, D.G.~Cerde\~no, J.M.~Moreno and J.~Quilis, \emph{{Reopening the
  Higgs portal for single scalar dark matter}},
  \href{https://doi.org/10.1007/JHEP05(2017)036}{\emph{JHEP} {\bfseries 05}
  (2017) 036} [\href{https://arxiv.org/abs/1701.08134}{{\ttfamily
  1701.08134}}].

\bibitem{Ellis:2017ndg}
J.~Ellis, A.~Fowlie, L.~Marzola and M.~Raidal, \emph{{Statistical Analyses of
  Higgs- and Z-Portal Dark Matter Models}},
  \href{https://doi.org/10.1103/PhysRevD.97.115014}{\emph{Phys. Rev. D}
  {\bfseries 97} (2018) 115014}
  [\href{https://arxiv.org/abs/1711.09912}{{\ttfamily 1711.09912}}].

\bibitem{Arcadi:2019lka}
G.~Arcadi, A.~Djouadi and M.~Raidal, \emph{{Dark Matter through the Higgs
  portal}}, \href{https://doi.org/10.1016/j.physrep.2019.11.003}{\emph{Phys.
  Rept.} {\bfseries 842} (2020) 1}
  [\href{https://arxiv.org/abs/1903.03616}{{\ttfamily 1903.03616}}].

\bibitem{Arcadi:2021mag}
G.~Arcadi, A.~Djouadi and M.~Kado, \emph{{The Higgs-portal for dark matter:
  effective field theories versus concrete realizations}},
  \href{https://doi.org/10.1140/epjc/s10052-021-09411-2}{\emph{Eur. Phys. J. C}
  {\bfseries 81} (2021) 653}
  [\href{https://arxiv.org/abs/2101.02507}{{\ttfamily 2101.02507}}].

\bibitem{Cirelli:2005uq}
M.~Cirelli, N.~Fornengo and A.~Strumia, \emph{{Minimal dark matter}},
  \href{https://doi.org/10.1016/j.nuclphysb.2006.07.012}{\emph{Nucl. Phys. B}
  {\bfseries 753} (2006) 178}
  [\href{https://arxiv.org/abs/hep-ph/0512090}{{\ttfamily hep-ph/0512090}}].

\bibitem{Bottaro:2021snn}
S.~Bottaro, D.~Buttazzo, M.~Costa, R.~Franceschini, P.~Panci, D.~Redigolo
  et~al., \emph{{Closing the window on WIMP Dark Matter}},
  \href{https://doi.org/10.1140/epjc/s10052-021-09917-9}{\emph{Eur. Phys. J. C}
  {\bfseries 82} (2022) 31} [\href{https://arxiv.org/abs/2107.09688}{{\ttfamily
  2107.09688}}].

\bibitem{Bottaro:2022one}
S.~Bottaro, D.~Buttazzo, M.~Costa, R.~Franceschini, P.~Panci, D.~Redigolo
  et~al., \emph{{The last complex WIMPs standing}},
  \href{https://doi.org/10.1140/epjc/s10052-022-10918-5}{\emph{Eur. Phys. J. C}
  {\bfseries 82} (2022) 992}
  [\href{https://arxiv.org/abs/2205.04486}{{\ttfamily 2205.04486}}].

\bibitem{Patt:2006fw}
B.~Patt and F.~Wilczek, \emph{{Higgs-field portal into hidden sectors}},
  \href{https://arxiv.org/abs/hep-ph/0605188}{{\ttfamily hep-ph/0605188}}.

\bibitem{Djouadi:2012zc}
A.~Djouadi, A.~Falkowski, Y.~Mambrini and J.~Quevillon, \emph{{Direct Detection
  of Higgs-Portal Dark Matter at the LHC}},
  \href{https://doi.org/10.1140/epjc/s10052-013-2455-1}{\emph{Eur. Phys. J. C}
  {\bfseries 73} (2013) 2455}
  [\href{https://arxiv.org/abs/1205.3169}{{\ttfamily 1205.3169}}].

\bibitem{CMS:2022qva}
{\scshape CMS} collaboration, \emph{{Search for invisible decays of the Higgs
  boson produced via vector boson fusion in proton-proton collisions at
  $\sqrt{s} =$ 13 TeV}},
  \href{https://doi.org/10.1103/PhysRevD.105.092007}{\emph{Phys. Rev. D}
  {\bfseries 105} (2022) 092007}
  [\href{https://arxiv.org/abs/2201.11585}{{\ttfamily 2201.11585}}].

\bibitem{CMS:2018yfx}
{\scshape CMS} collaboration, \emph{{Search for invisible decays of a Higgs
  boson produced through vector boson fusion in proton-proton collisions at
  $\sqrt{s} =$ 13 TeV}},
  \href{https://doi.org/10.1016/j.physletb.2019.04.025}{\emph{Phys. Lett. B}
  {\bfseries 793} (2019) 520}
  [\href{https://arxiv.org/abs/1809.05937}{{\ttfamily 1809.05937}}].

\bibitem{CMS:2017zts}
{\scshape CMS} collaboration, \emph{{Search for new physics in final states
  with an energetic jet or a hadronically decaying $W$ or $Z$ boson and
  transverse momentum imbalance at $\sqrt{s}=13\text{ }\text{ }\mathrm{TeV}$}},
  \href{https://doi.org/10.1103/PhysRevD.97.092005}{\emph{Phys. Rev. D}
  {\bfseries 97} (2018) 092005}
  [\href{https://arxiv.org/abs/1712.02345}{{\ttfamily 1712.02345}}].

\bibitem{CMS:2017nxf}
{\scshape CMS} collaboration, \emph{{Search for new physics in events with a
  leptonically decaying Z boson and a large transverse momentum imbalance in
  proton\textendash{}proton collisions at $\sqrt{s} $ = 13 $\,\text {TeV}$}},
  \href{https://doi.org/10.1140/epjc/s10052-018-5740-1}{\emph{Eur. Phys. J. C}
  {\bfseries 78} (2018) 291}
  [\href{https://arxiv.org/abs/1711.00431}{{\ttfamily 1711.00431}}].

\bibitem{ATLAS-CONF-2020-052}
{\scshape ATLAS} collaboration, \emph{{Combination of searches for invisible
  Higgs boson decays with the ATLAS experiment}},  Tech. Rep.
  \href{https://cds.cern.ch/record/2743055}{ATLAS-CONF-2020-052}, CERN, Geneva
  (Oct, 2020).

\bibitem{ATLAS:2019cid}
{\scshape ATLAS} collaboration, \emph{{Combination of searches for invisible
  Higgs boson decays with the ATLAS experiment}},
  \href{https://doi.org/10.1103/PhysRevLett.122.231801}{\emph{Phys. Rev. Lett.}
  {\bfseries 122} (2019) 231801}
  [\href{https://arxiv.org/abs/1904.05105}{{\ttfamily 1904.05105}}].

\bibitem{ATLAS:2022yvh}
{\scshape ATLAS} collaboration, \emph{{Search for invisible Higgs-boson decays
  in events with vector-boson fusion signatures using 139 fb$^{-1}$ of
  proton-proton data recorded by the ATLAS experiment}},
  \href{https://doi.org/10.1007/JHEP08(2022)104}{\emph{JHEP} {\bfseries 08}
  (2022) 104} [\href{https://arxiv.org/abs/2202.07953}{{\ttfamily
  2202.07953}}].

\bibitem{ATLAS:2017nyv}
{\scshape ATLAS} collaboration, \emph{{Search for an invisibly decaying Higgs
  boson or dark matter candidates produced in association with a $Z$ boson in
  $pp$ collisions at $\sqrt{s} =$ 13 TeV with the ATLAS detector}},
  \href{https://doi.org/10.1016/j.physletb.2017.11.049}{\emph{Phys. Lett. B}
  {\bfseries 776} (2018) 318}
  [\href{https://arxiv.org/abs/1708.09624}{{\ttfamily 1708.09624}}].

\bibitem{ATLAS:2018nda}
{\scshape ATLAS} collaboration, \emph{{Search for dark matter in events with a
  hadronically decaying vector boson and missing transverse momentum in $pp$
  collisions at $\sqrt{s} = 13$ TeV with the ATLAS detector}},
  \href{https://doi.org/10.1007/JHEP10(2018)180}{\emph{JHEP} {\bfseries 10}
  (2018) 180} [\href{https://arxiv.org/abs/1807.11471}{{\ttfamily
  1807.11471}}].

\bibitem{Yaguna:2021rds}
C.E.~Yaguna and O.~Zapata, \emph{{Fermion and scalar two-component dark matter
  from a Z4 symmetry}},
  \href{https://doi.org/10.1103/PhysRevD.105.095026}{\emph{Phys. Rev. D}
  {\bfseries 105} (2022) 095026}
  [\href{https://arxiv.org/abs/2112.07020}{{\ttfamily 2112.07020}}].

\bibitem{Okada:2020zxo}
N.~Okada, D.~Raut and Q.~Shafi, \emph{{Pseudo-Goldstone dark matter in a gauged
  $B-L$ extended standard model}},
  \href{https://doi.org/10.1103/PhysRevD.103.055024}{\emph{Phys. Rev. D}
  {\bfseries 103} (2021) 055024}
  [\href{https://arxiv.org/abs/2001.05910}{{\ttfamily 2001.05910}}].

\bibitem{Hara:2021lrj}
T.~Hara, S.~Kanemura and T.~Katayose, \emph{{Is light thermal scalar dark
  matter possible?}},
  \href{https://doi.org/10.1103/PhysRevD.105.035035}{\emph{Phys. Rev. D}
  {\bfseries 105} (2022) 035035}
  [\href{https://arxiv.org/abs/2109.03553}{{\ttfamily 2109.03553}}].

\bibitem{Espinosa:2012vu}
J.R.~Espinosa, M.~Muhlleitner, C.~Grojean and M.~Trott, \emph{{Probing for
  Invisible Higgs Decays with Global Fits}},
  \href{https://doi.org/10.1007/JHEP09(2012)126}{\emph{JHEP} {\bfseries 09}
  (2012) 126} [\href{https://arxiv.org/abs/1205.6790}{{\ttfamily 1205.6790}}].

\bibitem{Espinosa:2012im}
J.R.~Espinosa, C.~Grojean, M.~Muhlleitner and M.~Trott, \emph{{First Glimpses
  at Higgs' face}}, \href{https://doi.org/10.1007/JHEP12(2012)045}{\emph{JHEP}
  {\bfseries 12} (2012) 045} [\href{https://arxiv.org/abs/1207.1717}{{\ttfamily
  1207.1717}}].

\bibitem{Belanger:2013kya}
G.~Belanger, B.~Dumont, U.~Ellwanger, J.F.~Gunion and S.~Kraml, \emph{{Status
  of invisible Higgs decays}},
  \href{https://doi.org/10.1016/j.physletb.2013.05.024}{\emph{Phys. Lett. B}
  {\bfseries 723} (2013) 340}
  [\href{https://arxiv.org/abs/1302.5694}{{\ttfamily 1302.5694}}].

\bibitem{Bechtle:2014ewa}
P.~Bechtle, S.~Heinemeyer, O.~Stal, T.~Stefaniak and G.~Weiglein,
  \emph{{Probing the Standard Model with Higgs signal rates from the Tevatron,
  the LHC and a future ILC}},
  \href{https://doi.org/10.1007/JHEP11(2014)039}{\emph{JHEP} {\bfseries 11}
  (2014) 039} [\href{https://arxiv.org/abs/1403.1582}{{\ttfamily 1403.1582}}].

\bibitem{Kraml:2019sis}
S.~Kraml, T.Q.~Loc, D.T.~Nhung and L.D.~Ninh, \emph{{Constraining new physics
  from Higgs measurements with Lilith: update to LHC Run 2 results}},
  \href{https://doi.org/10.21468/SciPostPhys.7.4.052}{\emph{SciPost Phys.}
  {\bfseries 7} (2019) 052} [\href{https://arxiv.org/abs/1908.03952}{{\ttfamily
  1908.03952}}].

\bibitem{Bechtle:2020uwn}
P.~Bechtle, S.~Heinemeyer, T.~Klingl, T.~Stefaniak, G.~Weiglein and
  J.~Wittbrodt, \emph{{HiggsSignals-2: Probing new physics with precision Higgs
  measurements in the LHC 13 TeV era}},
  \href{https://doi.org/10.1140/epjc/s10052-021-08942-y}{\emph{Eur. Phys. J. C}
  {\bfseries 81} (2021) 145}
  [\href{https://arxiv.org/abs/2012.09197}{{\ttfamily 2012.09197}}].

\bibitem{hsnew}
H.~Bahl, T.~Biek\"otter, S.~Heinemeyer, C.~Li, S.~Paasch, G.~Weiglein et~al.,
  \emph{{HiggsTools: BSM scalar phenomenology with new versions of HiggsBounds
  and HiggsSignals}},  \href{https://arxiv.org/abs/2210.09332}{{\ttfamily
  2210.09332}}.

\bibitem{Belanger:2013xza}
G.~Belanger, B.~Dumont, U.~Ellwanger, J.F.~Gunion and S.~Kraml, \emph{{Global
  fit to Higgs signal strengths and couplings and implications for extended
  Higgs sectors}},
  \href{https://doi.org/10.1103/PhysRevD.88.075008}{\emph{Phys. Rev. D}
  {\bfseries 88} (2013) 075008}
  [\href{https://arxiv.org/abs/1306.2941}{{\ttfamily 1306.2941}}].

\bibitem{Bernon:2014vta}
J.~Bernon, B.~Dumont and S.~Kraml, \emph{{Status of Higgs couplings after run 1
  of the LHC}}, \href{https://doi.org/10.1103/PhysRevD.90.071301}{\emph{Phys.
  Rev. D} {\bfseries 90} (2014) 071301}
  [\href{https://arxiv.org/abs/1409.1588}{{\ttfamily 1409.1588}}].

\bibitem{Bechtle:2013xfa}
P.~Bechtle, S.~Heinemeyer, O.~Stal, T.~Stefaniak and G.~Weiglein,
  \emph{{$HiggsSignals$: Confronting arbitrary Higgs sectors with measurements
  at the Tevatron and the LHC}},
  \href{https://doi.org/10.1140/epjc/s10052-013-2711-4}{\emph{Eur. Phys. J. C}
  {\bfseries 74} (2014) 2711}
  [\href{https://arxiv.org/abs/1305.1933}{{\ttfamily 1305.1933}}].

\bibitem{Berger:2019wnu}
N.~Berger et~al., \emph{{Simplified Template Cross Sections - Stage 1.1}},
  \href{https://arxiv.org/abs/1906.02754}{{\ttfamily 1906.02754}}.

\bibitem{LHCHiggsCrossSectionWorkingGroup:2011wcg}
{\scshape LHC Higgs Cross Section Working Group} collaboration, \emph{{Handbook
  of LHC Higgs Cross Sections: 1. Inclusive Observables}},
  \href{https://arxiv.org/abs/1101.0593}{{\ttfamily 1101.0593}}.

\bibitem{ParticleDataGroup:2020ssz}
{\scshape Particle Data Group} collaboration, \emph{{Review of Particle
  Physics}}, \href{https://doi.org/10.1093/ptep/ptaa104}{\emph{PTEP} {\bfseries
  2020} (2020) 083C01}.

\bibitem{ATLAS-CONF-2019-045}
{\scshape ATLAS} collaboration, \emph{{Analysis of $t\bar{t}H$ and $t\bar{t}W$
  production in multilepton final states with the ATLAS detector}},  Tech. Rep.
  \href{http://cds.cern.ch/record/2693930}{ATLAS-CONF-2019-045}, CERN, Geneva
  (Oct, 2019).

\bibitem{ATLAS:2021qou}
{\scshape ATLAS} collaboration, \emph{{Measurement of Higgs boson decay into
  $b$-quarks in associated production with a top-quark pair in $pp$ collisions
  at $\sqrt{s}=13$ TeV with the ATLAS detector}},
  \href{https://doi.org/10.1007/JHEP06(2022)097}{\emph{JHEP} {\bfseries 06}
  (2022) 097} [\href{https://arxiv.org/abs/2111.06712}{{\ttfamily
  2111.06712}}].

\bibitem{CMS:2021kom}
{\scshape CMS} collaboration, \emph{{Measurements of Higgs boson production
  cross sections and couplings in the diphoton decay channel at $
  \sqrt{\mathrm{s}} $ = 13 TeV}},
  \href{https://doi.org/10.1007/JHEP07(2021)027}{\emph{JHEP} {\bfseries 07}
  (2021) 027} [\href{https://arxiv.org/abs/2103.06956}{{\ttfamily
  2103.06956}}].

\bibitem{ATLAS:2020pvn}
{\scshape ATLAS} collaboration, \emph{{Measurement of the properties of Higgs
  boson production at $\sqrt{s}$=13 TeV in the $H\to \gamma\gamma$ channel
  using 139 fb$^{-1}$ of $pp$ collision data with the ATLAS experiment}},
  {\emph{{\rm ATLAS-CONF-2020-026}} (2020) }.

\bibitem{Cepeda:2019klc}
M.~Cepeda et~al., \emph{{Report from Working Group 2}: {Higgs Physics at the
  HL-LHC and HE-LHC}},
  \href{https://doi.org/10.23731/CYRM-2019-007.221}{\emph{CERN Yellow Rep.
  Monogr.} {\bfseries 7} (2019) 221}
  [\href{https://arxiv.org/abs/1902.00134}{{\ttfamily 1902.00134}}].

\bibitem{ATLAS:2016neq}
{\scshape ATLAS, CMS} collaboration, \emph{{Measurements of the Higgs boson
  production and decay rates and constraints on its couplings from a combined
  ATLAS and CMS analysis of the LHC pp collision data at $ \sqrt{s}=7 $ and 8
  TeV}}, \href{https://doi.org/10.1007/JHEP08(2016)045}{\emph{JHEP} {\bfseries
  08} (2016) 045} [\href{https://arxiv.org/abs/1606.02266}{{\ttfamily
  1606.02266}}].

\bibitem{CMS:2018uag}
{\scshape CMS} collaboration, \emph{{Combined measurements of Higgs boson
  couplings in proton\textendash{}proton collisions at $\sqrt{s}=13\,\text
  {Te}\text {V} $}},
  \href{https://doi.org/10.1140/epjc/s10052-019-6909-y}{\emph{Eur. Phys. J. C}
  {\bfseries 79} (2019) 421}
  [\href{https://arxiv.org/abs/1809.10733}{{\ttfamily 1809.10733}}].

\bibitem{ATLAS:2019nkf}
{\scshape ATLAS} collaboration, \emph{{Combined measurements of Higgs boson
  production and decay using up to $80$ fb$^{-1}$ of proton-proton collision
  data at $\sqrt{s}=$ 13 TeV collected with the ATLAS experiment}},
  \href{https://doi.org/10.1103/PhysRevD.101.012002}{\emph{Phys. Rev. D}
  {\bfseries 101} (2020) 012002}
  [\href{https://arxiv.org/abs/1909.02845}{{\ttfamily 1909.02845}}].

\bibitem{FileviezPerez:2008bj}
P.~Fileviez~Perez, H.H.~Patel, M.J.~Ramsey-Musolf and K.~Wang, \emph{{Triplet
  Scalars and Dark Matter at the LHC}},
  \href{https://doi.org/10.1103/PhysRevD.79.055024}{\emph{Phys. Rev. D}
  {\bfseries 79} (2009) 055024}
  [\href{https://arxiv.org/abs/0811.3957}{{\ttfamily 0811.3957}}].

\bibitem{No:2015xqa}
J.M.~No, \emph{{Looking through the pseudoscalar portal into dark matter: Novel
  mono-Higgs and mono-Z signatures at the LHC}},
  \href{https://doi.org/10.1103/PhysRevD.93.031701}{\emph{Phys. Rev. D}
  {\bfseries 93} (2016) 031701}
  [\href{https://arxiv.org/abs/1509.01110}{{\ttfamily 1509.01110}}].

\bibitem{Arina:2019tib}
C.~Arina, A.~Beniwal, C.~Degrande, J.~Heisig and A.~Scaffidi, \emph{{Global fit
  of pseudo-Nambu-Goldstone Dark Matter}},
  \href{https://doi.org/10.1007/JHEP04(2020)015}{\emph{JHEP} {\bfseries 04}
  (2020) 015} [\href{https://arxiv.org/abs/1912.04008}{{\ttfamily
  1912.04008}}].

\bibitem{Biekotter:2021ovi}
T.~Biek\"otter and M.O.~Olea-Romacho, \emph{{Reconciling Higgs physics and
  pseudo-Nambu-Goldstone dark matter in the S2HDM using a genetic algorithm}},
  \href{https://doi.org/10.1007/JHEP10(2021)215}{\emph{JHEP} {\bfseries 10}
  (2021) 215} [\href{https://arxiv.org/abs/2108.10864}{{\ttfamily
  2108.10864}}].

\bibitem{CMS:2021sdq}
{\scshape CMS} collaboration, \emph{{Analysis of the $CP$ structure of the
  Yukawa coupling between the Higgs boson and $\tau$ leptons in proton-proton
  collisions at $ \sqrt{s} $ = 13 TeV}},
  \href{https://doi.org/10.1007/JHEP06(2022)012}{\emph{JHEP} {\bfseries 06}
  (2022) 012} [\href{https://arxiv.org/abs/2110.04836}{{\ttfamily
  2110.04836}}].

\bibitem{Aguilar-Saavedra:2021ijx}
J.A.~Aguilar-Saavedra, D.E.~L\'opez-Fogliani, C.~Mu\~noz and M.~Pierre,
  \emph{{WIMP dark matter in the U$\mu \nu$SSM}},
  \href{https://doi.org/10.1088/1475-7516/2022/05/004}{\emph{JCAP} {\bfseries
  05} (2022) 004} [\href{https://arxiv.org/abs/2111.07091}{{\ttfamily
  2111.07091}}].

\bibitem{Cline:2013gha}
J.M.~Cline, K.~Kainulainen, P.~Scott and C.~Weniger, \emph{{Update on scalar
  singlet dark matter}},
  \href{https://doi.org/10.1103/PhysRevD.88.055025}{\emph{Phys. Rev. D}
  {\bfseries 88} (2013) 055025}
  [\href{https://arxiv.org/abs/1306.4710}{{\ttfamily 1306.4710}}].

\bibitem{Lebedev:2011iq}
O.~Lebedev, H.M.~Lee and Y.~Mambrini, \emph{{Vector Higgs-portal dark matter
  and the invisible Higgs}},
  \href{https://doi.org/10.1016/j.physletb.2012.01.029}{\emph{Phys. Lett. B}
  {\bfseries 707} (2012) 570}
  [\href{https://arxiv.org/abs/1111.4482}{{\ttfamily 1111.4482}}].

\bibitem{Aalbers:2016jon}
{\scshape DARWIN} collaboration, \emph{{DARWIN: towards the ultimate dark
  matter detector}},
  \href{https://doi.org/10.1088/1475-7516/2016/11/017}{\emph{JCAP} {\bfseries
  1611} (2016) 017} [\href{https://arxiv.org/abs/1606.07001}{{\ttfamily
  1606.07001}}].

\bibitem{Billard:2013qya}
J.~Billard, L.~Strigari and E.~Figueroa-Feliciano, \emph{{Implication of
  neutrino backgrounds on the reach of next generation dark matter direct
  detection experiments}},
  \href{https://doi.org/10.1103/PhysRevD.89.023524}{\emph{Phys. Rev. D}
  {\bfseries 89} (2014) 023524}
  [\href{https://arxiv.org/abs/1307.5458}{{\ttfamily 1307.5458}}].

\bibitem{DelNobile:2013sia}
M.~Cirelli, E.~Del~Nobile and P.~Panci, \emph{{Tools for model-independent
  bounds in direct dark matter searches}},
  \href{https://doi.org/10.1088/1475-7516/2013/10/019}{\emph{JCAP} {\bfseries
  1310} (2013) 019} [\href{https://arxiv.org/abs/1307.5955}{{\ttfamily
  1307.5955}}].

\bibitem{Alloul:2013bka}
A.~Alloul, N.D.~Christensen, C.~Degrande, C.~Duhr and B.~Fuks, \emph{{FeynRules
  2.0 - A complete toolbox for tree-level phenomenology}},
  \href{https://doi.org/10.1016/j.cpc.2014.04.012}{\emph{Comput. Phys. Commun.}
  {\bfseries 185} (2014) 2250}
  [\href{https://arxiv.org/abs/1310.1921}{{\ttfamily 1310.1921}}].

\bibitem{Belanger:2018ccd}
G.~B\'elanger, F.~Boudjema, A.~Goudelis, A.~Pukhov and B.~Zaldivar,
  \emph{{micrOMEGAs5.0 : Freeze-in}},
  \href{https://doi.org/10.1016/j.cpc.2018.04.027}{\emph{Comput. Phys. Commun.}
  {\bfseries 231} (2018) 173}
  [\href{https://arxiv.org/abs/1801.03509}{{\ttfamily 1801.03509}}].

\bibitem{Belanger:2013oya}
G.~Belanger, F.~Boudjema, A.~Pukhov and A.~Semenov, \emph{{micrOMEGAs-3: A
  program for calculating dark matter observables}},
  \href{https://doi.org/10.1016/j.cpc.2013.10.016}{\emph{Comput. Phys. Commun.}
  {\bfseries 185} (2014) 960}
  [\href{https://arxiv.org/abs/1305.0237}{{\ttfamily 1305.0237}}].

\bibitem{Freitas:2015hsa}
A.~Freitas, S.~Westhoff and J.~Zupan, \emph{{Integrating in the Higgs Portal to
  Fermion Dark Matter}},
  \href{https://doi.org/10.1007/JHEP09(2015)015}{\emph{JHEP} {\bfseries 09}
  (2015) 015} [\href{https://arxiv.org/abs/1506.04149}{{\ttfamily
  1506.04149}}].

\bibitem{Bechtle:2008jh}
P.~Bechtle, O.~Brein, S.~Heinemeyer, G.~Weiglein and K.E.~Williams,
  \emph{{HiggsBounds: Confronting Arbitrary Higgs Sectors with Exclusion Bounds
  from LEP and the Tevatron}},
  \href{https://doi.org/10.1016/j.cpc.2009.09.003}{\emph{Comput. Phys. Commun.}
  {\bfseries 181} (2010) 138}
  [\href{https://arxiv.org/abs/0811.4169}{{\ttfamily 0811.4169}}].

\bibitem{Bechtle:2011sb}
P.~Bechtle, O.~Brein, S.~Heinemeyer, G.~Weiglein and K.E.~Williams,
  \emph{{HiggsBounds 2.0.0: Confronting Neutral and Charged Higgs Sector
  Predictions with Exclusion Bounds from LEP and the Tevatron}},
  \href{https://doi.org/10.1016/j.cpc.2011.07.015}{\emph{Comput. Phys. Commun.}
  {\bfseries 182} (2011) 2605}
  [\href{https://arxiv.org/abs/1102.1898}{{\ttfamily 1102.1898}}].

\bibitem{Bechtle:2013wla}
P.~Bechtle, O.~Brein, S.~Heinemeyer, O.~Stal, T.~Stefaniak, G.~Weiglein et~al.,
  \emph{{$\mathsf{HiggsBounds}-4$: Improved Tests of Extended Higgs Sectors
  against Exclusion Bounds from LEP, the Tevatron and the LHC}},
  \href{https://doi.org/10.1140/epjc/s10052-013-2693-2}{\emph{Eur. Phys. J. C}
  {\bfseries 74} (2014) 2693}
  [\href{https://arxiv.org/abs/1311.0055}{{\ttfamily 1311.0055}}].

\bibitem{Bechtle:2020pkv}
P.~Bechtle, D.~Dercks, S.~Heinemeyer, T.~Klingl, T.~Stefaniak, G.~Weiglein
  et~al., \emph{{HiggsBounds-5: Testing Higgs Sectors in the LHC 13 TeV Era}},
  \href{https://doi.org/10.1140/epjc/s10052-020-08557-9}{\emph{Eur. Phys. J. C}
  {\bfseries 80} (2020) 1211}
  [\href{https://arxiv.org/abs/2006.06007}{{\ttfamily 2006.06007}}].

\bibitem{ATLAS:2021hbr}
{\scshape ATLAS} collaboration, \emph{{Search for Higgs boson decays into a
  pair of pseudoscalar particles in the $bb\mu\mu$ final state with the ATLAS
  detector in $pp$ collisions at $\sqrt s$=13\,\,TeV}},
  \href{https://doi.org/10.1103/PhysRevD.105.012006}{\emph{Phys. Rev. D}
  {\bfseries 105} (2022) 012006}
  [\href{https://arxiv.org/abs/2110.00313}{{\ttfamily 2110.00313}}].

\bibitem{OPAL:2007qwz}
{\scshape OPAL} collaboration, \emph{{Search for invisibly decaying Higgs
  bosons in e+ e- ---\ensuremath{>} Z0 h0 production at s**(1/2) = 183-GeV -
  209-GeV}}, \href{https://doi.org/10.1016/j.physletb.2009.09.010}{\emph{Phys.
  Lett. B} {\bfseries 682} (2010) 381}
  [\href{https://arxiv.org/abs/0707.0373}{{\ttfamily 0707.0373}}].

\bibitem{LEPWorkingGroupforHiggsbosonsearches:2003ing}
{\scshape LEP Working Group for Higgs boson searches, ALEPH, DELPHI, L3, OPAL}
  collaboration, \emph{{Search for the standard model Higgs boson at LEP}},
  \href{https://doi.org/10.1016/S0370-2693(03)00614-2}{\emph{Phys. Lett. B}
  {\bfseries 565} (2003) 61}
  [\href{https://arxiv.org/abs/hep-ex/0306033}{{\ttfamily hep-ex/0306033}}].

\bibitem{L3:2004svb}
{\scshape L3} collaboration, \emph{{Search for an invisibly-decaying Higgs
  boson at LEP}},
  \href{https://doi.org/10.1016/j.physletb.2005.01.030}{\emph{Phys. Lett. B}
  {\bfseries 609} (2005) 35}
  [\href{https://arxiv.org/abs/hep-ex/0501033}{{\ttfamily hep-ex/0501033}}].

\bibitem{Weinberg:2013kea}
S.~Weinberg, \emph{{Goldstone Bosons as Fractional Cosmic Neutrinos}},
  \href{https://doi.org/10.1103/PhysRevLett.110.241301}{\emph{Phys. Rev. Lett.}
  {\bfseries 110} (2013) 241301}
  [\href{https://arxiv.org/abs/1305.1971}{{\ttfamily 1305.1971}}].

\bibitem{Fernandez-Martinez:2021ypo}
E.~Fernandez-Martinez, M.~Pierre, E.~Pinsard and S.~Rosauro-Alcaraz,
  \emph{{Inverse Seesaw, dark matter and the Hubble tension}},
  \href{https://doi.org/10.1140/epjc/s10052-021-09760-y}{\emph{Eur. Phys. J. C}
  {\bfseries 81} (2021) 954}
  [\href{https://arxiv.org/abs/2106.05298}{{\ttfamily 2106.05298}}].

\bibitem{Dawson:2017jja}
S.~Dawson and M.~Sullivan, \emph{{Enhanced di-Higgs boson production in the
  complex Higgs singlet model}},
  \href{https://doi.org/10.1103/PhysRevD.97.015022}{\emph{Phys. Rev. D}
  {\bfseries 97} (2018) 015022}
  [\href{https://arxiv.org/abs/1711.06683}{{\ttfamily 1711.06683}}].

\bibitem{Lee:1973iz}
T.D.~Lee, \emph{{A Theory of Spontaneous T Violation}},
  \href{https://doi.org/10.1103/PhysRevD.8.1226}{\emph{Phys. Rev. D} {\bfseries
  8} (1973) 1226}.

\bibitem{Kim:1979if}
J.E.~Kim, \emph{{Weak Interaction Singlet and Strong CP Invariance}},
  \href{https://doi.org/10.1103/PhysRevLett.43.103}{\emph{Phys. Rev. Lett.}
  {\bfseries 43} (1979) 103}.

\bibitem{Branco:2011iw}
G.C.~Branco, P.M.~Ferreira, L.~Lavoura, M.N.~Rebelo, M.~Sher and J.P.~Silva,
  \emph{{Theory and phenomenology of two-Higgs-doublet models}},
  \href{https://doi.org/10.1016/j.physrep.2012.02.002}{\emph{Phys. Rept.}
  {\bfseries 516} (2012) 1} [\href{https://arxiv.org/abs/1106.0034}{{\ttfamily
  1106.0034}}].

\bibitem{Grzadkowski:2009iz}
B.~Grzadkowski and P.~Osland, \emph{{Tempered Two-Higgs-Doublet Model}},
  \href{https://doi.org/10.1103/PhysRevD.82.125026}{\emph{Phys. Rev. D}
  {\bfseries 82} (2010) 125026}
  [\href{https://arxiv.org/abs/0910.4068}{{\ttfamily 0910.4068}}].

\bibitem{Biekotter:2022bxp}
T.~Biek\"otter, P.~Gabriel, M.O.~Olea-Romacho and R.~Santos, \emph{{Direct
  detection of pseudo-Nambu-Goldstone dark matter in a two Higgs doublet plus
  singlet extension of the SM}},
  \href{https://doi.org/10.1007/JHEP10(2022)126}{\emph{JHEP} {\bfseries 10}
  (2022) 126} [\href{https://arxiv.org/abs/2207.04973}{{\ttfamily
  2207.04973}}].

\bibitem{Bauer:2017ota}
M.~Bauer, U.~Haisch and F.~Kahlhoefer, \emph{{Simplified dark matter models
  with two Higgs doublets: I. Pseudoscalar mediators}},
  \href{https://doi.org/10.1007/JHEP05(2017)138}{\emph{JHEP} {\bfseries 05}
  (2017) 138} [\href{https://arxiv.org/abs/1701.07427}{{\ttfamily
  1701.07427}}].

\bibitem{Falkowski:2013dza}
A.~Falkowski, F.~Riva and A.~Urbano, \emph{{Higgs at last}},
  \href{https://doi.org/10.1007/JHEP11(2013)111}{\emph{JHEP} {\bfseries 11}
  (2013) 111} [\href{https://arxiv.org/abs/1303.1812}{{\ttfamily 1303.1812}}].

\bibitem{Ferreira:2014naa}
P.M.~Ferreira, J.F.~Gunion, H.E.~Haber and R.~Santos, \emph{{Probing wrong-sign
  Yukawa couplings at the LHC and a future linear collider}},
  \href{https://doi.org/10.1103/PhysRevD.89.115003}{\emph{Phys. Rev. D}
  {\bfseries 89} (2014) 115003}
  [\href{https://arxiv.org/abs/1403.4736}{{\ttfamily 1403.4736}}].

\bibitem{deSalas:2020hbh}
P.F.~de~Salas and A.~Widmark, \emph{{Dark matter local density determination:
  recent observations and future prospects}},
  \href{https://doi.org/10.1088/1361-6633/ac24e7}{\emph{Rept. Prog. Phys.}
  {\bfseries 84} (2021) 104901}
  [\href{https://arxiv.org/abs/2012.11477}{{\ttfamily 2012.11477}}].

\end{thebibliography}\endgroup

\end{document}